\PassOptionsToPackage{dvipsnames}{xcolor}
\documentclass{lmcs} % use [draft] to see the table of contents
\pdfoutput=1

\usepackage{lastpage}
\lmcsdoi{22}{4}{16}
\lmcsheading{}{\pageref{LastPage}}{}{}%
{May~13,~2025}{Nov.~05,~2026}{}

\pdfoutput=1

\usepackage[T1]{fontenc}
\usepackage[utf8]{inputenc}

\usepackage[
  breaklinks=true,
  final, % prevents the disappearance of hyperlinks in draft mode
]{hyperref}
\usepackage[all]{hypcap}

\usepackage{ifdraft}

\usepackage[style=alphabetic,natbib=true,maxalphanames=4,maxbibnames=9]{biblatex}
\newbibmacro{string+url}[1]{\iffieldundef{url}{#1}{\href{\thefield{url}}{#1}}}
\DeclareFieldFormat{title}{\usebibmacro{string+url}{\mkbibemph{#1}}}
\DeclareFieldFormat[article,inproceedings,unpublished,incollection]{title}{\usebibmacro{string+url}{\mkbibquote{#1}}}

\usepackage{cleveref}
\usepackage{moreverb}
\usepackage{xspace}
\input{ushyphex.tex}

\ifdraft{%
\usepackage{todonotes}
}{%
\usepackage[disable]{todonotes}
}

\usepackage{graphicx}
\usepackage{subcaption}
\usepackage{tikz}
\usetikzlibrary{angles, calc, decorations.pathreplacing}

\usepackage[scaled=0.85]{beramono}

\newcommand{\colorocaml}{\color{green!8}}
\newcommand{\colorcoq}{\color{yellow!16}}
\newcommand{\colorcoqbroken}{\color{red!8}}

\usepackage[final]{listings}
\newcommand{\ocamlcommentstyle}{\color{blue}}
\newcommand{\ocamldoccommentstyle}{\color{BlueGreen}}

\lstdefinelanguage{ocaml}[Objective]{Caml}{
  deletekeywords={closed,ref,parser},
  morekeywords={initializer,effect,perform,continue,discontinue},
  flexiblecolumns=false,
  showstringspaces=false,
  framesep=5pt,
  commentstyle=\ocamlcommentstyle,
  basicstyle=\ttfamily\small,
  numberstyle=\footnotesize,
  mathescape=true,
  escapeinside={(*@}{*)},
  rangeprefix=(*\ ,% opening comment plus space
  rangesuffix=\ *),% space plus closing comment
  morecomment=[s][\ocamldoccommentstyle]{(**}{*)},
}

\def\oc|#1|{\text{\lstinline[language=ocaml,basicstyle=\ttfamily,flexiblecolumns=true]|#1|}}
\def\ocplus+#1+{\text{\lstinline[language=ocaml,basicstyle=\ttfamily,flexiblecolumns=true]+#1+}}

\def\qoc|#1|{``\oc|#1|''}

\newcommand{\coqcommentstyle}{\color{blue}}

\lstdefinelanguage{coq}{
  mathescape=true,
  texcl=false,
  morekeywords=[1]{Section, Module, End, Require, Import, Export,
    Variable, Variables, Parameter, Parameters, Axiom, Hypothesis,
    Hypotheses, Notation, Local, Tactic, Reserved, Scope, Open, Close,
    Bind, Delimit, Definition, Equations, Let, Ltac, Fixpoint, CoFixpoint, Add,
    Morphism, Relation, Implicit, Arguments, Unset, Contextual,
    Strict, Prenex, Implicits, Inductive, CoInductive, Record,
    Structure, Canonical, Coercion, Context, Class, Global, Instance,
    Program, Infix, Theorem, Lemma, Corollary, Proposition, Fact, Fail,
    Remark, Example, Proof, Goal, Save, Qed, Defined, Hint, Resolve,
    Rewrite, View, Search, Show, Print, Printing, All, Eval, Check,
    Projections, inside, outside, Def, Obligation, Next},
  morekeywords=[2]{forall, exists, exists2, fun, fix, cofix, struct, by,
    match, with, end, as, in, return, let, if, is, then, else, for, of,
    nosimpl, when},
  morekeywords=[3]{Type, Prop, SProp, Set, true, false, option},
  morekeywords=[4]{pose, set, move, case, elim, apply, clear, hnf,
    intro, intros, generalize, rename, pattern, after, destruct,
    induction, using, refine, inversion, injection, rewrite, congr,
    unlock, compute, ring, field, fourier, replace, fold, unfold,
    change, cutrewrite, simpl, have, suff, wlog, suffices, without,
    loss, nat_norm, assert, cut, trivial, revert, bool_congr, nat_congr,
    symmetry, transitivity, auto, split, autorewrite},
  morekeywords=[5]{by, done, exact, reflexivity, tauto, romega, omega,
    assumption, solve, contradiction, discriminate},
  morekeywords=[6]{do, last, first, try, idtac, repeat},
  morecomment=[s]{(*}{*)},
  flexiblecolumns=false,
  showstringspaces=false,
  framesep=5pt,
  morestring=[b]",
  morestring=[d],
  tabsize=4,
  extendedchars=true,
  sensitive=true,
  breaklines=false,
  basicstyle=\ttfamily\small,
  captionpos=b,
  identifierstyle={\ttfamily\color{black}},
  keywordstyle=[1]{\ttfamily\bfseries\color{black}},
  keywordstyle=[2]{\ttfamily\bfseries\color{black}},
  keywordstyle=[3]{\ttfamily\color{black}},
  keywordstyle=[4]{\ttfamily\color{VioletRed}},
  keywordstyle=[5]{\ttfamily\color{black}},
  keywordstyle=[6]{\ttfamily\color{black}},
  stringstyle=\ttfamily,
  commentstyle={\ttfamily\coqcommentstyle},
}[keywords,comments,strings]

\lstnewenvironment{coq}{\lstset{language=coq}}{}

\def\coqe{\lstinline[language=coq, basicstyle=\ttfamily]}
\def\coqes{\lstinline[language=coq, basicstyle=\ttfamily\small]}

\let\origfootnote\footnote
\renewcommand{\footnote}[1]{\origfootnote{\let\coqe\coqes #1}}
\makeatletter
\def\@makefnmark{%
  \leavevmode
  \raise.9ex\hbox{%
    \begingroup\normalfont\normalsize\check@mathfonts\endgroup
    \fontsize\sf@size\z@\normalfont\@thefnmark}}
\makeatother

\newcommand{\sref}[1]{\cref{#1}} % LMCS recommends \autoref but tolerates \cref
\newcommand{\srefs}[2]{Sections~\ref{#1} and~\ref{#2}}
\newcommand{\fref}[1]{\cref{#1}}
\newcommand{\lineref}[1]{line~\ref{line:#1}}
\newcommand{\linerefs}[2]{lines~\ref{line:#1} and~\ref{line:#2}}

\def\etal.{\emph{et al.}}

\newcommand{\kt}{Kaplan and Tarjan\xspace}

\newcommand{\Coq}{Rocq\xspace}
\newcommand{\Equations}{\emph{Equations}\xspace}
\newcommand{\Hammer}{\emph{Hammer}\xspace}
\newcommand{\AACTactics}{\emph{AAC\_tactics}\xspace}
\newcommand{\Bush}{\textit{Bush}\xspace}

\newcommand{\Sdeques}{Deques\xspace}
\newcommand{\sdeque}{deque\xspace}
\newcommand{\sdeques}{deques\xspace}

\newcommand{\greenorred}{green or red\xspace}

\newcommand{\opush}{\emph{push}\xspace}
\newcommand{\opop}{\emph{pop}\xspace}
\newcommand{\oinject}{\emph{inject}\xspace}
\newcommand{\oeject}{\emph{eject}\xspace}
\newcommand{\oconcat}{\emph{concat}\xspace}

\newcommand{\semiopen}[2]{[#1, #2)}
\newcommand{\pow}[1]{2^{#1}}

\newcommand{\treerep}{tree representation\xspace}
\newcommand{\Treerep}{Tree representation\xspace}
\newcommand{\pointerrep}{pointer representation\xspace}
\newcommand{\Pointerrep}{Pointer representation\xspace}

\newcommand{\indexed}[1]{\index{#1}#1}
\newcommand{\emphindexed}[1]{\index{#1}\emph{#1}}

\newcommand{\gitrepo}{https://github.com/JulesViennotFranca/VerifiedCatenableDeque/}
\RequirePackage{fontawesome5}
\newcommand{\gitrepoGlobeURL}[1]{\href{\gitrepo #1}{\faGlobe}}
\newcommand{\gitrepoGlobeFile}[1]{\gitrepoGlobeURL{blob/main/#1}}
\newcommand{\gitrepoGlobeFileInMargin}[1]{\marginpar{\centering\gitrepoGlobeFile{#1}}}
\newcommand{\gitrepofile}[1]{\gitrepoGlobeFileInMargin{#1}}

\newcommand{\natnum}{Natural numbers\xspace}
\newcommand{\natnumocaml}{\natnum (OCaml)}
\newcommand{\natnumcoq}{\natnum (\Coq)}

\newcommand{\ncd}{Deques\xspace}
\newcommand{\ncdocaml}{\ncd (OCaml)}
\newcommand{\ncdcoq}{\ncd (\Coq)}

\newcommand{\cd}{Cadeques\xspace}
\newcommand{\cdocaml}{\cd (OCaml)}
\newcommand{\cdcoq}{\cd (\Coq)}

\newcommand{\listnil}{[]}
\newcommand{\listsingleton}[1]{[#1]}
\newcommand{\listcat}{\mathop{++}}

\newcommand{\xs}{\mathit{xs}}
\newcommand{\ys}{\mathit{ys}}

\newcommand{\bnum}[1]{\overline{#1}}

\newcommand{\successor}{\mathnormal{S}}
\newcommand{\incr}[1]{\successor \; (#1)}
\newcommand{\arrowsucc}{\stackrel{\smash\successor}{\rightarrow}}

\definecolor{myviolet}{rgb}{0.75, 0.2, 0.95}
\definecolor{myblue}{rgb}{0.1, 0.3, 0.9}
\definecolor{mygreen}{rgb}{0.0, 0.8, 0.35}
\definecolor{myyellow}{rgb}{1.0, 0.8, 0.3}
\definecolor{myorange}{rgb}{1.0, 0.55, 0.15}
\definecolor{myred}{rgb}{1.0, 0.25, 0.3}
\definecolor{mygray}{rgb}{0.9, 0.9, 0.9}

\newcommand{\bgreendot}{\textcolor{mygreen}{
  \raisebox{-0.45mm}{\scalebox{1.5}{$\bullet$}}
}}
\newcommand{\byellowdot}{\textcolor{myyellow}{
  \raisebox{-0.45mm}{\scalebox{1.5}{$\bullet$}}
}}
\newcommand{\yellowdot}{\textcolor{myyellow}{\bullet}}
\newcommand{\breddot}{\textcolor{myred}{
  \raisebox{-0.45mm}{\scalebox{1.5}{$\bullet$}}
}}

\newcommand{\drawleftbracket}[4][black]{
  \draw [draw=#1] (#2 + 0.3 * #4, #3 - 0.8 * #4) -- (#2, #3 - 0.8 * #4) --
        (#2, #3 + 0.8 * #4) -- (#2 + 0.3 * #4, #3 + 0.8 * #4); }

\newcommand{\drawrightbracket}[4][black]{
  \draw [draw=#1] (#2 - 0.3 * #4, #3 - 0.8 * #4) -- (#2, #3 - 0.8 * #4) --
        (#2, #3 + 0.8 * #4) -- (#2 - 0.3 * #4, #3 + 0.8 * #4); }

\newcommand{\drawtriple}[4][black]{
  \draw [draw=#1] (#2 - 4.25 * #4, #3 - 0.6 * #4) --
                  (#2 - 0.5 * #4, #3 - 0.6 * #4);
  \draw [draw=#1] (#2 + 0.5 * #4, #3 - 0.6 * #4) --
                  (#2 + 4.25 * #4, #3 - 0.6 * #4);}

\newcommand{\drawsdeque}[4][black]{
  \drawleftbracket[#1]{#2 - 5 * #4}{#3}{#4}
  \drawtriple[#1]{#2}{#3}{#4}
  \drawrightbracket[#1]{#2 + 5 * #4}{#3}{#4}}

\newcommand{\drawsending}[4][black]{
  \drawleftbracket[#1]{#2 - 2.625 * #4}{#3}{#4}
  \draw [draw=#1] (#2 - 1.875 * #4, #3 - 0.6 * #4) --
                  (#2 + 1.875 * #4, #3 - 0.6 * #4);
  \drawrightbracket[#1]{#2 + 2.625 * #4}{#3}{#4}}

\newcommand{\Cdeques}{Cadeques\xspace}
\newcommand{\cdeque}{cadeque\xspace}
\newcommand{\cdeques}{cadeques\xspace}

\usetikzlibrary{calc}

\newcommand{\tlpkt}[4][]{
  \draw[#1] (#2) ++(-0.707*#3, 0.707*#3) arc (135:-45:#3);
  \draw[#1] (#2) ++(-0.707*#3, 0.707*#3) -- +(-0.5*#4, -0.5*#4);
  \draw[#1] (#2) ++(0.707*#3, -0.707*#3) -- +(-0.5*#4, -0.5*#4);}

\newcommand{\tcpkt}[4][]{
  \draw[#1] (#2) ++(#3, 0) arc (0:180:#3);
  \draw[#1] (#2) ++(#3, 0) -- +(0, -0.5*#4);
  \draw[#1] (#2) ++(-#3, 0) -- +(0, -0.5*#4);}

\newcommand{\trpkt}[4][]{
  \draw[#1] (#2) ++(0.707*#3, 0.707*#3) arc (45:225:#3);
  \draw[#1] (#2) ++(0.707*#3, 0.707*#3) -- +(0.5*#4, -0.5*#4);
  \draw[#1] (#2) ++(-0.707*#3, -0.707*#3) -- +(0.5*#4, -0.5*#4);}

\newcommand{\tbpkt}[4][]{
  \draw[#1] (#2) circle (#3);}

\newcommand{\lbpkt}[4][]{
  \draw[#1] (#2) ++(-0.707*#3, -0.707*#3) arc (-135:45:#3);
  \draw[#1] (#2) ++(-0.707*#3, -0.707*#3) -- +(-0.5*#4, 0.5*#4);
  \draw[#1] (#2) ++(0.707*#3, 0.707*#3) -- +(-0.5*#4, 0.5*#4);}

\newcommand{\cbpkt}[4][]{
  \draw[#1] (#2) ++(-#3, 0) arc (180:360:#3);
  \draw[#1] (#2) ++(-#3, 0) -- +(0, 0.5*#4);
  \draw[#1] (#2) ++(+#3, 0) -- +(0, 0.5*#4);}

\newcommand{\rbpkt}[4][]{
  \draw[#1] (#2) ++(-0.707*#3, 0.707*#3) arc (135:315:#3);
  \draw[#1] (#2) ++(-0.707*#3, 0.707*#3) -- +(0.5*#4, 0.5*#4);
  \draw[#1] (#2) ++(0.707*#3, -0.707*#3) -- +(0.5*#4, 0.5*#4);}

\newcommand{\lrpkt}[4][]{
  \draw[#1] (#2) ++(0.707*#3-0.5*#4, 0.707*#3+0.5*#4) -- +(#4, -#4);
  \draw[#1] (#2) ++(-0.707*#3-0.5*#4, -0.707*#3+0.5*#4) -- +(#4, -#4);}

\newcommand{\llpkt}[4][]{
  \draw[#1] (#2) ++(0.707*#3-0.5*#4, 0.707*#3+0.5*#4) -- +(0.5*#4, -0.5*#4);
  \draw[#1] (#2) ++(0.707*#3, 0.707*#3) arc (45:-45:#3);
  \draw[#1] (#2) ++(0.707*#3, -0.707*#3) -- +(-0.5*#4, -0.5*#4);
  \draw[#1] (#2) ++(-0.707*#3-0.5*#4, -0.707*#3+0.5*#4)
             -- +(-0.707*#3+0.5*#4, 0.707*#3-0.5*#4) -- +(0, 1.414*#3-#4);}

\newcommand{\rlpkt}[4][]{
  \draw[#1] (#2) ++(-0.707*#3+0.5*#4, 0.707*#3+0.5*#4) -- +(-#4, -#4);
  \draw[#1] (#2) ++(0.707*#3+0.5*#4, -0.707*#3+0.5*#4) -- +(-#4, -#4);}

\newcommand{\rcpkt}[4][]{
  \draw[#1] (#2) ++(-0.707*#3+0.5*#4, 0.707*#3+0.5*#4) -- +(-0.5*#4, -0.5*#4);
  \draw[#1] (#2) ++(-#3, 0) arc (180:135:#3);
  \draw[#1] (#2) ++(-#3, 0) -- +(0, -0.5*#4);
  \draw[#1] (#2) ++(#3, -0.5*#4) -- +(0, -0.414*#3+0.5*#4)
                             -- +(-0.293*#3+0.5*#4, -0.707*#3+#4);}

\newcommand{\rrpkt}[4][]{
  \draw[#1] (#2) ++(-0.707*#3+0.5*#4, 0.707*#3+0.5*#4) -- +(-0.5*#4, -0.5*#4);
  \draw[#1] (#2) ++(-0.707*#3, 0.707*#3) arc (135:225:#3);
  \draw[#1] (#2) ++(-0.707*#3, -0.707*#3) -- +(0.5*#4, -0.5*#4);
  \draw[#1] (#2) ++(0.707*#3+0.5*#4, -0.707*#3+0.5*#4)
             -- +(0.707*#3-0.5*#4, 0.707*#3-0.5*#4) -- +(0, 1.414*#3-#4);}

\begin{document}

\title{Verified Purely Functional Catenable Real-Time Deques}

% Authors.

\author[J.~Viennot]{Jules Viennot\lmcsorcid{0009-0007-4401-9108}}[a]

\author[A.~Wendling]{Arthur Wendling}[b]

\author[A.~Guéneau]{Armaël Guéneau\lmcsorcid{0000-0003-3072-4045}}[c]

\author[F.~Pottier]{François Pottier\lmcsorcid{0000-0002-4069-1235}}[a]

% Institutions and emails.

\address{Inria Paris, France}
\email{jules.viennot@inria.fr, francois.pottier@inria.fr}

\address{Tarides, France}
\email{art.wendling@gmail.com}

\address{Université Paris-Saclay, CNRS, ENS Paris-Saclay, Inria, LMF, France}
\email{armael.gueneau@inria.fr}

\begin{abstract}
We present OCaml and \Coq implementations of \kt's
purely functional, real-time catenable deques.
The correctness of our \Coq code is machine-checked.

\end{abstract}

\maketitle

\ifdraft{%
  \setcounter{tocdepth}{3}
  \tableofcontents
}{}

% ------------------------------------------------------------------------------

\section{Introduction}

% ------------------------------------------------------------------------------

% Describe what \kt achieve.

Twenty-five years ago, \citet{kaplan-tarjan-99} established a striking result:
there exist
``purely functional, real-time deques with catenation''.
In other words,
there exists a~data structure that enjoys the following properties:
\begin{itemize}
\item This data structure represents a sequence of elements.
\item It supports \opush and \opop,
      which insert and extract one element at the front end, \\
      and \oinject and \oeject,
      which insert and extract one element at the rear end. \\
      In other words, it is a \emph{deque},
      a double-ended queue.
      % The word "deque" is due to Knuth (volume 1 of TAOCP).
\item It supports concatenation, \oconcat.
\item It is immutable,
      therefore persistent: \\
      none of the five operations modifies or destroys its argument.
\item Each of the five operations has worst-case time complexity $O(1)$.
\end{itemize}

% ------------------------------------------------------------------------------

% Describe what we achieve, in a short way (greater punch to say it now).

% Fix Indentation
\noindent
In this paper, we present \emph{the first implementation}
of \kt's catenable deques.
This implementation is expressed in the purely functional subset
of the OCaml programming language \cite{ocaml}.
Furthermore, we present \emph{the first verified implementation}
of \kt's catenable deques.
This implementation is expressed in \Coq \cite{rocq}.
Its correctness is stated and verified
within \Coq;
in other words, it is machine-checked.

% ------------------------------------------------------------------------------

% Avoid a possible misunderstanding,
% which I believe caused reviewer 2 to be disappointed.

To avoid any misunderstanding,
let us immediately point out
the main limitations of our results.
First, our OCaml code has been thoroughly tested, but is not verified. There
is no formal connection between our OCaml implementation and our verified \Coq
implementation. Although the latter is inspired by the former, there are some
differences between them, because one uses OCaml's generalized algebraic data
types (GADTs) while the other uses \Coq's indexed inductive families.
Second, although our OCaml code achieves worst-case time complexity $O(1)$ for
each of the five operations, our \Coq code currently does not. Indeed, \Coq
currently does not allow indices (which in our case are colors, levels, and
sizes) to be erased. Therefore, these indices, which ideally should be used
only at type-checking time to verify that the code is well-typed, are computed
and stored also at runtime. This is true both when our code is interpreted by
\Coq and when it is ``extracted'', that is, translated to OCaml code.
This is a~limitation of our work but can also be understood as a~limitation
of \Coq, which hopefully will be removed in the future.
We discuss these limitations, and possible workarounds, later on in this
introduction (\srefs{sec:intro:coq}{sec:intro:alternative}) and at the end of
the paper (\sref{sec:eval}).

% ------------------------------------------------------------------------------

% Explain the three structures in their paper.

\subsection{\kt's paper}

\kt proceed in several steps.
Section~4 of their paper presents a~\emph{deque},
which does not support \oconcat;
section~5 presents a~\emph{catenable steque},
which supports \oconcat but not \oeject;
section~6 presents the \emph{catenable deque},
which supports all five operations.
The catenable deque in section~6
uses the deque of section~4
as a~building block.
% (\S6.1, ``each buffer is a noncatenable deque'')
\begin{comment}
therefore,
structuring the code in two modules (deques; catenable deques)
is both pedagogical
and good engineering practice.
\end{comment}
%
The catenable steque of section~5,
on the other hand,
is~ultimately unused;
its main purpose seems pedagogical.

% ------------------------------------------------------------------------------

% Give some more detail about the way in which each data structure is described.
% Introduce the idea that for each structure,
% there is both a high-level view
% and a low-level view (with direct pointers).

% LATER should we show the diamond diagram here?

\newcommand{\footnoteterminology}{\footnote{%
In a purely functional programming language, every data structure is
``pointer-based'', that is, represented in memory
using memory blocks and pointers. In our OCaml and \Coq implementations,
\kt's ``\pointerrep'' is just an inductive data structure.}}
% I was going to write that every data structure is "tree-structured", but
% this is not quite true: because of sharing, the data structure can be a DAG
% in memory.

At each step,
\kt
give a high-level English description of the data structure,
and explain its design.
They typically begin with a sketch of
the general shape of
the data structure,
which forms a certain kind of~tree.
Then, they restrict this shape
by imposing several invariants.
%
% Each data structure is typically subject to two invariants.
%
One invariant imposes
bounds on the sizes
of certain ``buffers''.
Another invariant,
the \emph{regularity} invariant,
requires a~certain \emph{coloring} scheme to be respected,
where colors are dictated by buffer sizes.
We refer to this high-level description
as the \emph{\treerep}.
(\kt do not introduce a specific name for it.)

Although the \treerep enables
a broad understanding of the data structure
and is pedagogically helpful,
it~does~not allow achieving
% the desired time complexity, namely
worst-case time complexity~$O(1)$.
To~achieve this complexity,
direct pointers to subtrees % substructures
with certain color characteristics
are needed.
Thus,
a lower-level description,
which reflects how the data structure
must actually be laid~out in memory,
is needed.
\kt refer to it % this low-level description
as the~\emph{pointer representation}.\footnoteterminology
%
% Intentionally not using the macro \pointerrep,
% since this is Kaplan & Tarjan's terminology; it is fixed.
%
\begin{comment}
For example, in the case of non-catenable deques, on page 585, \kt write:
``For this reason, we do not represent a deque in the obvious way, as a stack
of prefix–suffix pairs. Instead, we break this stack up into substacks. [...]
An equivalent pointer-based representation is to use a node with four pointers
[...].'' The pointer representation of non-catenable deques is illustrated in
Figure~2 of \kt's paper.
\end{comment}

The \pointerrep
can be
significantly more complex
than the \treerep.
Indeed,
an~informal explanation
or a~formal~definition
of the \treerep
can be broken up in several stages:
first,
the shape of the data structure
is defined;
then
an~assignment of colors to nodes is introduced;
finally,
a~regularity invariant,
which imposes certain requirements
on the manner in which nodes are colored,
restricts the~shape of the data structure.
In~the~\pointerrep,
on the other hand,
the shape of the data structure,
the color assignment,
and the regularity invariant
influence each other:
therefore, they must be defined simultaneously.
This difficulty is not apparent in \kt's paper
because they use informal language.
% one could say that use an untyped informal language
%
However,
in a
% strongly
typed programming language,
such as OCaml or \Coq,
this difficulty is significant.
A large fraction of our paper is in fact devoted
to presenting and explaining our type definitions.

% ------------------------------------------------------------------------------

% \kt give correctness proofs (or proof sketches)
% but no type definitions and no code.

After describing each data structure and its two representations,
\kt devote their attention
to proving
that,
in the \pointerrep,
each operation \emph{can} be implemented.
That is,
they prove that
each operation preserves the regularity invariant,
which means that the result of this operation
can be expressed
in the \pointerrep.
This proof is expressed as English prose.
The fact
that each operation is correct
(that~is, produces a~data structure
that represents the desired sequence of elements)
and has worst-case time complexity $O(1)$
is not explicitly proved:
these checks are considered ``straightforward''.
%
\begin{comment}
\kt write: ``The only nontrivial part of the proof is to verify
that the regularization procedure is correct; it is then straightforward to
verify that each deque operation is performed correctly and that the time
bound is $O(1)$''.
\end{comment}
%
% Should we clarify whether *we* agree that these checks are straightforward?
% We believe that they can indeed be considered straightforward once the type
% definitions and the code have been made explicit, which is not the case in
% \kt's paper.

In summary,
although
\kt's result
represents a remarkable achievement,
their paper does not provide any type definitions or any code.
It is reasonable to assert that
a type definition is implicit in their description of each data structure
and that
an algorithm is implicit in their proof that the invariant \emph{can} be preserved by each operation.
However,
spelling out
these type definitions and this code
is extremely challenging---so much so that,
in the twenty-five years that have elapsed
since their paper appeared,
no implementation of
\kt's
catenable deques
has been published.
We are aware of several attempts,
which we discuss at the end of this paper (\sref{sec:related}).

% ------------------------------------------------------------------------------

% Applications.

\subsection{Applications}

The reader may wonder about the practical applications
of persistent catenable deques.
It seems difficult to provide compelling examples.
Persistent sequences of characters, or \emph{strings},
are obviously very useful, but a full-featured ``string'' data structure
must offer many operations that catenable deques do not support,
such as random access
(that~is, reading or updating the $i$\textsuperscript{th} element of a string)
or extracting substrings.
In~the~algorithms literature,
to the best of our knowledge,
only a~few papers need a persistent
catenable deque data structure.
One example is Demaine, Langerman, and Price's
confluently persistent tries~\cite{demaine-langerman-price-10}.
Even in situations where such a~data structure is needed,
instead of Kaplan and Tarjan's purely functional catenable deques,
one might prefer to use Kaplan, Okasaki, and Tarjan's
simple persistent catenable deques~\cite{kaplan-okasaki-tarjan-00},
which have internal mutable state,
and whose time complexity guarantee
is weaker,
as it is amortized over a~sequence of operations.
% and (we believe) is not valid in a concurrent setting.
%
These simpler deques have been implemented in OCaml
and verified using \Coq and Iris
by Ponsonnet and Pottier~\cite{ponsonnet-pottier-26}.

% ------------------------------------------------------------------------------

\subsection{This paper}

% This paper presents OCaml and \Coq implementations
% of deques and catenable deques.

% ------------------------------------------------------------------------------

% Explain how we rely on the OCaml type-checker.

\subsubsection{OCaml implementations}

Our OCaml implementations
take advantage of OCaml's strong type discipline
in several ways.
Where possible,
we exploit
generalized algebraic data types (GADTs)
to express data structure invariants:
for example,
we index several types with colors
and encode constraints on colors
within our type definitions.
The OCaml type-checker
is then able
to verify,
at compile time,
that the code respects these invariants.
Furthermore,
also at compile time,
it identifies provably dead branches in case analyses,
% (that is, branches where a contradiction arises)
and lets us omit these branches.

However,
the expressive power of GADTs is limited:
for example,
properties that involve size arithmetic
cannot easily be expressed.
In one such case (\sref{sec:cadeque:ocaml:buffers}),
we exploit
an~abstract type with a~phantom type parameter
to enforce a form of one-sided security:
inside the abstraction barrier,
runtime assertions of the form ``this cannot happen'' ensure
that the invariant is established;
outside of the abstraction barrier,
the OCaml type-checker
statically verifies
that the invariant is preserved.

% ------------------------------------------------------------------------------

% A remark on the absence of recursion in our OCaml code.

Quite remarkably,
our OCaml implementations
of deques and catenable deques
do not
involve
any recursive function definitions:
there is no occurrence of \oc|let rec| in the code.
%
% We used to have `let rec ensure_green` in lib/cadeque/core.ml.
% Jules eliminated it (at the cost of duplicating 4 lines of code).
%
Thus,
the fact that every operation has worst-case time complexity $O(1)$
is literally obvious.
This is rather mind-boggling:
although these data structures
are described by complex recursive types,
they
are updated
by non-recursive operations.
An intuitive explanation
for this phenomenon
is that,
by design,
these data structures
contain direct pointers
to the places
where work will be needed next.
(An analogous and much simpler example is an~implementation of
a~stack as a linked list. It is a recursive data structure. Yet, its \opush
and \opop operations are not recursive, because they access only the top of
the stack.)
This is not our contribution:
it is \kt's magic.
It is perhaps more clearly visible in our paper
than in theirs
because we offer executable code
whereas they
describe algorithms in prose.

% ------------------------------------------------------------------------------

% Explain how we rely on the Coq type-checker,
% and what challenges this involves.

\subsubsection{\Coq implementations}
\label{sec:intro:coq}

Our \Coq implementations
are inspired by our OCaml implementations,
but differ in substantial ways.
%
% Coq is more expressive:
%
On the one hand,
\Coq's type discipline offers much greater expressive power.
This lets us
statically
enforce invariants
and
omit dead branches;
thus
there is no need for
runtime assertions.
% or phantom types
%
% Coq is less expressive:
%
On the other hand,
to guarantee strong normalization,
\Coq
imposes restrictions that do not exist in OCaml.
In particular,
some of the
% (generalized)
algebraic data types
that we define in OCaml
cannot be literally ported to \Coq,
because
they are ``truly nested'' data types (\sref{sec:related:nested}),
which,
to the best of our knowledge,
no current proof assistant accepts.
In particular,
we believe that they are not supported by
\Coq, Agda, Lean, or Isabelle/HOL.
We work around this problem
by giving alternative type definitions,
which \Coq accepts.
These definitions carry extra ``level'' and ``size'' indices,
but have simpler recursive structure.

To state the correctness of our \Coq implementations,
a family of ``model'' functions is needed.
A model function
maps
a~data structure
to the mathematical sequence
of elements
that this data structure represents.
%
% Notational conventions for sequences:
We represent a mathematical sequence as a \Coq list.
We write
$\listnil$ for the empty list,
$\listsingleton{x}$ for a singleton list,
and $\xs\listcat\ys$ for the concatenation of the lists $\xs$ and $\ys$.
A~model function computes the \emph{fringe} of a~data structure.
Model functions
appear in statements of correctness.
% For example,
The correctness of a~\opush operation,
which inserts an element~$x$ into a~data structure~$s$,
and which is expected to insert~$x$ ``in front of'' the elements of~$s$,
is expressed by the equation
$\mathit{fringe}(\mathit{push}\;x\;s)=[x] \mathop{++} \mathit{fringe}(s)$.

Our \Coq implementation includes
the definition of a model function
for each data structure
and a statement of correctness
for each operation.
As claimed by \kt,
the proofs of these correctness statements
are ``straightforward''---a fairly simple matter
of proving equations that involve fringes,
empty lists, singleton lists, and
list concatenations.
Somewhat unexpectedly,
the main technical challenge that we encounter
lies not in these correctness proofs
but in the very definition of the model functions.
Indeed,
\Coq requires us to prove that these definitions,
which are recursive,
are in some sense well-founded.
Our first attempt is rejected;
fortunately,
we are able to find a reformulation that \Coq accepts.

Our work not only tests the limits of \Coq's expressiveness, but also stresses
the quality and robustness of its implementation. With Coq 8.19 and 8.20, it
took a very long time for the \Coq type-checker to accept some of our function
definitions. We now require \Coq~9.2, which we believe has received
improvements in the treatment of nested inductive types and in the guard
checker. We find it to be significantly faster: the time needed to check
our~project has come from roughly 10 minutes down to roughly 3~minutes.
% LATER revisit this with 9.3, once available (with aac-tactics and hammer)

% fp: I seemed to recall that several minutes are spent by Coq 8.20 on the
% definition of the main inductive types (\fref{fig:cdc_types}). However, I
% just checked again, and this is not the case; this definition is accepted
% immediately. Instead, it seems that a LOT of time (several minutes) is spent
% on Cadeque/Models.v and some time (between one and two minutes) is spent on
% Cadeque/Core.v.

% ------------------------------------------------------------------------------

% Summary of our findings about the Coq implementation.

In summary,
somewhat unexpectedly,
the main challenges
that we encounter and overcome
% in our \Coq implementation
are
formulating our inductive type definitions
and
the inductive definitions of our model functions
so that they are accepted by \Coq.
In other words, we find that
describing a~data structure
can be more difficult
than implementing operations on this data structure,
and that
stating a correctness property
can be more difficult
than proving this property!

% ------------------------------------------------------------------------------

% Discussion of complexity in Coq.

We do not prove that our \Coq implementation has worst-case time complexity
$O(1)$.
There are two main reasons why we do not do so.
First, such a claim cannot easily be stated,
because \Coq does not have a notion of computational cost.
Indeed,
as two computations that have the same result are deemed equal,
there is no way of distinguishing their costs.
Second, our \Coq implementation involves ``size'' indices,
which are natural numbers.
However, to the best of our knowledge,
\Coq does not allow the programmer to indicate
that indices are computationally irrelevant,
by which we mean that
indices can be erased at runtime
and that
index computations have no cost.
This issue is discussed in \sref{sec:eval:coq}.

% ------------------------------------------------------------------------------

% Discussion of using other styles or other proof assistants.

\subsubsection{Alternative approaches}
\label{sec:intro:alternative}

A~reader may wonder whether, by using rich type definitions, involving nested
algebraic data types as well as color, size, and level indices, we are making
life needlessly difficult for ourselves. An alternative approach, both in
OCaml and in \Coq, would be to define simpler, coarser types, where fewer
invariants are expressed a priori. Instead, the desired invariants would be
expressed a posteriori, by defining one or more well-formedness predicates,
and verified a posteriori.

Out of necessity,
such coarser types would contain junk: that~is, they would
have inhabitants that are not well-formed data structures. Therefore our code
would have to contain more dynamic tests and more dead branches where (in
OCaml) one places a~runtime assertion of the form ``this cannot happen'' or
(in \Coq) one returns an arbitrary inhabitant of the function's result type.
% In case no such inhabitant can be manufactured (because the~required type is
% abstract), the function's result type would have to be changed to an~option type.
% fp: I am removing this comment because it seems easy to assume that the
%     type A is inhabited. This adds one global type class constraint and
%     does not visibly pollute the code.
These changes in the code seem aesthetically unpleasant.
Furthermore, in~this approach,
it~is up to the~programmer
% while~writing the code,
to recognize which branches
are reachable (so~a~valid result must be returned)
and which branches are dead (so a~dummy result can be returned).
This can be quite difficult.

To some extent, this approach can be compared with working in an~untyped
language. The programmer cannot rely on static type-checking to eliminate
a~class of programming mistakes and to automatically recognize a~class of
dead branches; they must instead rely on their own discipline
and on runtime checks.
Thus, the question of which approach to prefer
is reminiscent of
the ``long and rather acrimonious debate''
between proponents of typed languages
and advocates of untyped languages
\cite{reynolds-85}.
In this case,
the debate is between simpler, weaker types
and richer, stronger types.

\citet[\S10.1.1]{okasaki-book-99}
discusses this question.
He offers the example of the nested data type
\ocplus+type 'a seq = Nil | Cons of 'a * ('a * 'a) seq+.
(We use OCaml syntax.)
Okasaki uses Standard ML,
where this data type can be defined,
but where many operations on this type cannot be defined,
because Standard ML does not have polymorphic recursion.
A~workaround is to collapse elements,
pairs of elements, pairs of pairs of elements, and~so~on,
into a~data type:
\ocplus+type 'a ep = Elem of 'a | Pair of 'a ep * 'a ep+.
This data type contains junk:
it contains not only perfectly balanced
pairs of pairs of {\ldots}\!\! of elements,
but also
arbitrary unbalanced binary trees.
Then, a coarse definition of \oc|'a seq|
as a~non-nested data type
can be given:
\ocplus+type 'a seq = Nil | Cons of 'a ep * 'a seq+.
Although this approach is workable,
Okasaki gives three reasons
for preferring the original nested data type:
elegance, % readability, conciseness,
efficiency,
and
static detection of programmer errors.
Therefore
% he prefers to use the richer~data type.
%
he pretends that Standard~ML supports polymorphic recursion,
and lets the reader translate to Standard~ML without polymorphic recursion,
if necessary.
A~modern reader is likely to prefer a~more expressive language,
such as Haskell, OCaml, or Scala. % or \Coq, Agda, Lean, etc.

In this paper,
we follow Okasaki's lead
and prefer to use richer, more accurate types.
During our first attempt at implementing
Kaplan and Tarjan's data structures,
we found the type-checker extremely helpful % an invaluable help
in detecting mistakes and dead branches.
% given the complexity of these data structures,

We do not claim that an alternative approach
based on simple (non-indexed) types is unworkable.
We do believe that it can work.
Certainly,
now that we have been able to implement Kaplan and Tarjan's data structures,
one could port our code to a simply-typed style.
However, we doubt that the simply-typed approach would
make verification significantly easier.
In fact,
prompted by one of the anonymous reviewers,
we have applied the simply-typed approach
to the~redundant binary numbers of \sref{sec:binary}
\gitrepofile{theory/BinCounting/CoreSimplyTyped.v}
and to the non-catenable deques of \sref{sec:deque}.
\gitrepofile{theory/Deque/DequeSimplyTyped.v}
(The globe icons in the margin are links to these files.)
We~find that the two approaches
require roughly the same number of lines of code.
According to the~tool \texttt{rocq wc},
the indexed-type approach involves longer definitions and statements
whereas the~simply-typed approach involves longer proofs.

Currently,
one~benefit of the simply-typed approach
over the indexed-type approach
lies in the~efficiency of the code that is obtained
by extraction of \Coq to OCaml (\sref{sec:eval:coq:extract}).
During extraction,
all proofs
% (of sort \coqe|Prop|)
are erased,
but indices (colors, levels, and sizes) remain,
requiring computations whose results are ultimately useless.
% The same is true also when computing inside \Coq.
That said, perhaps future versions of \Coq
will be able to erase indices (\sref{sec:lessons}).

% ------------------------------------------------------------------------------

% Explain the structure of the paper.

\subsubsection{Structure of this paper}

Following \kt,
we proceed in several stages.
First,
as a warm-up,
we present a~data structure
that represents a~natural number
in a~redundant binary numbering system,
and
we~implement this data structure
in OCaml and in \Coq
(\sref{sec:binary}).
Then, we present
\kt's deques,
and implement them
in~OCaml and in~\Coq
(\sref{sec:deque}).
Next, we move to
\kt's catenable deques.
We describe
the \treerep
and
the \pointerrep
of
this data structure
(\sref{sec:cadeque}).
We present our OCaml implementation
of it
(\sref{sec:cadeque:ocaml}).
We explain why
\Coq's restrictions on
inductive type definitions
% and
% inductive function definitions
prevent
a~direct port
of our OCaml code to \Coq
(\sref{sec:obstacles}).
%
% jv: only the type definitions of non-catenable deques are indexed
%     with levels AND sizes.
%     The inductive type definitions of catenable deques are
%     indexed with just a level.
%
Fortunately,
we are able to work around these restrictions
by introducing ``level'' and/or ``size'' indices
and by reformulating the definitions
of certain ``model'' functions.
Deques,
which serve as a building block
in the construction of catenable deques,
must be indexed with levels and sizes.
Therefore,
we revisit our \Coq implementation of deques
(\sref{sec:deque:coq:revisited}).
Then,
we formulate
inductive type definitions
for catenable deques,
indexed with levels,
which \Coq accepts
(\sref{sec:cadeque:coq}).
The paper ends
with
an~experimental evaluation
of the performance of our OCaml code,
a~discussion of
several ways
of executing our \Coq code
(\sref{sec:eval}),
and a~review of the related work
(\sref{sec:related}).

% ------------------------------------------------------------------------------

% Comments about our code.

\subsubsection{Structure of the code}
\label{sec:intro:code}

In this paper,
we show very little code.
Instead, we focus on the type definitions
that describe the structure of
\sdeques and catenable deques.
These type definitions, alone, are quite long and complex.
Once these type definitions are given,
the code is relatively straightforward.
% this is a benefit of the "rich type" approach
%
Our complete code is available online.
A~globe icon\marginpar{\gitrepoGlobeURL{}}
represents a~link
to our~repository
or to a~specific file in this repository.
In particular,
the file \texttt{Signatures.v}
\gitrepofile{theory/Signatures.v}
sums up the types and operations
that must be offered by an implementation of catenable deques,
as well as the correctness properties that these operations must satisfy;
and the file \texttt{Cadeque/Summary.v}
\gitrepofile{theory/Cadeque/Summary.v}
proves that we have implemented these operations
and established their correctness properties.

Although our code is split over several files, the paper is
meant to be understood without worrying about file names. As a~default rule,
the code that is shown in each section of the paper is self-contained and does
not refer to code that is shown in an earlier section.%
\footnote{Taking advantage of this convention, we allow distinct sections to
define distinct objects by the same name. For example, the OCaml type named
\oc|chain| in \sref{sec:binary}, defined in \fref{fig:bc_ocaml_types}, and
the OCaml type named \oc|chain| in \sref{sec:deque:ocaml}, defined in
\fref{fig:ncd_ocaml_types}, have nothing to do with one another.}
As an exception to this rule, because our implementations of catenable deques
depend on our implementations of non-catenable deques, we allow references
from the former to the latter.
These references are identified by the prefix \oc|Deque.|
For example,
% the module \oc|Buffer|,
% which is defined in \fref{fig:cd_ocaml_buffer}
% and is part of
our OCaml implementation of \cdeques (\sref{sec:cadeque:ocaml})
contains several references to the module \oc|Deque|,
which is our OCaml implementation of non-catenable deques
(\sref{sec:deque:ocaml}).
Similarly,
our \Coq implementation of catenable deques (\sref{sec:cadeque:coq})
contains several references to the module \oc|Deque|,
% see e.g. \fref{fig:ncdr_coq_types}
which is our \Coq implementation of non-catenable deques,
indexed with levels and sizes (\sref{sec:deque:coq:revisited}).

\section{Natural numbers: a redundant binary representation}
\label{sec:binary}

% ------------------------------------------------------------------------------

\subsection{Concept}
\label{sec:binary:concept}

To explain the fundamental mechanisms at play
in their representations of \emph{sequences},
\citet{kaplan-tarjan-99} point out a connection with
representations of natural \emph{numbers}.
This connection is made clear in \sref{sec:deque},
where non-catenable deques are introduced.
%
% Kaplan and Tarjan note that redundant number representations
% and applications can be found in several papers
% \citep{brodal-96,clancy-knuth-77,kaplan-tarjan-96}.
%
Through this lens,
there is a loose analogy
between
the operation of inserting an~element into a sequence
and
the operation of incrementing a number.
%
% fp: a loose analogy, not a perfect one;
%     the correspondence is not one-to-one as one might imagine.
%     Inserting an element into a sequence can be trivial
%     (if a certain buffer is not full, or sufficiently far from full).
%     It is nontrivial, and requires incrementing a number,
%     only this buffer is full or too close to being full.
%
Therefore, Kaplan and Tarjan ask: which representations of natural numbers
support constant-time incrementation?

In the following, let us decorate the binary representation of a number with
an overline, and let us write the least significant digit on the left. Thus,
for example, the decimal number~8 equals the binary number $\bnum{0001}$, and
the decimal number~94 equals the binary number $\bnum{0111101}$. The
incrementation operation, or \emph{successor}, is written $\successor$.

In the well-known \emph{binary} representation, where a digit is a bit (that
is, either~0 or~1), incrementing a number can require as little as a single
bit flip, in the best case; yet, due to carry propagation, it can also require
several bit flips. For example, incrementing 94 requires just one bit flip:
$\incr{\bnum{0111101}}$ is $\bnum{1111101}$; whereas incrementing 95
requires six bit flips: $\incr{\bnum{1111101}}$ is $\bnum{0000011}$.
In general, incrementing the binary representation of the number~$n$
has worst-case time complexity $O(\log n)$.
% It has amortized time complexity $O(1)$.

To guarantee constant-time incrementation, it is necessary to ensure that only
a constant number of digits are affected by incrementation.
\kt note that this is achieved by Clancy and Knuth's
\emphindexed{redundant binary representation} (\indexed{RBR})
\citep{clancy-knuth-77},
together with a~\emph{regularity} invariant.

In this representation, a natural number is still represented in base 2, but
there are \emph{three} possible digits, namely~0,~1, and~2.
For example,
$\bnum{021}$ is a representation
of the number $0 \times 2^0 + 2 \times 2^1 + 1 \times 2^2$,
which is~8.
A~number can have multiple representations: for~instance, 94 can
be written under the form $\bnum{0111101}$, $\bnum{011112}$, or $\bnum{222221}$.
Clearly, some representations are better than others. A representation that
begins with the~digit~0 or~1 can be incremented in constant time, by changing
the first digit: for example, $\incr{\bnum{1111101}}$ is $\bnum{2111101}$.
A~representation that begins with the digit~2 requires more work: for example,
$\incr{\bnum{222221}}$ is $\bnum{111112}$.
%
% Whereas \citet{clancy-knuth-77} propose a constant-time incrementation
% algorithm without explicitly defining
% which representations are problematic,
%
Kaplan and Tarjan identify \emph{regularity}, a property of RBRs that
enables constant-time incrementation. Incrementing a regular representation is
easy, and can be done in constant time. However, this operation can produce an
irregular result. Thus, a way of transforming such a result into a regular
representation, in constant time, is also needed.
% fp: I initially thought that Kaplan and Tarjan would propose an algorithm
%     that turns an *arbitrary* representation in a regular one,
%     but clearly this cannot be done in constant time.
%     It requires scanning the whole representation.
% fp: in fact, regularization turns a semiregular number into a regular one.

Kaplan and Tarjan define
regularity\index{regularity!natural numbers}
as follows.%
\footnote{\kt's redundant binary representation,
which involves the digits 0, 1, and~2, is reminiscent
of skew binary numbers~\cite{myers-83} \cite[\S9.3]{okasaki-book-99}.
However, their notion of regularity differs from
the canonical form that is imposed on skew binary numbers,
where only the lowest non-zero digit may be~2.}
Let us write
(a~representation~of) a~number as a list of digits
$d_0 d_1 \ldots d_n$ where $d_i \in \{0, 1, 2 \}$.
Such a representation is \emph{regular}
if
for every $j$ such that $d_j = 2$,
there exists an $i < j$
such that $d_i = 0$ and,
for every $k \in (i, j)$, $d_k = 1$.
This means that while scanning the list of digits
from right to left
(that is, from most significant to least significant digit),
after one has encountered the digit~2,
one must encounter the digit~0
before one encounters another 2
or runs out of digits.
For example,
the representations
$\bnum{0111101}$ and $\bnum{011112}$,
both of which denote the number 94,
are regular;
% and they are the only regular representations of 94
whereas $\bnum{222221}$ is not.

Clearly, if $d_0 d_1 \ldots d_n$ is regular,
then $d_0 \neq 2$.
That is,
the leftmost digit
of a regular representation cannot be~2.
Therefore, a regular representation can be incremented in constant time
just by incrementing its leftmost digit.
Naturally, this can break regularity.
For example,
incrementing $\bnum{011112}$ yields $\bnum{111112}$,
which is not regular,
as a right-to-left scan encounters the digit~2
but never encounters the digit~0.

Kaplan and Tarjan propose a simple way of transforming the result of an
incrementation, in constant time, so as to guarantee that it is regular.
They scan the digits from left to right,
looking for the least significant digit $d_i$ that is not~1.
If $d_i$ is~0,
then there is nothing to do; this representation is already regular.
If $d_i$ is 2,
then they set $d_i$ to~0 and increment the next digit, $d_{i+1}$.
% (If there is no next digit, imagine that the next digit is 0.)
% If $d_{i+1}$ is 0 then this yields a regular result.
% If $d_{i+1}$ is 1, then it becomes 2,
% and this may seem dangerous,
% as there could be another 2 further to the right;
% but this can be dangerous only if have a pattern of the form 1* 2 1* 2,
% which is impossible.
% (I am not sure I got this right; it is a bit tricky.)
%
We refer to this transformation as \emph{regularization}\index{regularization!natural numbers}.

For example, applying regularization to $\bnum{111112}$ yields $\bnum{1111101}$,
which is regular, as desired.
Incrementing $\bnum{1111101}$ yields $\bnum{2111101}$,
which is irregular:
regularizing it yields $\bnum{0211101}$,
which is regular.
In summary,
writing $\successor$ for the combination
of incrementation and regularization,
we have the following sequence of operations:
\[\begin{array}{ccccc}
  \bnum{011112}
  & \arrowsucc &
  \bnum{1111101}
  & \arrowsucc &
  \bnum{0211101}
\end{array}\]
%

% Fix Indentation
\noindent
It is not entirely obvious why regularization always produces a regular
result! We verify this fact by implementing regularization in \Coq in
\sref{sec:binary:coq}.

Kaplan and Tarjan note that regularization ``changes only a constant number of
digits, thus avoiding explicit carry propagation''. This suggests that
regularization can be performed in constant time. A~difficulty, however, is
that regularization requires scanning the digits from left to right, looking
for the least significant digit $d_i$ that is not~1. Thus, a way of skipping a
group of ``1'' digits, in constant time, is required.
The simple-minded representation of a~number as a linked list of digits,
which one may call the \emph{\treerep},
is not suitable.
Instead, a more elaborate representation,
the \emph{\pointerrep},
must be introduced.

% (Kaplan and Tarjan explain at the level of deques,
%  but not at the simpler level of numbers.)

In the \pointerrep,
consecutive ``1'' digits must be arranged into groups,
allowing a~group to be skipped in constant time.
To describe this data structure, let us introduce some terminology.
Following \kt, numeric digits are replaced with colors:
the digit~0 becomes the \emph{green} digit;
the digit~1 becomes the \emph{yellow} digit;
the digit~2 becomes the \emph{red} digit.
We refer to green and red digits, along with the very first digit $d_0$,
as \emph{heads}.\index{head!natural numbers}
Following a head, a series of consecutive yellow digits forms
a~\emph{body}.\index{body!natural numbers}
A~head and its body constitute a \emph{packet}.\index{packet!natural numbers}
A number,
or \emph{chain},\index{chain!natural numbers}
is a list of packets.
In other words,
a~chain is inductively defined as either
the empty chain
or a packet followed by a~chain.
Searching for the next non-``1'' digit
amounts to moving to the next packet:
this representation
allows this operation to be performed
in constant time.

Let us depict a digit as a colored dot,
whose color is green, yellow, or red.
We use a~large dot if the digit is a head,
and a small dot otherwise,
so every packet begins with a large dot.
We separate two consecutive packets with some space.
Thus, the numbers that we have used in the previous examples
are depicted as follows:
\bgroup
\renewcommand{\arraystretch}{1.5}
\[\begin{array}{ccccc}
  \bnum{011112}
  & \arrowsucc &
  \bnum{1111101}
  & \arrowsucc &
  \bnum{0211101} \\
  \hspace*{3pt}
  \bgreendot\yellowdot\yellowdot\yellowdot\yellowdot\;\; \breddot
  & \arrowsucc &
  \hspace*{3pt}
  \byellowdot\yellowdot\yellowdot\yellowdot\yellowdot\;\; \bgreendot\yellowdot
  & \arrowsucc &
  \hspace*{3pt}
  \bgreendot\;\; \breddot\yellowdot\yellowdot\yellowdot\;\; \bgreendot\yellowdot
\end{array}\]
\egroup

%Fix Indentation
\noindent
We assign colors not only to digits, but also to packets and chains.
A packet takes on the color of its head.
A chain takes on the color of its first packet.
(The empty chain is green.)
We remark that a head can be yellow only if it is the first digit, $d_0$.
Thus, if a chain contains a yellow packet,
then this must be the first packet.
Using these conventions,
regularity can be defined as a set of simple constraints on colors,
shown in \fref{fig:bc_reg}.

% Clarifications:
Because the color of a~chain
is the color of its first packet,
the constraint
``a green packet must be followed with a \greenorred chain''~\ref{bc_reg:green}
could also be written
``a green packet must be followed with a \greenorred packet''.
A~similar remark can be made about constraints~\ref{bc_reg:yellow}
and~\ref{bc_reg:red}.
Similarly, \ref{bc_reg:top} could be written under the form
``the top packet must be green or yellow''.
By the ``top packet'', we mean the root packet,
or, in the above picture, the leftmost packet.

\begin{figure}[t]
\begin{enumerate}[label=(R\arabic*)]
  \item \label{bc_reg:green}
    A green packet must be followed with a \greenorred chain.
  \item \label{bc_reg:yellow}
    A yellow packet must be followed with a green chain.
  \item \label{bc_reg:red}
    A red packet must be followed with a green chain.
  \item \label{bc_reg:top}
    The top chain must be green or yellow.
\end{enumerate}
\caption{\natnum: regularity constraints}
\label{fig:bc_reg}
\end{figure}

% ------------------------------------------------------------------------------

\subsection{OCaml implementation}
\label{sec:binary:ocaml}

We now describe the redundant binary representation of natural numbers,
together with the regularity invariant, as OCaml type definitions. Then
we~implement incrementation and regularization as OCaml functions.

So far,
% (\sref{sec:binary:concept}),
we have described a packet as a pair of a head and a body, where a~body is a~possibly
empty list of yellow digits. So, a packet was never empty, and its color was
green, yellow, or red. Here, we prefer to merge ``body'' and ``packet'' into a
single type. So, a packet is a~possibly empty list of digits, and the empty
packet must be viewed as uncolored.

% fp: commented out, because I prefer to avoid showing a dead end.
%     instead, we will explain in a positive way why we choose to
%     index our types with colors and impose the invariant a priori.
\begin{comment}
One could describe the representation of natural numbers,
\emph{without} the regularity invariant, using simple types.
For example,
one could define
a~``yellow digit'' as a unit value,
a~``body'' as a list of yellow digits,
a~``head'' as a green, yellow, or red digit,
and
a~``packet'' as a pair of a head and a body.
%
% \begin{figure}[ht]
% \begin{lstlisting}[language=ocaml]
% type body =
%   | Hole : body
%   | YDigit : body -> body
%
% type packet =
%   | Green_head  : body -> packet
%   | Yellow_head : body -> packet
%   | Red_head    : body -> packet
% \end{lstlisting}
% \caption{OCaml simple packet implementation}
% \label{fig:bc_ocaml_simple_pkt}
% \end{figure}
%
However, in so doing, one would miss a key opportunity: the OCaml type-checker
would be unable to statically check that the regularity invariant is
maintained, and would be unable to exploit this invariant to statically verify
that certain branches in the code are dead.
\end{comment}

As discussed in the introduction (\sref{sec:intro:alternative}),
we wish to build the regularity constraints of \fref{fig:bc_reg} into our type
definitions, so as to let the OCaml type-checker statically verify that the
regularity invariant is maintained and that certain branches in the code are
dead.
An~alternative approach,
which is to use simpler and coarser types,
is discussed in \sref{sec:binary:coq}.

For this purpose, we must encode color information in our type definitions.
The type of packets must be indexed with a color, so we can talk about packets
of type \oc|green packet|, \oc|yellow packet|, and \oc|red packet|.
Furthermore, we must express our regularity constraints,
some of which involve disjunctions: for example, a green
packet must be followed with ``a~\greenorred chain'' \ref{bc_reg:green}.
%
% It might seem natural to define \oc|red|, \oc|yellow| and
% \oc|green| at the type level as three distinct types.
% But such an approach would not allow us to encode disjunctions.
%
This is somewhat challenging, as the OCaml type system is based on the notion
of type equality, and does not offer a natural way of expressing disjunction.
To encode a~disjunction of colors,
or a~set of colors,
at the type level,
we can imagine two approaches.
One approach relies on OCaml's polymorphic variants,%
\footnote{For documentation on polymorphic variants,
please see
\href{https://ocaml.org/manual/5.2/polyvariant.html}{the OCaml manual}
\citep{ocaml}
or the book
\href{https://dev.realworldocaml.org/variants.html\#polymorphic-variants}{Real World OCaml}
\citep{real-world-ocaml}.}
whose static type discipline can express certain
superset and subset constraints.
However, polymorphic variants are a rather uncommon and advanced feature
of OCaml, and do not have a~counterpart in \Coq's type theory.
The second approach,
which we prefer,
involves encoding colors using 3~bits:
we use one bit per hue,
where a hue is green, yellow, or~red.
Thus, the color ``green''  is encoded as 100,
the color ``yellow'' as 010,
and the color ``red''  as 001.
The absence of a color can be encoded as 000.
The disjunction ``green or~red or~uncolored'' can be understood as ``not yellow''
and can be encoded as ?0?, where ``?'' is a wildcard,
that~is,
an~anonymous type variable.
These ideas result in the code
of \fref{fig:bc_ocaml_gyr}.
\gitrepofile{lib/color_GYR.ml}

Building on this encoding of colors, we define a type of packets that is
indexed with a~color. In short, the type \oc|'c packet| is the type of
a~packet whose color is \oc|'c|. Thus, for example, a~packet of type \oc|red
packet| must be a~red packet, and a packet of type \oc|uncolored packet| must
be an~uncolored packet.
Furthermore, by taking advantage of type variables,
it is possible to describe disjunctions of colors.
In particular,
as suggested above,
the constraint ``\greenorred or uncolored'',
which is synonymous with ``not yellow'',
can be encoded as \oc|_ * noyellow * _|.
Thus, a packet of type
\oc|(_ * noyellow * _) packet|
must be \greenorred or uncolored.
Similarly, the constraint ``yellow or uncolored'',
which is synonymous with ``not~green and not red'',
can be encoded by \oc|nogreen * _ * nored|.
Thus, a packet of type
\oc|(nogreen * _ * nored) packet|
must be yellow or uncolored.
This constraint appears in our definition
of the type \oc|packet|.

\begin{figure}[t]
\begin{lstlisting}[language=ocaml,backgroundcolor=\colorocaml]
(* A group of distinct types encode the presence or absence of a hue. *)
type somegreen  = SOME_GREEN
type nogreen    = NO_GREEN
type someyellow = SOME_YELLOW
type noyellow   = NO_YELLOW
type somered    = SOME_RED
type nored      = NO_RED
(* A color is a triple of three hue bits. *)
type green      =  somegreen *   noyellow *   nored
type yellow     =    nogreen * someyellow *   nored
type red        =    nogreen *   noyellow * somered
type uncolored  =    nogreen *   noyellow *   nored
\end{lstlisting}
\caption{\natnumocaml: type-level colors}
\label{fig:bc_ocaml_gyr}
\end{figure}

\begin{figure}[t]
\begin{lstlisting}[language=ocaml,backgroundcolor=\colorocaml]
type 'c packet =
  | Hole   :                              uncolored packet
  | GDigit : (nogreen * _ * nored) packet -> green  packet
  | YDigit : (nogreen * _ * nored) packet -> yellow packet
  | RDigit : (nogreen * _ * nored) packet -> red    packet
                        (* in each case, next packet must be yellow or uncolored *)

type (_, _) regularity =             (* parameters: packet color and chain color *)
  | G : (green , _ * noyellow * _) regularity (* next chain must be green or red *)(*@\label{line:G}*)
  | Y : (yellow,            green) regularity
  | R : (red   ,            green) regularity

type 'c chain =
  | Empty :                                                 green chain
  | Chain : ('c1, 'c2) regularity * 'c1 packet * 'c2 chain -> 'c1 chain

type number =
  | T : (_ * _ * nored) chain -> number   (* a number is a chain that is not red *)
\end{lstlisting}
\caption{\natnumocaml: types for packets, chains, and numbers}
\label{fig:bc_ocaml_types}
\end{figure}

The definitions of the types \oc|packet|, \oc|chain|, and \oc|number|
appear in \fref{fig:bc_ocaml_types}.
They are generalized algebraic data types (GADTs)~\citep{xi-chen-chen-03}.

The type \oc|packet| defines a packet as a list of colored digits. There is a
constructor for each kind of digit (green, yellow, and red) and a constructor
for the empty packet, also known as a \emph{hole}.
An empty packet is uncolored; a nonempty packet takes on the color of its
first digit.
Only the head of a packet can be a green digit or a red digit; every
subsequent digit must be yellow. This is expressed by letting the
constructors \oc|GDigit|, \oc|YDigit|, and \oc|RDigit|
require a yellow or uncolored packet.

The type \oc|regularity| defines the relationship between the color of a
packet and the color of the subsequent chain. The three constructors that
appear in its definition reflect the regularity constraints
\ref{bc_reg:green}--\ref{bc_reg:red}.
% in \fref{fig:bc_reg}.

The type \oc|chain| defines a chain as a list of packets. The empty chain is
green. In a~nonempty chain, the color \oc|'c1| of the packet and the
color \oc|'c2| of the subsequent chain must obey a regularity constraint.
A look at the definition of the type \oc|regularity| shows that the color
\oc|'c1| must be be green, yellow, or red: it cannot be uncolored.
The color of a~nonempty chain is the color of its first packet.
As a result, a chain is always green, yellow, or red; it cannot be uncolored.
This explains our comment on line~\ref{line:G}.
Although the constraint \oc|_ * noyellow * _|
may seem to allow the next chain to be ``green or red or uncolored'',
we~know that a chain cannot be uncolored.
Therefore we write that ``the next chain must be green or red''.

The type \oc|number| defines a number as a chain that is not red. Therefore, a
number is a~green or yellow chain.
This encodes the regularity constraint \ref{bc_reg:top}.
% in \fref{fig:bc_reg}.

\begin{figure}[t]
\begin{lstlisting}[language=ocaml,backgroundcolor=\colorocaml]
(* Turn a red chain into a green chain. *)
let green_of_red : red chain -> green chain = function
  | Chain (R, RDigit Hole, Empty) ->                          (*@ \label{green_of_red_pat1} *)
      Chain (G, GDigit (YDigit Hole), Empty)          (*@ \label{green_of_red_res1} *)
  | Chain (R, RDigit Hole, Chain (G, GDigit body, c)) -> (*@ \label{green_of_red_pat2} *)
      Chain (G, GDigit (YDigit body), c)              (*@ \label{green_of_red_res2} *)
  | Chain (R, RDigit (YDigit body), c) ->               (*@ \label{green_of_red_pat3} *)
      Chain (G, GDigit Hole, Chain (R, RDigit body, c))  (*@ \label{green_of_red_res3} *)

(* Turn a green or red chain into a green chain. *)
let ensure_green : type g r. (g * noyellow * r) chain -> green chain = fun c ->
  match c with
  | Empty           -> Empty
  | Chain (G, _, _) -> c
  | Chain (R, _, _) -> green_of_red c

(* Increment a number and perform regularization. *)
let succ : number -> number = function
  | T Empty ->
      T (Chain (Y, YDigit Hole, Empty))
  | T (Chain (G, GDigit body, c)) ->
      T (Chain (Y, YDigit body, ensure_green c))
  | T (Chain (Y, YDigit body, c)) ->
      T (green_of_red (Chain (R, RDigit body, c)))
\end{lstlisting}
\caption{\natnumocaml: regularization and incrementation}
\label{fig:bc_ocaml_func}
\end{figure}

Based on these type definitions, regularization and incrementation can be
implemented as OCaml functions.
\gitrepofile{lib/binCounting.ml}
These functions appear in \fref{fig:bc_ocaml_func}.

The function \oc|green\_of\_red| converts a red chain to a green one
in the way described by Kaplan and Tarjan (\sref{sec:binary:concept}),
that is,
by turning the first digit (which must be red) into a~green one
and by incrementing the following digit.
Three cases arise,
as the following digit might be absent, green, or yellow.
The first case
(lines~\ref{green_of_red_pat1}--\ref{green_of_red_res1})
is straightforward.
In the second case
(lines~\ref{green_of_red_pat2}--\ref{green_of_red_res2}),
two packets are merged into one.
The OCaml type-checker verifies that
\oc|GDigit (YDigit body)|
is a well-formed green packet
and that the chain~\oc|c|
is allowed to follow a green packet.
In the third case
(lines~\ref{green_of_red_pat3}--\ref{green_of_red_res3}),
one packet is split into two.
The OCaml type-checker verifies that
\oc|GDigit Hole| and \oc|RDigit body|
are well-formed packets
and that the chain~\oc|c|
is allowed to follow a red packet.

We are fortunate that the type-checker performs these checks
(so we cannot inadvertently break regularity) and that it performs
them silently for us (so the presence of these checks does not
clutter the code).
Furthermore, the OCaml type-checker verifies that all cases are covered:
that is, the patterns on lines
\ref{green_of_red_pat1},
\ref{green_of_red_pat2},
and
\ref{green_of_red_pat3}
cover all possible forms of red chains.
%
% fp: we could explain, in English, why this is true here.
%     maybe this is not useful.
%
In other words,
the type-checker verifies that all omitted cases are dead (contradictory).
This is also an invaluable feature:
here and especially in the forthcoming sections of this paper,
it would be extremely difficult
(in fact, we dare say, almost impossible)
to ensure coverage
without mechanical help.

The function \oc|ensure_green| transforms a \greenorred chain into a green
one. Its type involves a universal quantification on the type variables \oc|g|
and \oc|r|. Unfortunately, wildcards, which would be slightly more readable,
cannot be used here.
The function \oc|succ| increments a number and performs regularization, if
necessary, to maintain regularity.

It is worth noting that none of the functions in \fref{fig:bc_ocaml_func} is
recursive: no \oc|let rec| keyword appears. Therefore, the worst-case time
complexity of these functions is constant. The same striking remark holds
about the more complex data structures presented in this paper: \emph{no
recursive functions} appear in the implementation of these data structures.

% ------------------------------------------------------------------------------

\subsection{\Coq implementation}
\label{sec:binary:coq}

We now repeat the exercise of the previous section,
this time inside \Coq. That is, we encode the redundant binary representation
of natural numbers and the regularity invariant as \Coq type definitions,
and we implement incrementation and regularization as \Coq functions.

We begin by encoding colors in \Coq.
\gitrepofile{theory/Color/GYR.v}
A direct transliteration of the encoding that was~used in the previous
section (\sref{sec:binary:ocaml}) would lead us to define
a hue bit (``some green'', ``no~green'', and so on)
as an object of type \oc|Type|.
A color, which is a tuple of three hue bits,
would then have type \coqe|Type * Type * Type|.
However, that would be unnatural and needlessly complex. Because \Coq has true
dependent types, there is no need for hue bits to be encoded as objects of
type \coqe|Type|.
Instead, we define each of the three hues, separately,
as an inductive type with two constructors (\fref{fig:bc_coq_gyr}).
Then, we define a color as a combination of three hue~bits,
namely, a green bit, a yellow bit, and a red bit.
Instead of a tuple, we use an inductive type
with one constructor, \coqe|Mix|.

% fp: Why not just define \coqe|color| as an inductive type with 4 cases,
% ``uncolored or green or yellow or red'', and use propositions to describe our
% constraints on colors? Is there still something to be gained by defining a
% color as a triple of three hue bits? I suppose that the answer is positive: by
% encoding color constraints as equations, we allow \Equations to automatically
% identify some branches as dead (by unification), whereas otherwise we would
% have to explicitly prove that these branches are dead (by reasoning). Besides,
% as argued elsewhere, we need our color constraints (that is, the type
% \coqe|regularity|) to inhabit \coqe|Type|, because we sometimes perform a case
% analysis on this type. For now, we avoid these questions, and take it for
% granted that our \Coq encoding of colors should have the same structure as our
% OCaml encoding of colors.

\begin{figure}[tp]
\begin{lstlisting}[language=coq,backgroundcolor=\colorcoq]
(* Each of the three hues is an inductive type with two constructors. *)
Inductive green_hue  := SomeGreen  | NoGreen.
Inductive yellow_hue := SomeYellow | NoYellow.
Inductive red_hue    := SomeRed    | NoRed.

(* A color is a triple of three hue bits. *)
Inductive color      := Mix : green_hue -> yellow_hue -> red_hue -> color.
Notation green       := (Mix SomeGreen   NoYellow   NoRed).
Notation yellow      := (Mix   NoGreen SomeYellow   NoRed).
Notation red         := (Mix   NoGreen   NoYellow SomeRed).
Notation uncolored   := (Mix   NoGreen   NoYellow   NoRed).
\end{lstlisting}
\caption{\natnumcoq: types for colors}
\label{fig:bc_coq_gyr}
\end{figure}

\begin{figure}[tp]
\begin{lstlisting}[language=coq,backgroundcolor=\colorcoq]
Inductive packet : color -> Type :=
  | Hole       : packet uncolored
  | GDigit {y} : packet (Mix NoGreen y NoRed) -> packet green
  | YDigit {y} : packet (Mix NoGreen y NoRed) -> packet yellow
  | RDigit {y} : packet (Mix NoGreen y NoRed) -> packet red.
         (* next packet must be yellow or uncolored *)

Inductive regularity : color -> color -> Type :=(*@\label{line:regularity:Type}*)
  | G {g r} : regularity green  (Mix g NoYellow r)
                                (* next chain must be green or red *)
  | Y       : regularity yellow green
  | R       : regularity red    green.

Inductive chain : color -> Type :=
  | Empty                                                              : chain green
  | Chain {C1 C2 : color} : regularity C1 C2 -> packet C1 -> chain C2 -> chain C1.

Inductive number : Type :=
  | T {g y} : chain (Mix g y NoRed) -> number.
                    (* a number is a chain that is not red *)
\end{lstlisting}
\caption{\natnumcoq: types for packets, chains, and numbers}
\label{fig:bc_coq_types}
\end{figure}

Then, we define packets, regularity, chains, and numbers, in \Coq,
as shown in \fref{fig:bc_coq_types}.
These definitions are isomorphic to our earlier OCaml definitions
(\fref{fig:bc_ocaml_types}),
but colors now have type \coqe|color|
instead of being encoded as types.
We let the type \oc|regularity| inhabit the sort \coqe|Type|,
as opposed to \coqe|Prop|
(line~\ref{line:regularity:Type}).
This means that (during the extraction of \Coq to OCaml)
a~regularity witness cannot be erased;
it must exist at runtime.
In return,
this allows us
to perform case analyses on regularity witnesses,
not just in proofs,
but also in our~code.
Although we currently exploit this ability,
we believe that, with some extra effort,
we could avoid depending on~it,
thereby allowing regularity witnesses to be erased.
% LATER it would be nice to make this improvement

As discussed in the introduction (\sref{sec:intro:alternative}),
we choose to define packets, chains, and numbers as indexed types, and we
build a~regularity requirement into the definitions of these types. In other
words, we impose regularity \emph{a priori}: a data structure that violates
regularity cannot be constructed.

An~alternative approach
is to use simpler, coarser types,
that is,
to define packets, chains,
and numbers as simple (non-indexed) inductive types, and to impose regularity
\emph{a~posteriori}, as a property of packets, chains, and numbers.
We have explored this approach,
but do not describe it in the paper.
(For the reader's benefit, a~link is provided.)
\gitrepofile{theory/BinCounting/CoreSimplyTyped.v}
In terms of overall conciseness,
we find that neither approach seems
significantly superior to the other.
As a~subjective judgment, we find the use of indexed types more elegant and
comfortable than the use of simple types together with well-formedness
predicates.
One downside of using indexed types is that \Coq currently does not allow
indices to be erased at runtime. We come back to this point in
\sref{sec:eval}.

In either approach,
we find it extremely helpful
% essential
to set things up so that
(by exploiting regularity)
\Coq
% either the type-checker, or Equations
is able
to identify certain branches as dead
and lets us omit dead branches.
%
% fp: see CoreSimplyTyped.v: if regularity is imposed a posteriori (via well-formedness predicates)
%     then one can still use Equations and write pre- and postconditions; but Equations does not
%     allow us to omit dead branches. (This seems to be an accidental limitation of Equations.)
%
Furthermore,
we find it essential
to set things up so that
\Coq
does not require us to write a~case analysis
whose structure
mimics the structure of the code.
To achieve these two goals,
we use \Coq's \Equations facility
\citep{sozeau-mangin-19,equations}.
\Equations is so named because it lets the user
write a~function definition
in the form of a~set of equations.
However, this is not its most important aspect.
More crucially,
it allows pattern matching on indexed types,
allows dead branches to be omitted,%
\footnote{As far as we can tell, as of today, \Equations can ``see'' that
a~branch is dead, and allows this~branch to be omitted, only if an
unsatisfiable equation between indices appears in this branch. We heavily rely
on this feature. If one can prove that a~branch is dead via some other
argument then this branch must still be provided. Its body can be a~wildcard
\coqe|_|. This gives rise to a proof obligation, whose proof can begin with
\texttt{exfalso}. The user is then expected to provide a~proof that this
branch is dead.}
and is able to extract ``proof obligations'' out of each equation
separately,
saving the user the trouble of writing a~case analysis
that mimics the structure of the equations.

% \begin{figure}[ht]
% \begin{lstlisting}[language=coq]
% Definition green_of_red (c : chain red) : chain green :=
%   match c with
%   | Chain R (RDigit Hole) Empty =>
%       Chain G (GDigit (YDigit Hole)) Empty
%   | Chain R (RDigit Hole) (Chain G (GDigit body) c) =>
%       Chain G (GDigit (YDigit body)) c
%   | Chain R (RDigit (YDigit body)) c =>
%       Chain G (GDigit Hole) (Chain R (RDigit body) c)
%   end.
% (* Error : Non exhaustive pattern-matching:
%            no clause found for pattern Chain G _ _ *)
% \end{lstlisting}
% \caption{\Coq first \texttt{green\_of\_red} implementation}
% \label{fig:bc_coq_fst_gor}
% \end{figure}

\begin{figure}[t]
\begin{lstlisting}[language=coq,backgroundcolor=\colorcoq]
(* Turn a red chain into a green chain. *)
Equations green_of_red : chain red -> chain green :=
green_of_red (Chain R (RDigit Hole) Empty) :=
  Chain G (GDigit (YDigit Hole)) Empty;
green_of_red (Chain R (RDigit Hole) (Chain G (GDigit body) c)) :=
  Chain G (GDigit (YDigit body)) c;
green_of_red (Chain R (RDigit (YDigit body)) c) :=
  Chain G (GDigit Hole) (Chain R (RDigit body) c).
\end{lstlisting}
\caption{\natnumcoq: regularization without correctness claim}
\label{fig:bc_coq_gor}
\end{figure}

The use of \Equations is illustrated in \fref{fig:bc_coq_gor}, which
presents a definition of the function \coqe|green_of_red|. We explicitly
declare that this function expects a red chain as an argument. As a result,
\Equations recognizes that the three cases that are explicitly considered
cover all possible situations; there is no need to explicitly consider and
eliminate the remaining cases.
Furthermore, we explicitly declare that \coqe|green_of_red| returns a green
chain. This claim, too, is automatically verified by \Coq.

A fact that we have not expressed yet is the \emph{correctness} of \coqe|green_of_red|.
% and other functions like it, namely \coqe|ensure_green|, and \coqe|succ|
%
We~are not content to know that the chain \coqe|green_of_red c| is green:
we~want to know that it~represents the same natural number as the chain~\coqe|c|.
To express this fact, we~must define which natural number is represented by
a packet, a chain, or a number.
In other words, we must define \emph{model functions}\index{model functions!natural numbers}
that map each data structure to its model, a natural number.

\begin{figure}[tp]
\begin{lstlisting}[language=coq,backgroundcolor=\colorcoq]
Equations packet_nat {C : color} : packet C -> nat -> nat :=
packet_nat Hole n := n;
packet_nat (GDigit body) n := 0 + 2 * packet_nat body n;
packet_nat (YDigit body) n := 1 + 2 * packet_nat body n;
packet_nat (RDigit body) n := 2 + 2 * packet_nat body n.

Equations chain_nat {C : color} : chain C -> nat :=
chain_nat Empty := 0;
chain_nat (Chain _ pkt c) := packet_nat pkt (chain_nat c). (*@ \label{chain_nat_Chain} *)

Equations number_nat : number -> nat :=
number_nat (T c) := chain_nat c.
\end{lstlisting}
\caption{\natnumcoq: model functions}
\label{fig:bc_coq_models}
\end{figure}

These definitions
appear in \fref{fig:bc_coq_models}.
The function \coqe|packet_nat| takes the model of the hole as an extra
parameter: that is, \coqe|packet_nat pkt n| computes the model of the
packet~\coqe|pkt|, a~natural number, under the assumption that the model of
the hole is the natural number~\coqe|n|.
The~definition of \coqe|packet_nat pkt n|
is formulated in such a way that
the number \coqe|n| is multiplied
by~$2^k$
where $k$ is the length of the packet~\coqe|pkt|.
In the definition of \coqe|chain_nat|, at line~\ref{chain_nat_Chain},
the~parameter~\coqe|n| is instantiated with \coqe|(chain_nat c)|:
that~is,
the model of a chain is
the model of its first packet,
under the assumption that the hole
represents
% the model of
the remainder of the chain.

It is worth noting that the model functions \coqe|packet_nat|,
\coqe|chain_nat|, and \coqe|number_nat| are inductively defined. Although
recursion is not needed in the code that operates on packets, chains, and
numbers, induction is needed in the specification and proof of this code.

% a priori / a posteriori
% intrinsic / extrinsic

Once the model function \coqe{chain_nat} has been defined,
the correctness of the function \coqe|green_of_red|
can be stated.
Two approaches come to mind:
this correctness statement can be expressed
either \emph{a posteriori},
as a stand-alone lemma,
or \emph{a priori},
as part of the definition of \coqe|green_of_red|.
In the first approach, the correctness statement
would take the following form:
\begin{lstlisting}[language=coq, backgroundcolor=\colorcoq]
Lemma green_of_red_correct :
  forall (c : chain red),
  chain_nat (green_of_red c) = chain_nat c.
Proof. ... Qed.
\end{lstlisting}
The proof must mimic the structure of the code; fortunately, a custom case
analysis principle is automatically generated by \Equations and helps
reduce the redundancy between the code and the proof.
In the second approach, which we adopt, no lemma is required:
instead,
at definition time,
the function \coqe|green_of_red|
is ascribed the dependent function type
\coqe+forall (c : chain red), { c' : chain green | chain_nat c' = chain_nat c }+.
This~type claims that \coqe|green_of_red|
transforms a red chain~\coqe|c|
into a green chain~\coqe|c'|
such that
\coqe|c|~and~\coqe|c'| represent the same number.
We refer to the equation
\coqe|chain_nat c' = chain_nat c|
as the \emph{postcondition}
of the function \coqe|green_of_red|.
This approach integrates well with \Equations:
in each case, \Equations automatically generates
a proof obligation (that~is, a~goal) of the form
\coqe|chain_nat c' = chain_nat c|,
where \coqe|c| and~\coqe|c'|
are suitably instantiated.
The~proofs of these goals
are provided by the user
below the function's definition
and (in~many cases) can be automated
via tactics.%
\footnote{The command \coqe|Obligation Tactic := <tactic>.| lets the user
provide a~tactic that is applied by \Equations to every proof obligation. If
this user-provided tactic is able to solve a proof obligation then this
obligation is never presented to the user. Otherwise this obligation is
presented to the user, who is expected to write a proof, delimited by
\coqe|Next Obligation ... Qed.| In \fref{fig:bc_coq_func}, every proof
obligation is automatically solved by \Equations's default obligation tactic,
so neither \coqe|Obligation Tactic| nor \coqe|Next Obligation| need be used.}

\begin{figure}[tp]
\begin{lstlisting}[language=coq,backgroundcolor=\colorcoq]
(* ? x denotes an application of the subset type constructor to the witness x. *)
Notation "? x" := (@exist _ _ x _) (at level 100).

(* Turn a red chain into a green chain that represents the same number. *)
Equations green_of_red (c : chain red) :
  { c' : chain green | chain_nat c' = chain_nat c } :=
green_of_red (Chain R (RDigit Hole) Empty) :=
  ? Chain G (GDigit (YDigit Hole)) Empty;
green_of_red (Chain R (RDigit Hole) (Chain G (GDigit body) c)) :=
  ? Chain G (GDigit (YDigit body)) c;
green_of_red (Chain R (RDigit (YDigit body)) c) :=
  ? Chain G (GDigit Hole) (Chain R (RDigit body) c).

(* Turn a green or red chain into a green chain that represents the same number. *)
Equations ensure_green {g r} (c : chain (Mix g NoYellow r)) :
  { c' : chain green | chain_nat c' = chain_nat c } :=
ensure_green Empty :=
  ? Empty;
ensure_green (Chain G pkt c) :=
  ? Chain G pkt c;
ensure_green (Chain R pkt c) :=
  green_of_red (Chain R pkt c).

(* Turn a number n into a number n' that represents the successor of n. *)
Equations succ (n : number) :
  { n' : number | number_nat n' = S (number_nat n) } :=
succ (T Empty) :=
  ? T (Chain Y (YDigit Hole) Empty);
succ (T (Chain G (GDigit body) c))
  with ensure_green c => { | ? c' :=
  ? T (Chain Y (YDigit body) c') };
succ (T (Chain Y (YDigit body) c))
  with ensure_green (Chain R (RDigit body) c) => { | ? c' :=(*@ \label{with:example} *)
  ? T c' }.
\end{lstlisting}
\caption{\natnumcoq: regularization and incrementation,\\
with correctness claims}
\label{fig:bc_coq_func}
\end{figure}

Our \Coq code
for regularization and incrementation,
with correctness claims,
is shown in \fref{fig:bc_coq_func}.
\gitrepofile{theory/BinCounting/Core.v}
Up to minor differences in syntax,
the code is essentially the same as the OCaml code
of \fref{fig:bc_ocaml_func}.
Each function carries a postcondition,
which expresses its correctness property.
The functions \coqe|green_of_red|
and \coqe|ensure_green|
have postconditions
of the form \coqe|chain_nat c' = chain_nat c|.
% which mean that the chains \coqe|c| and \coqe|c'|
% represent the same natural number.
%
The function \coqe|succ|
has the postcondition
\coqe|number_nat n' = S (number_nat n)|,
where \coqe|S : nat -> nat|
is the successor function for natural numbers:
this means that \coqe|succ|
correctly computes the successor of its argument.

Because the codomain of each function is a subset type
\coqe+{ x : T | P x }+,
the code must contain construction and deconstruction operations
for the subset type.
(\textit{Program}~\citep{sozeau-finger-07}, a~predecessor of
\Equations, allowed implicit construction and deconstruction of subset types.
\Equations does not seem to allow this.)
We write \coqe|? x| for an application of the constructor of the subset
type, \coqe|exist|, to the witness~\coqe|x| and to a wildcard~\coqe|_|.
When this notation is used for construction, the wildcard gives rise to
a~proof obligation: a~proof of \coqe|P x| must be provided. When it is used
for deconstruction, the wildcard stands for an anonymous proof of \coqe|P x|.
% fp: I wanted to write that this proof is discarded,
%     but it must not be discarded; it can be useful
%     in a proof obligation. Something slightly weird
%     is going on here.

The \coqe|with| notation allows calling an auxiliary function and
deconstructing its result. For example, on line~\ref{with:example},
the variable~\coqe|c'|
is bound to the chain
produced by the function call~\coqe|ensure_green (Chain R (RDigit body) c)|.
This can be understood as an~alternative notation for
a~local definition.
% \coqe|let '(? c') := ensure_green (Chain R (RDigit body) c) in ...|
% but \Equations deals with it better
% than it would deal with a let binding

\section{Non-catenable deques}
\label{sec:deque}

% ------------------------------------------------------------------------------

\subsection{Concept}

\citet[\S4]{kaplan-tarjan-99} introduce
\emph{non-catenable deques},
which we refer to as just \emph{\sdeques}.
They support the four operations
\opush, \opop, \oinject, and \oeject,
with worst-case time complexity~$O(1)$.
\Sdeques serve as a stepping stone between natural numbers (\sref{sec:binary})
and catenable deques (\sref{sec:cadeque}).
They are more complex than natural numbers,
yet simpler than catenable deques,
so, from a pedagogical point of view,
it is useful to understand them
before attempting to understand catenable deques.
Furthermore, \sdeques actually serve as a~building block in the construction
of catenable deques.

\subsubsection{\Treerep, without buffers}

When a \sdeque's elements and buffers are abstracted away,
the structure that remains
is
just
a natural number in redundant binary representation,
that is,
a list of digits
(\sref{sec:binary}).
This is the \sdeque's \emph{\treerep},
without buffers.
We also refer to it as the \sdeque's \emph{skeleton}.
To~grow or shrink this skeleton,
one uses
incrementation, decrementation, and regularization
of natural numbers in RBR.

The skeleton of a \sdeque of $n$~elements
is a~representation of \emph{some} natural number~$n'$
which may differ (and usually differs) from~$n$.
Indeed,
as explained shortly after,
each digit
is expected to carry one or two buffers,
which together can contain between~0 and 10~elements.

\subsubsection{\Treerep, with buffers}

Onto the skeleton,
a number of elements are attached.
More precisely,
elements are stored in \emph{buffers},
which are attached to the skeleton.
A~buffer is a sequence % or bounded-size tuple
of~0 to~5~elements.
We refer to the number of elements of a~buffer
as the \emph{size} of this buffer.
Because a \sdeque must support inserting or extracting elements
at either end,
%
% one cannot simply attach one buffer to each digit of the skeleton
% fp: again, better avoid negative sentences.
%
\emph{two} buffers,
a~\emph{prefix buffer}
and
a~\emph{suffix buffer},
are attached onto each digit.
The~last digit is special: it~carries only one buffer.
(We deviate slightly from \kt's presentation,
which does not have such a special treatment of the last digit.
We find that this deviation simplifies our code.)
This is the \sdeque's \emph{\treerep},
with buffers.

\begin{figure}
\centering
\newcommand{\niveau}[1]{\small\emph{level~$#1$}}
\newcommand{\leftX}{-3.3}
\begin{tikzpicture}
  \node at (\leftX,3.2) {\niveau{0}};
  \drawsdeque{0}{3.2}{0.3}
  \node at (2.8,3.2) {$A$};

  \draw[->] (0, 3) -- (0, 2.6);

  \node at (\leftX,2.4) {\niveau{1}};
  \drawsdeque{0}{2.4}{0.3}
  \node at (3.2,2.4) {$A \times A$};

  \draw[->] (0, 2.2) -- (0, 1.9);

  \foreach \p in {(\leftX,1.75),(\leftX,1.6),(\leftX,1.45)}
    \fill \p circle(.03);
  \foreach \p in {(0,1.75),(0,1.6),(0,1.45)}
    \fill \p circle(.03);
  \foreach \p in {(2.8,1.75),(2.8,1.6),(2.8,1.45)}
    \fill \p circle(.03);

  \draw[->] (0, 1.3) -- (0, 1);

  \node at (\leftX,0.8) {\niveau{k-1}};
  \drawsdeque{0}{0.8}{0.3}
  \node at (3.1,0.8) {$A^{2^{k-1}}$};

  \draw[->] (0, 0.6) -- (0, 0.2);

  \node at (\leftX,0) {\niveau{k}};
  \drawsending{0}{0}{0.3}
  \node at (2.9,0) {$A^{2^k}$};

  \draw[thick] (-2,3.45) -- (-2,1);
  \draw[thick] (-2,1) arc (-180:0:2);
  \draw[thick,->] (2,1) -- (2,3.45);
\end{tikzpicture}
\caption{%
  \Treerep of a \sdeque, with buffers.
  The curved arrow indicates the logical order in which buffers are read.
  On the right are the types of the elements stored in the buffers.%
}
\label{fig:ncd_read_stack}
\end{figure}

An illustration of this structure
appears in \fref{fig:ncd_read_stack}.
Following \kt,
we organize this drawing
in a vertical manner.
Each level corresponds to a colored digit.
(In~the~drawing, the~colors are omitted.)
The~top~level is the least significant digit;
the bottom level is the most significant digit.
%
% A~\sdeque is a stack of colored levels.
%
In~the~drawing,
a~horizontal line segment
denotes a~buffer.
Each level
(except the last level)
% enclosed within square brackets
consists of two~buffers
and a~pointer to the next level.
% denoted by a vertical arrow.
%
The last level
consists of just one~buffer.
%
\begin{comment}
% fp: I think we can omit this
Vertical arrows are reminiscent
of the inductive definition
of \sdeques
given by \kt.
%
In their definition,
a \sdeque is a triple
made of a prefix buffer,
a child \sdeque,
and a suffix buffer.
\end{comment}

\begin{comment}
% fp: removing this paragraph, because
%     1- \kt's definition is not exactly the same as ours
%        (we have a special treatment of the last node)
%     2- This paragraph gives *two* definitions,
%        the direct inductive definition of \sdeques,
%        and the definition as a stack of nodes.
%        This is confusing, and *none* of these definitions is actually needed
%        in the following.
The abstract view of a \sdeque can be inductively defined.
\kt recursively define a nonempty \sdeque as a triple of a
prefix buffer, a potentially empty child \sdeque, and a suffix buffer.
%
The linear structure of a \sdeque
arises from the child relationship:
a \sdeque~$d$ can be viewed as
a stack of nodes $n_0, n_1, \ldots, n_k$,
where a \emph{node}~$n$ is a pair of a~prefix and a~suffix.
\end{comment}
%
To find out which sequence of elements is represented by a \sdeque,
% fp: removed this sentence because it relies on the inductive description of \sdeques:
\begin{comment}
one reads its elements from left to right,
starting with the prefix,
continuing with the child deque,
and ending with the suffix.
\end{comment}
the prefix buffers,
which appear on the left-hand side of the illustration,
must be read first,
from top to bottom;
then, the suffix buffers,
which appear on the right-hand side,
must be read from bottom to~top.
This is suggested by the curved arrow in \fref{fig:ncd_read_stack}.
%
% fp: the following informal sentences could be removed if desired:
%
Thus,
one can think of
the sequence of elements as
``folded in half''
over the skeleton.
The front and rear ends of the sequence
are easily accessible,
while the middle part of the sequence
is deeply buried.

\subsubsection{Recursive slowdown}

As one traverses the binary representation of a natural number,
from least significant toward most significant digit,
each digit
carries a weight
that is twice the weight
of the previous digit.
\kt
apply
a similar principle,
known as \emph{recursive slowdown},
to the elements
that are
stored in a \sdeque's buffers.
In a \sdeque that holds elements of type~$A$,
the~two buffers at the top level contain elements of type~$A$;
the~two buffers at the next level contain elements of type~$A\times A$;
% the~two buffers at the following level contain elements of type~$(A\times A)\times(A\times A)$;
and so on.
%
% fp: removed because we have not given this inductive view:
% the child \sdeque holds \emph{pairs} of elements of type~$A$,
% that~is, elements of type $A\times A$.
%
In general,
the elements stored in the buffers at level~$i$
have type $A^{2^i}$.
%
\begin{comment}
% fp: these sentences are OK, but add nothing to the figure.
They
can be thought of as
complete binary trees of~depth~$i$.
Each such tree has $2^i$ leaves,
where each leaf carries a value of type~$A$.
\end{comment}
%
This is illustrated in the rightmost column of \fref{fig:ncd_read_stack}.
To interpret a buffer as a~sequence of elements of type~$A$,
one reads it from left to right and
one expands each composite element of type $A^{2^i}$
% also from left to right
into a~sequence of $2^i$ basic elements of type~$A$.
%
% fp: We could say that the sequence associated with a \sdeque
%     is the concatenation of the sequences associated with the
%     buffers. But 1- this is kind of obvious and 2- it should
%     be said in the previous subsubsection.

\subsubsection{Color assignment}

In \sref{sec:binary}, a color (green, yellow, or red) was assigned
to each digit
in the representation of a natural number,
based on the value of this digit.
Here, each level in a \sdeque is assigned a color,
based on the sizes of its buffers,
as follows.
\kt
define a mapping of sizes to colors:
\begin{itemize}
\item
Sizes~2 and~3 are green;
sizes~1 and~4 are yellow;
sizes~0 and~5~are red.
\end{itemize}
This mapping allows assigning a color to a buffer,
based on its size,
that is,
based on the number of elements that it contains.

An ordering on colors is defined: $\text{red} < \text{yellow} < \text{green}$.
This means
that red is worse than yellow,
which itself is worse than green.
Both extreme sizes,~0 and~5, are bad, because
an element cannot be extracted out of an empty buffer
and
an element cannot be inserted into a buffer that is full.
%
% In the case of catenable deques,
% this will be different,
% because they use unbounded buffers.

Then, \kt
assign colors to levels,
based on the colors
% (sizes)
of their buffers.
The color of a level is defined
as the worst color of its two buffers,
that is,
the minimum color
among the colors
of its prefix buffer and suffix buffer.
The last level, which contains just one buffer, is always green,
regardless of the color of its buffer.

\subsubsection{Regularity}

With deques now rigorously defined,
we return to the more general tree representation,
corresponding to a natural number in RBR.
By virtue of this representation,
levels are grouped into packets
and regularity constraints are imposed
in the same way as in \sref{sec:binary}.
The regularity constraints are unchanged;
they can be found in \fref{fig:bc_reg}.

\subsubsection{\Pointerrep}

In the \pointerrep,
a
\sdeque is a~\emph{chain}\index{chain!deques},
that is,
a list of~\emph{packets}\index{packet!deques}.
This representation
allows moving from one packet to its successor
in constant time.
Our OCaml and \Coq implementations
(\sref{sec:deque:ocaml,sec:deque:coq})
are based on~it.

\subsubsection{Operations}

\opush and \oinject are analogous to
incrementation in \sref{sec:binary},
while \opop and \oeject are
analogous to decrementation.
Each of the four operations
first naively inserts or removes one element,
potentially making the chain irregular.
% actually, semi-regular; but we have not introduced this word yet
%
Then, the chain is regularized.
Both steps are carried out in constant time.

% ------------------------------------------------------------------------------

\subsection{OCaml implementation}
\label{sec:deque:ocaml}

We first define the type of buffers, \oc|('a, 'c) buffer|,
as a~generalized algebraic data type,
with one constructor for each buffer size,
from~0 to~5.
This type carries two parameters:
\oc|'a| is the type of the elements of the buffer;
\oc|'c| is the color of this buffer.
Its definition appears in \fref{fig:ncd_ocaml_types}.

We assign colors to buffers in a way that differs slightly from
\kt's scheme, and is more flexible. Here is why.
Thinking ahead about the way levels are colored,
we wish to avoid the need to compute the minimum of two colors,
because such a~computation is not easily expressed, in OCaml,
at the level of types.
Instead, we make the type \oc|buffer| covariant
in the parameter \oc|'c|.
That is, we allow a buffer to be assigned a worse color
than its true color.
In other words,
we allow a~buffer that contains 1~or~4 elements
to be viewed not only as a yellow buffer,
but also, if desired, as a red buffer;
and we allow a~buffer that contains 2~or~3 elements
to be viewed not only as a green buffer,
but also, if desired,
as~a~yellow buffer or a~red buffer.
This coloring scheme is more flexible than \kt's scheme,
yet still lets us perform the four operations on \sdeques
in constant time in the worst case.

\begin{figure}
\begin{lstlisting}[language=ocaml,backgroundcolor=\colorocaml]
type ('a, 'c) buffer =
  | B0 :                           ('a, red            ) buffer (* red *)
  | B1 : 'a                     -> ('a, nogreen * _ * _) buffer (* yellow or red *)
  | B2 : 'a * 'a                -> ('a, _              ) buffer (* any color *)
  | B3 : 'a * 'a * 'a           -> ('a, _              ) buffer (* any color *)
  | B4 : 'a * 'a * 'a * 'a      -> ('a, nogreen * _ * _) buffer (* yellow or red *)
  | B5 : 'a * 'a * 'a * 'a * 'a -> ('a, red            ) buffer (* red *)

type ('a, 'b, 'c) packet =
  | Hole   :
      ('a, 'a, uncolored) packet                                    (*@ \label{hole} *)
  | Packet :
      ('a, 'c) buffer *                           (* prefix buffer *) (*@ \label{prefix_buffer} *)
      ('a * 'a, 'b, nogreen * _ * nored) packet * (* child packet *)  (*@ \label{child_packet} *)
      ('a, 'c) buffer ->                          (* suffix buffer *) (*@ \label{suffix_buffer} *)
      ('a, 'b, 'c) packet                                           (*@ \label{packet} *)

type (_, _) regularity =             (* parameters: packet color and chain color *)
  | G : (green , _ * noyellow * _) regularity
  | Y : (yellow,            green) regularity
  | R : (red   ,            green) regularity

type ('a, 'c) chain =
  | Ending :
      ('a, _) buffer ->
      ('a, green) chain                           (* last level is green *)
  | Chain  :
      ('c1, 'c2) regularity * ('a, 'b, 'c1) packet * ('b, 'c2) chain ->
      ('a, 'c1) chain

type 'a deque =
  | T : ('a, _ * _ * nored) chain -> 'a deque
\end{lstlisting}
\caption{\ncdocaml: types for buffers, packets, chains, and deques}
\label{fig:ncd_ocaml_types}
\end{figure}

The type \oc|('a, 'b, 'c) packet| (\fref{fig:ncd_ocaml_types})
is defined in a way that is reminiscent of the previous section:
for comparison, see \fref{fig:bc_ocaml_types} in \sref{sec:binary}.
There is still a constructor for the \emph{hole},
that is,
for the end of a packet.
There is now a single constructor, \oc|Packet|,
for non-hole packets,
irrespective of their color.
% fp:
%      actually, the code sometimes performs a case analysis on
%      a "regularity" -- this suggests that one constructor for
%      all kinds of packets is not always enough, but we work
%      around its absence by analyzing the regularity instead.
%      Perhaps, if we had three constructors for packets, we
%      could have "regularity : Prop".
%
This constructor carries
a~prefix buffer,
a~child packet,
and a~suffix buffer.
As~before, the child packet must be yellow or uncolored.
As explained earlier,
instead of
computing
the minimum of the colors of the two buffers,
we require the two buffers to have the same color,
namely \oc|'c|
(\fref{fig:ncd_ocaml_types}, lines~\ref{prefix_buffer} and~\ref{suffix_buffer}).
This color becomes the color of this packet
(\fref{fig:ncd_ocaml_types}, line~\ref{packet}).

In \oc|('a, 'b, 'c) packet|,
the type parameter \oc|'a|
is the type of the elements of this packet.
Thus, in a packet whose elements have type \oc|'a|,
the child packet has elements of type \oc|'a * 'a|
(\fref{fig:ncd_ocaml_types}, line~\ref{child_packet}).
The type parameter \oc|'b| is the type of the elements
of the next packet in the chain.
This is why the constructor \oc|Hole|
imposes an equality between \oc|'a| and~\oc|'b|
(\fref{fig:ncd_ocaml_types}, line~\ref{hole}):
once the end of the current packet is reached,
the type~\oc|'a| of the elements of the current packet
must coincide with the type~\oc|'b| of the elements
of the next packet in the chain.
The constructor \oc|Packet| propagates \oc|'b|.

% Finally, a chain is composed of a packet followed by a chain.
% The elements held in the following chain are pairs of elements stored in the
% packet's last buffers.
% To determine the type of elements in the first node of the subsequent chain, we
% need to know the type of elements at the end of the packet.
% Since the following chain is directly connected to the packet's hole, the type
% parameter \oc|'a| associated with the \oc|Hole| must be remembered.
% We pass this type parameter through the recursive packet definition as \oc|'b|.

The type \oc|regularity| is the same as in \fref{fig:bc_ocaml_types}.
The type \oc|chain|
exhibits the same structure
as the type \oc|chain|
in \fref{fig:bc_ocaml_types},
except that the constructor \oc|Empty| is replaced with
the constructor \oc|Ending|,
which carries one buffer,
and forms a green chain.
% An ending chain is always green.
The type \oc|deque| is defined in the same way as the type \oc|number| in
\fref{fig:bc_ocaml_types}.

We implement
the main four operations on \sdeques,
as well as a number of auxiliary functions,
by translating
\kt's indications into OCaml code.
In this endeavor,
we crucially rely
on the fact that OCaml's type-checker
allows dead branches to be omitted
and guarantees complete coverage
(that is, verifies that no cases are inadvertently omitted).
For example, the function \oc|green_push|, which pushes an element
onto a~green buffer and returns a yellow buffer, is shown in
\fref{fig:ncd_ocaml_ex}.
Because a green buffer must contain two or three elements,
only the cases \oc|B2| and \oc|B3| must be considered;
all other cases can be omitted.

\begin{figure}
\begin{lstlisting}[language=ocaml,backgroundcolor=\colorocaml]
let green_push
: type a. a -> (a, green) buffer -> (a, yellow) buffer
= fun x buf ->
  match buf with
  | B2 (a, b)    -> B3 (x, a, b)
  | B3 (a, b, c) -> B4 (x, a, b, c)
\end{lstlisting}
\caption{\ncdocaml: the auxiliary function \oc|green_push|}
\label{fig:ncd_ocaml_ex}
\end{figure}

We write a total of 29 intermediate functions to support the main
regularization function, \oc|green_of_red|,
which transforms a red chain into a green chain,
while preserving the sequence of elements that this chain represents.
Once \oc|green_of_red| is available, the four main operations,
namely \opush, \opop, \oinject, and \oeject,
are obtained by composing a basic operation,
which may produce a red \sdeque or sub\sdeque,
with regularization.
Our complete OCaml code
contains 30 lines of type definitions
and 310 lines of code.
\gitrepofile{lib/deque_core.ml}
Here and elsewhere, we count only non-blank, non-comment lines of code, and round up to the nearest ten.

% ------------------------------------------------------------------------------

\subsection{\Coq implementation}
\label{sec:deque:coq}

\begin{figure}
\begin{lstlisting}[language=coq,backgroundcolor=\colorcoq]
Inductive buffer : Type -> color -> Type :=
  | B0 {A}       :                          buffer A red
  | B1 {A y r}   : A                     -> buffer A (Mix NoGreen y r)
  | B2 {A g y r} : A -> A                -> buffer A (Mix g y r)
  | B3 {A g y r} : A -> A -> A           -> buffer A (Mix g y r)
  | B4 {A y r}   : A -> A -> A -> A      -> buffer A (Mix NoGreen y r)
  | B5 {A}       : A -> A -> A -> A -> A -> buffer A red.

Inductive packet : Type -> Type -> color -> Type :=
  | Hole {A}         :
      packet A A uncolored
  | Packet {A B C y} :
      buffer A C ->
      packet (A * A) B (Mix NoGreen y NoRed) ->
      buffer A C ->
      packet A B C.

Inductive regularity : color -> color -> Type :=
  | G {g r} : regularity green  (Mix g NoYellow r)
  | Y       : regularity yellow green
  | R       : regularity red    green.

Inductive chain : Type -> color -> Type :=
  | Ending {A C}      :
      buffer A C ->
      chain A green
  | Chain {A B C1 C2} :
      regularity C1 C2 -> packet A B C1 -> chain B C2 -> (*@ \label{line:hole_type_match} *)
      chain A C1.

Inductive deque : Type -> Type :=
  | T {A g y} : chain A (Mix g y NoRed) -> deque A.
\end{lstlisting}
\caption{\ncdcoq: types for buffers, packets, chains, and deques}
\label{fig:ncd_coq_types}
\end{figure}

Translating these type definitions from OCaml to \Coq
is straightforward.
The result appears in \fref{fig:ncd_coq_types}.

Next, for each data structure,
we define a model function.
Whereas the model functions defined in \sref{sec:binary}
have result type \coqe|nat|,
because there the data structures
represent natural numbers,
the model functions defined in this section
have result type \coqe|list A|,
because the data structures in this section
represent sequences of elements.
The definitions of these model functions
appear in \fref{fig:ncd_coq_models}.
%
% Cases for \coqe|B3|, \coqe|B4|, and \coqe|B5| are not explicitly written,
% as they can be deduced from the previous buffer cases.
%
Their definitions are straightforward.
The auxiliary function \coqe|flattenp| flattens a list of pairs of elements
into a list of elements.
The use of \coqe|flattenp| in \coqe|packet_seq| is analogous to
the use of multiplication by 2 in \coqe|packet_nat| (\fref{fig:bc_coq_models}).
The~parameter~\coqe|l| in \coqe|packet_seq| plays the same role
as the parameter~\coqe|n| in \coqe|packet_nat| (\fref{fig:bc_coq_models}):
it is the model of the hole.
The definition of the model of a non-hole packet
(\fref{fig:ncd_coq_models}, line~\ref{packet_seq_Packet})
indicates that the sequence of elements of a packet
is obtained by concatenating
the sequence of elements of the prefix buffer,
the sequence of elements of the child packet,
and
the sequence of elements of the suffix buffer,
in this order.

\begin{figure}
\begin{lstlisting}[language=coq,backgroundcolor=\colorcoq]
Equations buffer_seq {A C} : buffer A C -> list A :=
buffer_seq B0       := [];
buffer_seq (B1 a)   := [a];
buffer_seq (B2 a b) := [a] ++ [b];
(* three more cases omitted *).

Equations flattenp {A} (l : list (A * A)) : list A :=
flattenp []            := [];
flattenp ((x, y) :: l) := [x] ++ [y] ++ flattenp l.

Equations packet_seq {A B C} : packet A B C -> list B -> list A :=
packet_seq Hole l             := l;
packet_seq (Packet p pkt s) l :=
  buffer_seq p ++ flattenp (packet_seq pkt l) ++ buffer_seq s. (*@ \label{packet_seq_Packet} *)

Equations chain_seq {A C} : chain A C -> list A :=
chain_seq (Ending b)       := buffer_seq b;
chain_seq (Chain _ pkt cd) := packet_seq pkt (chain_seq cd).

Equations deque_seq {A} : deque A -> list A :=
deque_seq (T c) := chain_seq c.
\end{lstlisting}
\caption{\ncdcoq: model functions for buffers, packets, and chains}
\label{fig:ncd_coq_models}
\end{figure}

% fp: the following has been explained for \coqe|chain_nat| already,
%     so I don't think there is a need to repeat.

% The model function \coqe|packet_seq| is slightly different from the others.
% The sequence of a chain is constructed by reading the packet's prefixes,
% followed by the subsequent chain sequence, and then the packet's suffixes.
% The sequence corresponding to the chain following a packet must be inserted
% between the packet's prefixes sequence and its suffixes sequence.
% To accomplish this, we pass an additional argument to the \coqe|packet_seq|
% function, representing the sequence of the subsequent chain.
% This sequence is treated as the one of the hole, ensuring that the resulting
% sequence accurately represents both the packet and its following chain.

\begin{figure}
\begin{lstlisting}[language=coq,backgroundcolor=\colorcoq]
Equations green_push {A : Type} (x : A) (b : buffer A green) :
  { b' : buffer A yellow | buffer_seq b' = [x] ++ buffer_seq b } :=
green_push x (B2 a b)   := ? B3 x a b;
green_push x (B3 a b c) := ? B4 x a b c.
\end{lstlisting}
\caption{\ncdcoq: the auxiliary function \coqe|green_push|}
\label{fig:ncd_coq_ex}
\end{figure}

As an illustration of the style in which our code is written, \Coq code for
the auxiliary function \coqe|green_push| is shown in \fref{fig:ncd_coq_ex}.
It is analogous to our OCaml code (\fref{fig:ncd_ocaml_ex}), but includes
a correctness statement:
the postcondition
\coqe|buffer_seq b' = [x] ++ buffer_seq b|
guarantees that the sequence of elements of the new buffer \coqe|b'|
is the concatenation of
the singleton sequence \coqe|[x]|
with the sequence of elements of the original buffer \coqe|b|.
%
% As the question mark notation \coqe|?| is used twice in this code,

This creates two proof obligations
of the form
\coqe|buffer_seq b' = [x] ++ buffer_seq b|
where \oc|b'| and \oc|b| are suitably instantiated.
In general, our code gives rise to a large number of proof obligations. In
every proof obligation, the goal is an equality between two sequences, which
we represent in \Coq as lists. In some proof obligations, one or more
equalities between sequences are available as hypotheses.
Thus, in general, our proof obligations are entailments between equalities on
sequences, where sequences are built out of universally quantified variables,
the empty list \coqe|[]|, singleton lists \coqe|[x]|, and list concatenation
\coqe|++|, and can also be applications of model functions such as
\coqe|buffer_seq|, \coqe|packet_seq|, \coqe|chain_seq|, and so on.

% Numerous list equalities must be proven as obligations, and not all of these
% can be resolved by the base automation provided by \Equations.

Although we do not know of a decision procedure that automatically solves this
class of proof obligations, we are able to deal with these proof obligations,
in a mostly automated way, thanks to two third-party \Coq modules.
First,
\Hammer~\citep{czajka-20},
which provides the tactic \texttt{hauto},
is particularly effective.
Like \texttt{eauto},
it performs goal-directed proof search
% https://coqhammer.github.io/#papers-about-coqhammer
and can exploit a database of user-provided hints.
It is able to automatically discharge many proof obligations.
We deal with the remaining proof obligations by hand.
Second,
to facilitate the remaining manual proofs,
we exploit \AACTactics~\citep{braibant-pous-11},
which provides tactics
for deciding entailment modulo AC
% (\coqe|aac_reflexivity|)
and
for rewriting modulo AC.
% (\coqe|aac_rewrite|)
%
This helps us exploit the fact
that list concatenation~\coqe|++|
is associative and commutative
and admits the empty list~\coqe|[]|
as a~unit.
%
% Hence, we use singletons and \emph{app} rather than the usual \emph{cons} of
% lists in model functions definition in \fref{fig:ncd_coq_models}.

Thanks to these third-party tactics,
we are able to verify the correctness of all functions
while performing only four manual proofs.
Our \Coq file for \sdeques
contains
30~lines of type definitions,
40~lines of definitions of model functions,
and~390 lines of definitions of operations. % on \sdeques.
\gitrepofile{theory/Deque/Deque_plain.v}
The result is a verified implementation of \kt's
non-catenable deques.

\section{Catenable deques: presentation}
\label{sec:cadeque}

\citet{kaplan-tarjan-99} design \emph{catenable deques},
a data structure that supports the full set of operations%
---namely
  \opush,
  \opop,
  \oinject,
  \oeject,
  and
  \oconcat---%
in worst-case constant time.
We refer to catenable deques as \emph{\cdeques}\index{\cdeque},
for short.

In this section,
we first describe
the \treerep
of \cdeques,
without buffers
(\sref{sec:cadeque:treerep:nobuffers}).
This representation
involves trees
% and forests
whose nodes carry a~color.
Based on this representation,
we decompose trees
% and forests
into packets
and impose regularity constraints.
This leads us to
the \pointerrep
of \cdeques,
still without buffers
(\sref{sec:cadeque:pointerrep:nobuffers}).
In~this representation,
packets and chains are primitive concepts,
and the regularity invariant holds by construction.
We are then ready to describe how buffers are set up and used.
As in the previous section,
each node is expected to carry
a prefix buffer and a~suffix buffer.
This~time, though,
a~more complex form of recursive slowdown
is involved
(\sref{sec:cadeque:buffers}).
We~revisit
the \treerep of \cdeques,
this time with buffers
(\sref{sec:cadeque:treerep:buffers}).
We give the size constraints that bear on buffers
% they are not the same as in the previous section
and indicate
how the sizes of the buffers
determine
the assignment of colors to nodes.

The \treerep, with buffers,
is a~complete description
of the data structure.
In an implementation,
in order to achieve worst-case time complexity $O(1)$,
the \pointerrep, with buffers,
must be used
instead.
We do not explicitly describe this representation in this section,
because it is visible
in our OCaml implementation of \cdeques,
which is the subject of \sref{sec:cadeque:ocaml}.

Implementing \cdeques in \Coq turns out to be somewhat challenging.
A~discussion of the issues
and a presentation of our \Coq implementation
appear
in \sref{sec:obstacles,sec:deque:coq:revisited,sec:cadeque:coq}.

% ------------------------------------------------------------------------------
% ------------------------------------------------------------------------------

% Trees of colored nodes.

\subsection{\Treerep, without buffers}
\label{sec:cadeque:treerep:nobuffers}

Whereas the skeleton of a non-catenable deque has linear structure,
the skeleton of a catenable deque has nontrivial tree structure.
Briefly put,
it~is a~forest of~trees of colored nodes,
where each node carries zero, one, or two subtrees.
Because of the~move
from a~setting where a~node has at most one successor
to a~setting where a~node has at most two successors,
a new color, orange, must~be introduced.
%
% fp: the ordering of colors is defined only later
%     because it is not relevant here.
\begin{itemize}
\item
  A \emph{color} is now one of
  red, orange, yellow, and green.
\end{itemize}

% ------------------------------------------------------------------------------

\begin{figure}[t]
\centering
\begin{tikzpicture}
  \node[circle, fill=myyellow, draw=black, minimum size=16pt, inner sep=0pt, outer sep=0pt] (0) at (-1, 6) {\footnotesize$Y$};
  \node[circle, fill=myorange, draw=black, minimum size=16pt, inner sep=0pt, outer sep=0pt] (a) at ( 1, 6) {\footnotesize$O_1$};
  \node[circle, fill=myyellow, draw=black, minimum size=16pt, inner sep=0pt, outer sep=0pt] (b) at (-2, 5) {\footnotesize$Y$};
  \node[circle, fill=myorange, draw=black, minimum size=16pt, inner sep=0pt, outer sep=0pt] (c) at ( 2, 5) {\footnotesize$O_2$};
  \node[circle, fill=myorange, draw=black, minimum size=16pt, inner sep=0pt, outer sep=0pt] (d) at (-3, 4) {\footnotesize$O$};
  \node[circle, fill=mygreen , draw=black, minimum size=16pt, inner sep=0pt, outer sep=0pt] (e) at ( 1, 4) {\footnotesize$G_1$};
  \node[circle, fill=mygreen , draw=black, minimum size=16pt, inner sep=0pt, outer sep=0pt] (f) at ( 3, 4) {\footnotesize$G_2$};
  \node[circle, fill=mygreen , draw=black, minimum size=16pt, inner sep=0pt, outer sep=0pt] (g) at (-4, 3) {\footnotesize$G$};
  \node[circle, fill=myyellow, draw=black, minimum size=16pt, inner sep=0pt, outer sep=0pt] (h) at (-2, 3) {\footnotesize$Y$};
  \node[circle, fill=myyellow, draw=black, minimum size=16pt, inner sep=0pt, outer sep=0pt] (i) at ( 1, 3) {\footnotesize$Y$};
  \node[circle, fill=myyellow, draw=black, minimum size=16pt, inner sep=0pt, outer sep=0pt] (j) at ( 4, 3) {\footnotesize$Y$};
  \node[circle, fill=mygreen , draw=black, minimum size=16pt, inner sep=0pt, outer sep=0pt] (k) at (-3, 2) {\footnotesize$G$};
  \node[circle, fill=mygreen , draw=black, minimum size=16pt, inner sep=0pt, outer sep=0pt] (l) at (-1, 2) {\footnotesize$G$};
  \node[circle, fill=myorange, draw=black, minimum size=16pt, inner sep=0pt, outer sep=0pt] (m) at ( 0, 2) {\footnotesize$O$};
  \node[circle, fill=myred   , draw=black, minimum size=16pt, inner sep=0pt, outer sep=0pt] (n) at ( 2, 2) {\footnotesize$R$};
  \node[circle, fill=mygreen , draw=black, minimum size=16pt, inner sep=0pt, outer sep=0pt] (o) at ( 3, 3) {\footnotesize$G$};
  \node[circle, fill=myred   , draw=black, minimum size=16pt, inner sep=0pt, outer sep=0pt] (p) at ( 4, 2) {\footnotesize$R$};
  \node[circle, fill=myorange, draw=black, minimum size=16pt, inner sep=0pt, outer sep=0pt] (q) at (-4, 1) {\footnotesize$O$};
  \node[circle, fill=myyellow, draw=black, minimum size=16pt, inner sep=0pt, outer sep=0pt] (r) at (-2, 1) {\footnotesize$Y$};
  \node[circle, fill=myred   , draw=black, minimum size=16pt, inner sep=0pt, outer sep=0pt] (s) at ( 0, 1) {\footnotesize$R$};
  \node[circle, fill=mygreen , draw=black, minimum size=16pt, inner sep=0pt, outer sep=0pt] (t) at ( 2, 1) {\footnotesize$G$};
  \node[circle, fill=myorange, draw=black, minimum size=16pt, inner sep=0pt, outer sep=0pt] (u) at ( 3, 1) {\footnotesize$O$};
  \node[circle, fill=mygreen , draw=black, minimum size=16pt, inner sep=0pt, outer sep=0pt] (v) at ( 5, 1) {\footnotesize$G$};
  \node[circle, fill=mygreen , draw=black, minimum size=16pt, inner sep=0pt, outer sep=0pt] (w) at (-4, 0) {\footnotesize$G$};
  \node[circle, fill=mygreen , draw=black, minimum size=16pt, inner sep=0pt, outer sep=0pt] (x) at (-2, 0) {\footnotesize$G$};
  \node[circle, fill=mygreen , draw=black, minimum size=16pt, inner sep=0pt, outer sep=0pt] (y) at ( 0, 0) {\footnotesize$G$};
  \node[circle, fill=mygreen , draw=black, minimum size=16pt, inner sep=0pt, outer sep=0pt] (z) at ( 3, 0) {\footnotesize$G$};

  \draw (0)--(b); \draw (a)--(c);
  \draw (b)--(d); \draw (c)--(e); \draw (c)--(f);
  \draw (d)--(g); \draw (d)--(h); \draw (e)--(i);
                                  \draw (f)--(j); \draw (f)--(o);
  \draw (h)--(k); \draw (h)--(l); \draw (i)--(m); \draw (i)--(n);
                                  \draw (j)--(p);
  \draw (k)--(q); \draw (k)--(r); \draw (m)--(s); \draw (n)--(t);
                                  \draw (p)--(u); \draw (p)--(v);
  \draw (q)--(w); \draw (r)--(x); \draw (s)--(y); \draw (u)--(z);

  \def\size{0.4}

  \tlpkt{0}{\size}{1}
  \trpkt{a}{\size}{1}
  \rlpkt{b}{\size}{1}
  \lrpkt{c}{\size}{1}
  \rrpkt{d}{\size}{1}
  \tbpkt{e}{\size}{1}
  \lbpkt{f}{\size}{1}
  \tbpkt{g}{\size}{1}
  \llpkt{h}{\size}{1}
  \tlpkt{i}{\size}{1}
  \tbpkt{o}{\size}{1}
  \tcpkt{j}{\size}{1}
  \rbpkt{k}{\size}{1}
  \tbpkt{l}{\size}{1}
  \rcpkt{m}{\size}{1}
  \tbpkt{n}{\size}{1}
  \cbpkt{p}{\size}{1}
  \tcpkt{q}{\size}{1}
  \tcpkt{r}{\size}{1}
  \cbpkt{s}{\size}{1}
  \tbpkt{t}{\size}{1}
  \tcpkt{u}{\size}{1}
  \tbpkt{v}{\size}{1}
  \cbpkt{w}{\size}{1}
  \cbpkt{x}{\size}{1}
  \tbpkt{y}{\size}{1}
  \cbpkt{z}{\size}{1}

\end{tikzpicture}
\caption{\cd: abstract structure and decomposition into packets}
\label{fig:cd_packets}
\end{figure}

% ------------------------------------------------------------------------------

\subsubsection{Trees and forests}
\label{sec:cadeque:treerep:nobuffers:trees}

The trees of interest
are given
by two mutually inductive definitions:
\begin{itemize}
\item
  A \emph{tree}
  consists of
  a color
  and a forest.
\item
  A \emph{forest}
  consists of zero, one, or two trees.
\end{itemize}

% Fix Indentation
\noindent
This can be visually understood as~follows:
a~tree consists of a root node, which carries a~color,
and a number of subtrees,
which together form a forest.
The number of subtrees,
which is known as the \emph{arity} of the root node,
must be~0,~1, or~2.

On top of this definition,
one well-formedness condition is imposed:

\begin{itemize}
\item
  A node of arity~0 must be green.
\end{itemize}

% ------------------------------------------------------------------------------

\subsubsection{Decomposing trees and forests into packets}
\label{sec:cadeque:treerep:nobuffers:packets}

A well-formed tree or forest can be decomposed into
a collection of disjoint
packets,
where a \emph{packet} is
a downward path
that traverses
a~(possibly null) number of orange and yellow nodes
and ends at a \greenorred node.
This decomposition exists and is unique;
it is determined by the colors and arities
of the nodes.
It is described by the following rules:
\begin{itemize}
\item
  A packet that reaches a~\greenorred node ends at this node. \\
  Each subtree
  % (of which there may be zero, one, or two)
  is independently decomposed into packets.
\item
  A packet that reaches
  an orange or yellow node of arity~1 \\
  extends down
  to this node's single child.
\item
  A packet that reaches
  an orange or yellow node of arity~2 \\
  extends down
  to this node's \emph{preferred child}.
  \begin{itemize}
  \item An orange node of arity~2 prefers its right-hand child.
  \item A yellow node of arity~2 prefers its left-hand child.
  \end{itemize}
  In either case, the non-preferred subtree
  is independently decomposed into packets.
\end{itemize}
%Fix Indentation
\noindent
In \kt's paper,
our packets are known as \emph{preferred paths}.
A packet goes through zero or more
orange or yellow
nodes
(whose arity is 1 or~2)
and
ends at a~\greenorred node.
We refer to the sequence of orange or yellow nodes
as the \emph{body} of the packet;
we~refer to the final \greenorred node
as the \emph{tail} of the packet.%

  In the previous
  sections~(\sref{sec:binary}, \sref{sec:deque}),
  a~packet \emph{begins}
  with a distinguished
  \greenorred node,
  followed with a number of yellow nodes.
  In contrast,
  in this section,
  a~packet \emph{ends}
  with a distinguished
  \greenorred node,
  preceded by a number of orange or yellow nodes.
  This design seems forced upon us
  by the following two facts:
  (a)~packets have linear structure,
  that is, they are paths;
  (b)~a \greenorred node can have arity~2,
  and does not have a preferred child.
%
% One might wish to give a better explanation,
% but we simply follow \kt.
%

An example of a forest,
and its decomposition into disjoint packets,
are shown in \fref{fig:cd_packets}.
A~node is represented by a colored disc.
A~child relationship between two nodes is represented by a~straight edge.
A~packet is represented by a~curved line that surrounds a group of nodes.
The forest in \fref{fig:cd_packets} is composed of two trees.
Let us explain
the structure of
the packet that begins at the root
of the second tree.
This packet begins at an orange node of arity~1, named~$O_1$.
It~extends down to its child,~$O_2$,
an orange node of arity~2.
It then extends down to the~preferred child of node~$O_2$,
% an orange node of arity~2 prefers its right-hand child
the~green node~$G_2$,
and ends there.
The non-preferred child of node $O_2$,
the green node~$G_1$,
forms the beginning of a new packet.
Because $G_1$ is a green node,
this new packet ends there;
the body of this packet is empty.
The children of the green nodes~$G_1$ and~$G_2$
form the beginning of new packets;
and so on.

% ------------------------------------------------------------------------------

\subsubsection{Regularity}
\label{sec:cadeque:treerep:nobuffers:regularity}

\begin{figure}[t]
\begin{enumerate}[label=(RC\arabic*)]
% The trivial constraint RC1 is included to make comparison
% with the earlier figure easier. The footnote explains this.
\item \label{cd_reg:trivial}
      Trivial.% \textsuperscript{\ref{footnote:regularity:constraints}}
\item \label{cd_reg:orange}
      A packet that begins at the left child of an orange node
      % (and is therefore the non-preferred child)
      must be green.
\item \label{cd_reg:red}
      A packet that begins at a child of a red node must be green.
\item \label{cd_reg:top}
      The top packets must be green. % root packet or top packet?
\end{enumerate}
\caption{\cd: regularity constraints}
\label{fig:cd_reg}
\end{figure}

We now assign colors to packets, as follows:
\begin{itemize}
\item The color of a packet is the color of its tail node.
\end{itemize}
Because a tail node is \greenorred,
the color of a packet is \greenorred.

% fp:
% We could use the word \emph{chain} as a synonym for \emph{tree} or
% \emph{subtree}, and try to define "the color of a chain", in order
% to state the (semi-)regularity constraints in a way that resembles
% \fref{fig:bc_reg}. However, because a chain can be formed of zero,
% one, or two trees, "the color of a chain" does not work. I think
% that it is easier to abandon this idea and express the constraints
% in terms of packets only.

Furthermore, we impose constraints,
shown in \fref{fig:cd_reg},
on the coloring of packets.
We~refer to constraints
\ref{cd_reg:orange} and \ref{cd_reg:red}
as the \emph{semi-regularity} constraints.
We refer to the constraints
\ref{cd_reg:orange}, \ref{cd_reg:red}, and \ref{cd_reg:top}
together
as the \emph{regularity} constraints.
%
% \label{footnote:regularity:constraints}
These constraints can be loosely compared with the earlier regularity
constraints of \fref{fig:bc_reg}, which concern the redundant binary
representation of natural numbers, as follows.
Because every packet is \greenorred,
constraint~\ref{bc_reg:green}
becomes
the vacuously true
constraint~\ref{cd_reg:trivial}.
Constraint~\ref{bc_reg:yellow}
corresponds loosely to constraint~\ref{cd_reg:orange}.
Constraint~\ref{bc_reg:red}
becomes constraint~\ref{cd_reg:red}.
Constraint~\ref{bc_reg:top}
becomes constraint~\ref{cd_reg:top}.

% ------------------------------------------------------------------------------

\subsection{\Pointerrep, without buffers}
\label{sec:cadeque:pointerrep:nobuffers}

In the \treerep,
trees and forests
have extremely simple structure
(\sref{sec:cadeque:treerep:nobuffers:trees}).
\emph{A~posteriori},
they
can be decomposed
into packets
(\sref{sec:cadeque:treerep:nobuffers:packets})
and
subjected to regularity constraints
(\sref{sec:cadeque:treerep:nobuffers:regularity}).
In~the \pointerrep,
in~contrast,
the decomposition into packets
and the regularity constraints
are imposed \emph{a priori}.

% ------------------------------------------------------------------------------

\subsubsection{Packets and chains}
\label{sec:cadeque:pointerrep:nobuffers:packets}

In~the \pointerrep,
instead of working at the granularity of individual nodes,
one works at the granularity of packets.
Furthermore,
whereas the \treerep is based on trees and forests,
the \pointerrep is based on \emph{chains}
whose arity can be 0,~1,~or~2.

One can think of
the arity of a chain
as the number of trees
that this chain represents.
A~chain of arity~1,
which has a root packet and a child chain,
corresponds to a~tree.
A~chain of arity 0,~1,~or~2
corresponds to a forest of 0,~1,~or~2 trees.

In the following,
we also speak of
the arity of a node (or packet).
One can think of it as
the~number of children
that this node (or packet)
expects.
In other words, one can think of~it
as a~constraint
on the arity of the child chain
that follows this node (or packet).

The \pointerrep is defined as follows.
We give a definition of \emph{nodes},
followed with
mutually inductive definitions of
\emph{chains of arity~$a$},  % fp: this clarifies that chains and packets are indexed
\emph{packets of arity~$a$},
and \emph{bodies}.

\begin{itemize}
\item A \emph{node} is a pair of a~color and an~arity,
      where an arity $a$ is a member of $\{ 0, 1, 2 \}$. \\
      A~node's arity can be~0 only if its color is green.
\item A \emph{chain} of arity $a$ is defined as follows:
  \begin{itemize}
  \item A chain of arity~0 is empty. % trivial. % It carries no data. % It consists of nothing.
  \item A chain of arity~1 consists of
        a packet of arity~$a'$
        (the \emph{root packet}) \\
        and
        a chain of arity~$a'$
        (the \emph{child chain}),
        for some $a'$.
        \\
        If the root packet is red then the child chain must be green.
        \\
        (\emph{A chain is green} if its zero, one, or two root packets are green.)
        % fp: yes, this is a bit subtle, as one must read the whole definition
        %     in order to understand what we mean by "the root packets of a chain".
        %     But I think this can work.
  \item A chain of arity~2 consists of
        two chains of arity~1.
  \end{itemize}
\item A \emph{packet} of arity~$a$
      consists of
      a body
      and
      a~\greenorred
      node of arity~$a$ (the \emph{tail node}).
      \\
      The color of a packet is the color of its tail node.
\item A \emph{body} is one of the following:
  \begin{itemize}
  \item Empty.
  \item An orange or yellow node of arity~1 \\
        and a body
        (the \emph{continuation}).
        % (the \emph{single child})
  \item An orange node of arity~2, \\
        a chain of arity~1 whose root packet is green, \\
        % (the \emph{left child}; the \emph{non-preferred child})
        and
        a body
        (the \emph{continuation}).
        % (the \emph{right child}; the \emph{preferred} child)
  \item A yellow node of arity~2, \\
        a body
        (the \emph{continuation}), \\
        % (the \emph{left child}; the \emph{preferred} child)
        and
        a chain of arity~1.
        % (the \emph{right child}; the \emph{non-preferred child}).
  \end{itemize}
\end{itemize}

% Fix Indentation
\noindent
This representation allows
moving in constant time
``from a~packet to its children'',
that~is,
from a chain of arity~1 to its child chain.
It also allows
moving in constant time
from a~packet
to its tail node.
%
% The~concrete runtime representation of \cdeques,
% which appears in \sref{sec:cadeque:ocaml},
% is based on this representation,
% with buffers.

% ------------------------------------------------------------------------------

\subsubsection{Correspondence between \treerep and \pointerrep}

The \treerep
(\sref{sec:cadeque:treerep:nobuffers})
and the \pointerrep
(above)
are
isomorphic:
they are alternative representations of the same data.
Indeed,
a~chain, as defined above, can be decoded into a well-formed forest
(\sref{sec:cadeque:treerep:nobuffers:trees}),
which,
once decomposed into packets
as per the rules of \sref{sec:cadeque:treerep:nobuffers:packets},
satisfies the semi-regularity constraints
of \sref{sec:cadeque:treerep:nobuffers:regularity}.
If this chain is green, then the resulting forest
also satisfies the regularity constraint
of \sref{sec:cadeque:treerep:nobuffers:regularity}.
Conversely, every forest that is well-formed and satisfies
the semi-regularity constraints
can be encoded as a chain, as defined above.

\input{figures/cd_chains}

The process of
gradually
decoding
a~chain
into a forest
is illustrated in \fref{fig:cd_chains}.
A~chain
that has not yet been decoded
is depicted as an~opaque area.
Its arity is indicated by the number~1~or~2
inside this area.
(Chains of arity~0 are not shown, as they are empty.)
%
% All other notations remain consistent
% with \fref{fig:cd_packets}.
%
At~stage~$(a)$,
a chain of arity~2
stands for the entire forest
shown in \fref{fig:cd_packets}.
At stage~$(b)$,
this chain of arity~2 has been unfolded % split
into two chains of arity~1.
%
% Thus, the Pair constructor in the inductive definition of chains has been unfolded.
%
At stage~$(c)$,
each of these chains of arity~1
has been unfolded into a root packet and a child chain.
Furthermore,
the root packet itself
has been unfolded
into a body and a tail node,
and the body has been unfolded
into a sequence of orange or yellow nodes,
some of which
carry chains.
In the following stages,
the process continues
until the entire forest
(and its decomposition into packets)
emerge.

The remarks
in the previous paragraphs
are not essential.
They are provided only
in the hope that
they may help the reader
understand the mutually inductive structure
of chains, packets, and bodies.

% ------------------------------------------------------------------------------
% ------------------------------------------------------------------------------

\subsection{Buffers and recursive slowdown}
\label{sec:cadeque:buffers}

The structure that we have just described is abstract, or~incomplete:
it is just a~skeleton.
To obtain
a complete description of \cdeques,
one must enrich
this structure
with
elements and~buffers.
Both the \treerep
and the \pointerrep
can be enriched
in this way.

The introduction of buffers can be briefly and partially described as follows.
\begin{enumerate}
\item
Every data structure is parameterized with a type~$A$,
% which is
the type of its elements.
\item
Every node carries
a~\emph{prefix buffer} and a~\emph{suffix buffer}
that contain elements of type~$A$.
\item
A~buffer is a~non-catenable \sdeque,
as~defined in \sref{sec:deque}.
\item
As one goes from a node down to a~child,
the parameter~$A$ is changed to ``stored triple of~$A$'',
where ``stored triple'' is one of the data types
involved in the definition of \cdeques.
\end{enumerate}

% Fix Indentation
\noindent
In comparison with \sref{sec:deque},
the~nature of buffers changes.
There,
a~buffer was a tuple of bounded size.
% its size was at most 5
Here,
it is a non-catenable deque,
whose size is unbounded.
(By the \emph{size} of a data structure,
we mean the number of elements that it contains.)
% This definition may be somewhat ambiguous,
% because when a subdeque contains elements
% of type A^{2^k}, it is unclear whether one
% such composite element counts as 1 or as 2^k.
% Let's hope that this ambiguity is unnoticed or tolerable.

Recursive slowdown is still present,
but follows a~more complex and more deeply recursive pattern
than in \sref{sec:deque}.
There,
as one went from a node down to its child,
the parameter~$A$ was transformed to $A\times A$.
Here,
it~is changed to ``stored triple of~$A$'',
where ``stored triple''
is one of the data types
that we are in the process of defining.
This makes
\cdeques
a~\emph{truly nested data type}
\cite{abel-matthes-uustalu-05}.

To complete the description of buffers,
a connection between buffer sizes and colors must be established.
Very roughly,
buffer sizes
must be subjected
to certain~constraints,
and
the color of a~node
must be related with
% fp: avoid "is determined by"
%     because the node kind also plays a role
the sizes of the buffers that it~carries.
%
% We could try to say that
% + we distinguish several kinds of nodes: only|right|left|stored
% + a node's kind implies
%   certain lower bounds on the sizes of its buffers;
% + a node's kind
%   and the sizes of its buffers
%   determine
%   its color.
%
% But it seems simpler to just move on to the next subsection.
% Otherwise, we create redundancy and confusion.
%
This is spelled out in the next subsection.

% ------------------------------------------------------------------------------
% ------------------------------------------------------------------------------

\subsection{\Treerep, with buffers}
\label{sec:cadeque:treerep:buffers}

To describe buffers in full~detail,
it is preferable
to do so
in the setting of the \treerep
first.
This description can later be adapted
to the setting of the \pointerrep.

% ------------------------------------------------------------------------------

\subsubsection{Triples and \cdeques} % corresponds to: trees and forests
\label{sec:cadeque:treerep:buffers:trees}

The \treerep of \cdeques
involves two mutually recursive types,
namely trees and forests
(\sref{sec:cadeque:treerep:nobuffers:trees}).
Once this representation is
enriched with buffers,
it is still described by two mutually recursive types,
namely triples and \cdeques.
A~triple is an enriched~tree;
a~\cdeque is an enriched~forest.
Here is a definition of these types \cite[\S6.1]{kaplan-tarjan-99}:

\begin{itemize}
\item
A \emph{triple} of elements of type~$A$
is composed of:
\begin{itemize}
\item a \sdeque, also known as a \emph{prefix buffer}, \\
whose elements have type~$A$;
\item a \cdeque, also known as the \emph{child \cdeque}, \\
whose elements are triples of elements of type~$A$, \\
which \kt refer to as
\emph{stored triples};
and
\item a \sdeque, also known as a \emph{suffix buffer}, \\
whose elements have type~$A$.
\end{itemize}
\item
A \emph{\cdeque} is either:
\begin{itemize}
\item empty;
\item composed of a single triple,
      which \kt refer to as
      an \emph{only triple};
or
\item composed of two triples,
      which they refer to as
      a \emph{left triple} and a \emph{right triple}.
\end{itemize}
\end{itemize}

% We do not yet explain
% how one assigns a color to a triple.
% We do so later on (\sref{sec:cadeque:treerep:buffers:colors}).
%Fix Indentation
\noindent
The four kinds of triples
mentioned above
% (namely
% stored~triples,
% only~triples,
% left~triples,
% and
% right~triples)
can be divided in two independent categories.
On the one hand,
stored~triples
exist \emph{inside} buffers:
they serve as elements of buffers.
When buffers are abstracted away,
they disappear altogether.
On the other hand,
only~triples,
left~triples,
and
right~triples
exist \emph{outside} buffers.
When buffers are abstracted away,
they
form the~skeleton
that remains.
We refer to
them
collectively as
\emph{ordinary triples}.
In our OCaml and \Coq implementations,
stored~triples
and
ordinary~triples
form
two distinct~types.
%
% fp:
% I have wondered whether only triples, left triples, and right triples
% could form three distinct types. The answer is negative. There is one
% place in the code, in \oc|ensure_green|, where one looks at the root
% triple of a tree and asks: is this an only, left, or right triple?
%

% ------------------------------------------------------------------------------

\subsubsection{Kinds}
\label{sec:cadeque:treerep:buffers:kinds}

% This is a concept that could have been introduced earlier,
% at the level of the \treerep with buffers.
% But it would have been useless there;
% it was simpler, there, to just assume that a node has a color.

To distinguish between the three varieties of ordinary triples,
we say that the \emph{kind} of an~ordinary triple
is ``only'', ``left'', or ``right''.

Above,
we have indicated
that
an~ordinary triple
corresponds
to a subtree in the skeleton.
%
% Based on this idea,
The kind of an ordinary triple
records
the relationship
between this subtree and its parent:
it~indicates
whether this subtree
is
the only child,
the left child,
or
the right child
of its~parent.

% In our implementations,
% kinds will exist both
% at the type level
%   (we sometimes statically keep track of a node's kind)
% and at runtime
%   (we tag a node with its kind).

% ------------------------------------------------------------------------------

\subsubsection{Size constraints}
\label{sec:cadeque:treerep:buffers:size}

\kt impose size constraints
on buffers.
In the case of
stored~triples,
the constraints are as follows:
\begin{enumerate}[label=(ST\arabic*)]
\item
\label{size:stored:big}
In a stored triple,
both buffers must have
at least~3 elements,
\item
\label{size:stored:small}
unless the child \cdeque
and one of the buffers are empty, \\
in which case the other buffer must have
at least~3 elements.
\end{enumerate}
In the case of
ordinary~triples,
the constraints are as follows:
\begin{enumerate}[label=(OT\arabic*)]
\item
\label{size:ordinary:only}
In an only triple,
both buffers must have
at least~5 elements,
\item
\label{size:ordinary:onlyend}
unless the child \cdeque
and one of the buffers are empty, \\
in which case the other buffer must contain
at least~1 element.
\item
\label{size:ordinary:left}
In a left triple,
the prefix buffer must have
at least~5 elements, \\
while the suffix buffer
must have exactly~2 elements.
\item
\label{size:ordinary:right}
% Symmetrically,
In a right triple,
the prefix buffer must have exactly~2 elements, \\
while the suffix buffer must have
at least~5 elements.
\end{enumerate}

% ------------------------------------------------------------------------------

\subsubsection{Color assignment}
\label{sec:cadeque:treerep:buffers:colors}

\kt
define a mapping of sizes to colors:
\begin{enumerate}[label=(C\arabic*)]
\setcounter{enumi}{4}
\item
\label{color5}
size~5~is red;
\item
\label{color6}
size~6~is orange;
\item
\label{color7}
size~7~is yellow;
\item
\label{color8}
size~8 and greater sizes are green.
\end{enumerate}
%Fix Indentation
\noindent
This mapping allows assigning a color to a buffer,
based on the size of this buffer,
provided this size is at least~5.
Then, \kt
assign colors to ordinary triples,
%
% fp: I have removed the rule: ``A stored~triple is green'',
%     because it seems irrelevant. Jules agrees.
%
based on their kind
and on the colors
% (sizes)
of some of their buffers:
\begin{enumerate}[label=(BC\arabic*)]
\item
\label{buffer:color:empty}
An ordinary triple
whose child \cdeque is empty
is green.
\item
An ordinary triple
whose child \cdeque is nonempty
obeys the following rules:
  \begin{enumerate}[label=(BC2\alph*)]
  \item
  \label{buffer:color:nonempty:only}
  The color of an only triple
  % with a nonempty child \cdeque
  is the worst color
  of its two buffers.
  % In other words,
  % it is the color that corresponds to the minimum
  % of the sizes of its two buffers.
  \item
  \label{buffer:color:nonempty:left}
  The color of a left triple
  % with a nonempty child \cdeque
  is the color of its prefix buffer.
  \item
  \label{buffer:color:nonempty:right}
  The color of a right triple
  % with a nonempty child \cdeque
  is the color of its suffix buffer.
  \end{enumerate}
\end{enumerate}
The~ordering of colors,
from worst to best,
is
$\text{red} < \text{orange} < \text{yellow} < \text{green}$.

% ------------------------------------------------------------------------------

\subsubsection{Regularity}
\label{sec:cadeque:treerep:buffers:regularity}

As each ordinary triple
now implicitly carries a color,
one can decompose
triples and \cdeques
into packets,
and impose regularity constraints,
in the same manner as earlier
(\sref{sec:cadeque:treerep:nobuffers:packets,sec:cadeque:treerep:nobuffers:regularity}).

% ------------------------------------------------------------------------------

\subsection{\Pointerrep, with buffers}
\label{sec:cadeque:pointerrep:buffers}

Of four ways of looking at \kt's \cdeques,
we have just presented three.
%
% We have presented three of the four vertices of the diamond.
%
Indeed,
we have presented
a skeletal structure,
without buffers,
under two equivalent representations,
namely
the \treerep
(\sref{sec:cadeque:treerep:nobuffers})
and
the \pointerrep
(\sref{sec:cadeque:pointerrep:nobuffers}).
Then, we have introduced buffers
% (\sref{sec:cadeque:buffers})
and
have given a~complete description
of \cdeques
under the \treerep
(\sref{sec:cadeque:treerep:buffers}).
At~this point,
we are ready to describe
the fourth and last view, % the last vertex of the diamond
that~is,
the \pointerrep of \cdeques,
with buffers.
We~do not give an English definition of~it.
Instead, we move directly to our OCaml definition.

\section{Catenable deques: OCaml implementation}
\label{sec:cadeque:ocaml}

% fp: are there sum types where every `match` construct has only one branch
% (which means that the tag is actually never used at runtime)?
% If so, document it.
% e.g., the type `node`?
%   -- still need to distinguish Only versus Only_end

We now transcribe the ideas of the previous section
into OCaml type definitions and code.

% ------------------------------------------------------------------------------

% Buffers.

\subsection{Buffers: size-indexed \sdeques}
\label{sec:cadeque:ocaml:buffers}

As explained earlier (\sref{sec:cadeque:buffers}),
within a~catenable deque,
a~\emph{buffer}
is a~non-catenable deque,
subject to certain
size constraints.
% (\sref{sec:cadeque:treerep:buffers:size}).
%
Thus,
our implementation of \sdeques
(\sref{sec:deque:ocaml})
can be used
as a~building block
in our implementation of \cdeques.
% and this is desirable.

However,
as far as static type-checking is concerned,
the~API
% or: type signature
offered by
our
% OCaml implementation of
\sdeques
is not quite as expressive
as we would like.
This~API consists of
a~parameterized type \oc|'a deque|~(\fref{fig:ncd_ocaml_types}),
which is equipped
with a number of constants and operations,
such as \oc|empty|, \oc|push|, \oc|pop|, \oc|inject|, and \oc|eject|.
It offers no way of referring
at the type~level
to~the~size of a \sdeque.%
\footnote{This API could offer a \oc|size| function,
which computes and returns the size of a \sdeque.
However, an~expression of the form \oc|size(d)| cannot appear
within~a~type: OCaml does not have dependent types.}
Therefore,
if we chose to build our implementation of \cdeques
directly on~top of this~API,
we would be unable to
express
at the type~level
the size constraints
of \sref{sec:cadeque:treerep:buffers:size}.
Thus, the OCaml type-checker
would be unable to statically verify
that our implementation of \cdeques
respects these size constraints.
Yet,
% as noted earlier,
such a~static check
is highly desirable:
it~is a~crucial help,
and a guide,
while developing the code.

\newcommand{\footnotewhylowerbound}{\footnote{%
It would seem more natural for this index
to represent the size of the \sdeque,
as opposed to a~lower bound on
the size of the \sdeque.
However, some operations in the deque API,
such as \oc|push_vector|
(not~shown in \fref{fig:cd_ocaml_buffer}),
insert an undetermined number of elements into a deque.
% LATER check: beside push_vector and its variants,
%             are there other functions that explain
%             why (only) a lower bound must be tracked?
%
For this reason,
only a~lower bound on the size of the \sdeque
can be statically known.}\xspace}

\newcommand{\footnoteexactlytwo}{\footnote{%
This API does not let us express the~constraint:
``this buffer must have exactly two elements''.
Fortunately,
we are able to remove the need for this constraint:
where a~buffer of exactly two elements is required,
we instead use two naked elements.
This is visible in the definition of the type~\oc|node|
(\sref{sec:cadeque:ocaml:nodes:nodes}).}\xspace}

To address this problem,
on top of the \sdeque~API of \sref{sec:deque:ocaml},
we develop a~richer~API
where the type of \sdeques
is indexed with a type-level natural number~\oc|'n|.%
\footnote{Since OCaml 5.5,
the bizarre declarations \oc|type z = private Z|
and \oc|type 'n s = private S| are needed so as to ensure
that the OCaml type-checker views \oc|z| and \oc|'n s|
as two \emph{provably distinct} abstract types.
The data constructors \oc|Z| and \oc|S| are never used.}
The index \oc|'n| represents
a~lower bound\footnotewhylowerbound
on the size of the deque.
This API
lets us express the constraints
that impose a~constant lower bound on the size of a buffer,
that is,
all constraints
of the form:
``this buffer must have at least $k$ elements'',
where~$k$ is a constant.\footnoteexactlytwo

\begin{figure}[t]
\begin{lstlisting}[language=ocaml,backgroundcolor=\colorocaml]
type    z     = private Z      (* The number 0   *)
type 'n s     = private S      (* The number n+1 *)

type      eq0 = z              (* The number 0   *)
type      eq1 = z s            (* The number 1   *)
type      eq2 = z s s          (* The number 2   *)

type 'n plus1 = 'n s           (* The number n+1 *)
type 'n plus2 = 'n s plus1     (* The number n+2 *)
type 'n plus3 = 'n s plus2     (* The number n+3 *)
type 'n plus4 = 'n s plus3     (* The number n+4 *)
type 'n plus5 = 'n s plus4     (* The number n+5 *)
\end{lstlisting}
\caption{\ncdocaml: type-level natural numbers}
\label{fig:cd_ocaml_nat}
\end{figure}

\begin{figure}[t]
\begin{lstlisting}[language=ocaml,backgroundcolor=\colorocaml]
module Buffer : sig
  type ('a, 'n) t
  val empty : ('a, z) t
  val push   : 'a -> ('a, 'n) t -> ('a, 'n s) t (*@ \label{line:buffer:push} *)
  val inject : ('a, 'n) t -> 'a -> ('a, 'n s) t (*@ \label{line:buffer:inject} *)
  val pop   : ('a, 'n s) t -> 'a * ('a, 'n) t   (*@ \label{line:buffer:pop} *)
  val eject : ('a, 'n s) t -> ('a, 'n) t * 'a   (*@ \label{line:buffer:eject} *)
  (* Additional functions omitted *)
end = struct
  type ('a, 'n) t = 'a Deque.t (*@ \label{line:buffer:t} *)
  let empty  = Deque.empty
  let push   = Deque.push
  let inject = Deque.inject
  let pop t = match Deque.pop t with
    | None -> assert false (* this cannot happen *) (*@ \label{line:buffer:assertfalse1} *)
    | Some (x, t') -> (x, t')
  let eject t = match Deque.eject t with
    | None -> assert false (* this cannot happen *) (*@ \label{line:buffer:assertfalse2} *)
    | Some (t', x) -> (t', x)
  (* Additional functions omitted *)
end
\end{lstlisting}
\caption{\cdocaml: size-indexed \sdeques, also known as buffers}
\label{fig:cd_ocaml_buffer}
\end{figure}

This richer API and its construction
appear in \fref{fig:cd_ocaml_nat,fig:cd_ocaml_buffer}.
\gitrepofile{lib/cadeque_core.ml}
The construction relies on type abstraction.
Outside of the abstraction barrier
(that is, from a~user's point of~view),
the~type of buffers,
\oc|('a, 'n) Buffer.t|,
is~an~abstract type,
whose parameter~\oc|'n|
represents
a~lower bound on the size of the buffer.
Inside the abstraction barrier,
the parameter~\oc|'n|
is irrelevant:
indeed,
the type~\oc|('a, 'n) t| % LATER or: Buffer.t? (question of qualified names)
is defined as a synonym for~\oc|'a deque| % LATER: or: Deque.deque? (question of qualified names)
(\lineref{buffer:t} in \fref{fig:cd_ocaml_buffer}).
This yields a form of one-sided security.
Inside the abstraction barrier,
runtime assertions % of the form ``this cannot happen''
(\linerefs{buffer:assertfalse1}{buffer:assertfalse2})
ensure
that the parameter~\oc|'n|
is indeed a~lower bound on the size of the buffer.
Outside the abstraction barrier,
this parameter
is meaningful and
can be used to express size constraints, % lower bounds only
which
the OCaml type-checker
statically enforces.
This idiom is known in the literature
as a~phantom type parameter~\cite{leijen-meijer-99}.

The operations \oc|push| and \oc|inject|
increase the size of a buffer by one.
Therefore,
a~lower bound on the size of the buffer
can be safely increased by one, as well.
The types assigned to \oc|push| and \oc|inject|
(\linerefs{buffer:push}{buffer:inject} in \fref{fig:cd_ocaml_buffer})
reflect this idea.%
\footnote{It would be safe for \oc|push| to have result type
\oc|('a, 'n) t| instead of \oc|('a, 'n s) t|. This would encode
a~weaker lower bound, namely~$n$, instead of $n+1$.
Some operations not shown in \fref{fig:cd_ocaml_buffer},
such as \oc|push_vector|,
introduce an under-approximation in this way.}
The operations \oc|pop| and \oc|eject|
must be applied only to a nonempty buffer,
and
decrease the size of this buffer by one.
Therefore,
a~lower bound on the size of the buffer
must be decreased by one.
The types assigned to \oc|pop| and \oc|eject|
(\linerefs{buffer:pop}{buffer:eject})
reflect these two requirements.
The fact that \oc|pop| and \oc|eject| can be applied
only to a nonempty buffer
guarantees that the runtime assertions
at \linerefs{buffer:assertfalse1}{buffer:assertfalse2}
cannot fail;
however, this informal argument
is not machine-checked.

% ------------------------------------------------------------------------------

% Nodes.

\subsection{Nodes} % and their arities, kinds, and colors
\label{sec:cadeque:ocaml:nodes}

We now wish to define a type of \emph{nodes}.
At a~concrete level,
we want
a~node to be
an~ordinary~triple,
as defined in~\sref{sec:cadeque:treerep:buffers:trees},
deprived of its child cadeque.
Thus, roughly speaking,
a~node
should~be
a~pair of buffers.
Furthermore,
a~node should be subject to the size constraints
that bear on ordinary triples.
Four cases,
\ref{size:ordinary:only}
to~\ref{size:ordinary:right},
have been listed
in~\sref{sec:cadeque:treerep:buffers:size}.
Every node must match exactly one of these cases.
Thus,
nodes
should~inhabit a sum type
% (a GADT)
whose four constructors
correspond to these four cases.

% ------------------------------------------------------------------------------

\subsubsection{Type-level properties of nodes}
\label{sec:cadeque:ocaml:nodes:parameters}

We need to keep track,
at the type level,
of four properties of a node,
namely
the type of its elements,
its arity,
its kind,
and
its color.
An~arity is 0,~1,~or~2
(\sref{sec:cadeque:pointerrep:nobuffers:packets}).
A \emph{kind} is ``only'', ``left'', or ``right''
(\sref{sec:cadeque:treerep:buffers:kinds}).%
\footnote{A~node whose kind is ``only'' is an \emph{only node}.
Similarly, a~node whose kind is ``left'' is a \emph{left node},
and a~node whose kind is ``right'' is a \emph{right node}.}
Therefore,
we parameterize the type \oc|node|
with four parameters:
this~type takes the form
\oc|('a, 'arity, 'kind, 'c) node|.

The parameter~\oc|'a|
is the type of the elements
of this node. % fp: is it clear what we mean? LATER

The parameter \oc|'arity|
is meant to range over type-level arities.%
\footnote{This cannot be explicitly expressed in OCaml.}
%
% An arity is 0, 1, or~2.
%
A type-level arity is a type-level natural number
among 0, 1, and~2.
We introduce
\oc|empty|, \oc|single|, and \oc|pair|
as~convenient synonyms
for these three type-level natural numbers
(\fref{fig:cd_ocaml_kinds}).
% LATER revoir si/quand on introduit ces mots

The parameter \oc|'kind|
is meant to range over type-level kinds.
We introduce
\oc|only|, \oc|left|, and \oc|right|
as three distinct type-level kinds
(\fref{fig:cd_ocaml_kinds}).

The parameter \oc|'c|
is meant to range over type-level colors,
whose definition appears in \fref{fig:cd_ocaml_gyor}.
\gitrepofile{lib/color_GYOR.ml}
This definition obeys the same scheme
as our earlier definition
(\fref{fig:bc_ocaml_gyr}).
This time,
a new orange hue is added;
colors are now composed of four hue bits.

% ------------------------------------------------------------------------------

% Colors.

\subsubsection{Color assignment}
\label{sec:cadeque:ocaml:nodes:coloring}

% ------------------------------------------------------------------------------

% Assigning colors and arities to nodes.

\begin{figure}[t]
\begin{lstlisting}[language=ocaml,backgroundcolor=\colorocaml]
(* An arity is 0, 1, or 2. *)
type empty  = eq0
type single = eq1
type pair   = eq2

(* A kind is only, left, or right. *)
type only   = private Only
type left   = private Left
type right  = private Right
\end{lstlisting}
\caption{\cdocaml: type-level arities and kinds}
\label{fig:cd_ocaml_kinds}
\end{figure}

\begin{figure}[t]
\begin{lstlisting}[language=ocaml,backgroundcolor=\colorocaml]
(* A group of distinct types encode the presence or absence of a hue. *)
type somegreen  = SOME_GREEN
type nogreen    = NO_GREEN
type someyellow = SOME_YELLOW
type noyellow   = NO_YELLOW
type someorange = SOME_ORANGE
type noorange   = NO_ORANGE
type somered    = SOME_RED
type nored      = NO_RED
(* A color is a quadruple of four hue bits. *)
type green      =  somegreen *   noyellow *   noorange *   nored
type yellow     =    nogreen * someyellow *   noorange *   nored
type orange     =    nogreen *   noyellow * someorange *   nored
type red        =    nogreen *   noyellow *   noorange * somered
\end{lstlisting}
\caption{\cdocaml: type-level colors}
\label{fig:cd_ocaml_gyor}
\end{figure}

\begin{figure}
\begin{lstlisting}[language=ocaml,backgroundcolor=\colorocaml]
(* Convenient synonyms. *)
type ('a, 'n) prefix = ('a, 'n) Buffer.t
type ('a, 'n) suffix = ('a, 'n) Buffer.t

(* Constraints on buffer sizes, arity, and color. *)
(* Delta is size minus 5. *)
type ('prefix_delta, 'suffix_delta, 'arity, 'c) node_coloring =
  | EN : (      _,       _,     eq0, green ) node_coloring
  | GN : (_ plus3, _ plus3, _ plus1, green ) node_coloring
  | YN : (_ plus2, _ plus2, _ plus1, yellow) node_coloring
  | ON : (_ plus1, _ plus1, _ plus1, orange) node_coloring
  | RN : (      _,       _, _ plus1, red   ) node_coloring

(* Nodes. *)
type ('a, 'arity, 'kind, 'c) node =
  | Only      : ('pdelta, 'sdelta, 'n plus1, 'c) node_coloring
              * ('a, 'pdelta plus5) prefix
              * ('a, 'sdelta plus5) suffix
             -> ('a, 'n plus1, only, 'c) node
  | Only_end  : ('a, _ plus1) Buffer.t
             -> ('a, eq0, only, green) node
  | Left      : ('pdelta, _, 'arity, 'c) node_coloring
              * ('a, 'pdelta plus5) prefix
              * ('a * 'a)
             -> ('a, 'arity, left, 'c) node
  | Right     : (_, 'sdelta, 'arity, 'c) node_coloring
              * ('a * 'a)
              * ('a, 'sdelta plus5) suffix
             -> ('a, 'arity, right, 'c) node
\end{lstlisting}
\caption{\cdocaml: types for nodes}
\label{fig:cd_ocaml_node}
\end{figure}

To express the coloring rules
of~\sref{sec:cadeque:treerep:buffers:colors},
we define the type
\oc|('prefix_delta, 'suffix_delta, 'arity, 'c) node_coloring|.
% It is named node_coloring
% because there is another type triple_coloring
% in the repository.
%
This type encodes a relation
between
the sizes of a node's prefix and suffix buffers,
its arity,
and
its color.
An~inhabitant of this type is a witness that the relation holds.

The type variables \oc|'prefix_delta| and \oc|'suffix_delta|
are meant to
range over type-level natural numbers.
In the case of an only node,
where the sizes of both buffers determine the node's color
\ref{buffer:color:nonempty:only},
we intend to instantiate
\oc|'prefix_delta|
and~\oc|'suffix_delta|
respectively
with
the size of the prefix buffer minus~5
and
with
the size of the suffix buffer minus~5,
thereby constraining the sizes of both buffers.
In the case of a left node,
where only the size of the prefix buffer matters
\ref{buffer:color:nonempty:left},
we intend to instantiate
\oc|'prefix_delta|
and~\oc|'suffix_delta|
respectively
with
the size of the prefix buffer minus 5
and
with
a dummy variable,
thereby constraining only the size of the prefix buffer.
Symmetrically,
in the case of a right node~\ref{buffer:color:nonempty:right},
we intend to constrain
only the size of the suffix buffer.
This will be visible
in~\sref{sec:cadeque:ocaml:nodes:nodes},
where the \oc|node_coloring| relation
is used in the definition of the type~\oc|node|.
%
% fp: we are anticipating a bit,
%     but it seems difficult to *not* explain the meaning of
%     prefix_delta and suffix_delta in detail at this point.

The definition of \oc|node_coloring|
is shown in~\fref{fig:cd_ocaml_node}.
The constructor \oc|EN| concerns a~node of arity~0;
the other four constructors concern a node of nonzero arity.

``A node of arity~0''
can also be understood as
``a node whose child \cdeque has arity~0'',
that~is,
``a node whose child \cdeque is empty''.
Therefore,
the constructor \oc|EN|,
which states that a node of arity~0 is green,
corresponds to~\ref{buffer:color:empty}.
% in~\sref{sec:cadeque:treerep:buffers:colors}.

Regarding nodes of nonzero arity,
the constructors~\oc|GN|, \oc|YN|, \oc|ON|, and \oc|RN|
encode four implications:
if the node's color is green
then the two ``deltas''
must be at least~3;
if its color is yellow
then they
must be at least~2;
if it is orange
then they
must be at least~1;
and
if it is red
then fine.
This encodes
Rules~\ref{color5}--\ref{color8}.
In fact,
this encodes a relaxed version of these rules:
as in the case of non-catenable deques (\sref{sec:deque:ocaml}),
we allow a node to receive
a suboptimal color.
% a worse color than its true color

% ------------------------------------------------------------------------------

% Nodes.

\subsubsection{Nodes}
\label{sec:cadeque:ocaml:nodes:nodes}

The definition of the type~\oc|node|
appears in \fref{fig:cd_ocaml_node}.
As announced earlier,
this is a sum type
whose four constructors
correspond to~\ref{size:ordinary:only}--\ref{size:ordinary:right}.

The constructors \oc|Only| and \oc|Only_end|
construct an only node,
that is,
a node whose kind is \oc|only|.
The constructor \oc|Left|
constructs a left node,
whose kind is \oc|left|.
The constructor \oc|Right|
constructs a right node,
whose kind is \oc|right|.
\begin{itemize}

\item
The constructor \oc|Only| corresponds
to~\ref{size:ordinary:only}.
It constructs a node of nonzero arity. % \oc|'n plus1|
It~carries two buffers.
The sizes of these buffers
as well as the node's arity and color
are subject to a~coloring constraint.

\item
The constructor \oc|Only_end| corresponds
to~\ref{size:ordinary:onlyend}.
It~constructs a~node of arity~0. % \oc|eq0|
It~carries only one buffer,
which must contain at least one element.
Its~color is green.
No coloring constraint is needed in this case.

\item
The constructor \oc|Left| corresponds
to~\ref{size:ordinary:left}.
It~carries a prefix buffer of size at least~5
and a~suffix buffer of size~2.
The size of the prefix buffer
as well as the node's arity and color
are subject to a coloring constraint.
The suffix buffer
is represented
simply as a pair of elements.
% This is simpler and more efficient
% than using a buffer of size exactly 2.

\item
The constructor \oc|Right| corresponds
to~\ref{size:ordinary:right}.
It is symmetric
with~\oc|Left|.

\end{itemize}

% ------------------------------------------------------------------------------

% Chains and packets.

\subsection{Chains and packets}
\label{sec:cadeque:ocaml:chains}

\begin{figure}
\begin{lstlisting}[language=ocaml,backgroundcolor=\colorocaml]
type 'a stored =
| Big :
     ('a, _ plus3) prefix                      (* prefix buffer *)
   * ('a stored, _, only, _, _) chain          (* child chain *)
   * ('a, _ plus3) suffix                      (* suffix buffer *)
  -> 'a stored
| Small :
     ('a, _ plus3) Buffer.t                    (* just one buffer *)
  -> 'a stored

and ('a, 'b, 'hk, 'tk) body =
| Hole :
     ('a, 'a, 'k, 'k) body
| Single_child :
     ('a, eq1, 'hk, nogreen*_*_*nored) node    (* orange or yellow node of arity 1 *)
   * ('a stored, 'b, only, 'tk) body               (* child: continuation of body *)
  -> ('a, 'b, 'hk, 'tk) body
| Pair_orange :
     ('a, eq2, 'hk, orange) node               (* orange node of arity 2 *)
   * ('a stored, single, left, green, green) chain (* left: green chain of arity 1 *)
   * ('a stored, 'b, right, 'tk) body              (* right: continuation of body *)
  -> ('a, 'b, 'hk, 'tk) body
| Pair_yellow :
     ('a, eq2, 'hk, yellow) node               (* yellow node of arity 2 *)
   * ('a stored, 'b, left, 'tk) body               (* left: continuation of body *)
   * ('a stored, single, right, 'c, 'c) chain      (* right: a chain of arity 1 *)
  -> ('a, 'b, 'hk, 'tk) body

and ('a, 'b, 'arity, 'kind, 'c) packet =
| Packet :
     ('a, 'b, 'kind, 'tk) body                 (* body *)
   * ('b, 'arity, 'tk, _*noyellow*noorange*_ as 'c) node (* green or red tail node *)
  -> ('a, 'b stored, 'arity, 'kind, 'c) packet

and ('a, 'arity, 'kind, 'lc, 'rc) chain =
| Empty :
     ('a, empty, _, _, _) chain
| Single :
     ('c, 'lc1, 'rc1) regularity               (* regularity constraint *)(*@ \label{line:Single:regularity} *)
   * ('a, 'b, 'arity1, 'kind, 'c) packet       (* root packet *)
   * ('b , 'arity1, only, 'lc1, 'rc1) chain    (* child chain *)
  -> ('a, single, 'kind, 'c, 'c) chain
| Pair :
     ('a, single, left , 'lc, 'lc) chain       (* left single chain *)
   * ('a, single, right, 'rc, 'rc) chain       (* right single chain *)
  -> ('a, pair, _, 'lc, 'rc) chain
\end{lstlisting}
\caption{\cdocaml: stored triples, bodies, packets, and chains}
% I have omitted "types for" so that the caption fits on one line.
\label{fig:cd_ocaml_stored}
\label{fig:cd_ocaml_body}
\label{fig:cd_ocaml_pkt}
\label{fig:cd_ocaml_chain}
\label{fig:cd_ocaml_central}
\end{figure}

We now present
the definitions of the types
\oc|stored|,
\oc|body|,
\oc|packet|,
and \oc|chain|.
These four definitions are mutually recursive.
They appear in \fref{fig:cd_ocaml_central}.
In the following,
we first explain
the five parameters of the type~\oc|chain|.
Then,
we describe each of these four definitions
in turn.
Finally,
we present the type
of \cdeques.

\subsubsection{Type-level properties of chains}

We keep track,
at the type~level,
of five properties of a~chain,
namely
the type of its elements,
its arity,
the kind of its root node
(when it has a~single root node),
and two colors
that describe its root node or root nodes.
Thus,
we parameterize the type \oc|chain|
with five parameters:
this~type takes the form
\oc|('a, 'arity, 'kind, 'lc, 'rc) chain|.

The parameter \oc|'a| is the type of the elements of this chain.

The parameter \oc|'arity|
ranges over type-level arities.
In~\sref{sec:cadeque:ocaml:nodes:parameters},
we have introduced
\oc|empty|, \oc|single|, and \oc|pair|
as~synonyms
for the type-level numbers 0, 1, and~2.
The arity of a~chain
indicates
how many trees this chain contains,
when it is viewed as a forest.

The parameter \oc|'kind| represents the kind of the chain's
root node.
%
% (The kind of a~node has been defined
% in~\sref{sec:cadeque:ocaml:nodes:parameters}).
%
This parameter is meaningful
only in the case of single chains
(that is, chains of arity~1),
which have a single root node.
In the cases of empty chains and pair chains,
it is unconstrained.
% fp: Jules set it to \oc|only|, but that was arbitrary.
%     It seems more natural to leave it unconstrained.
%     I have made the change in the OCaml code; nothing breaks.

The parameters \oc|'lc| and \oc|'rc|
% range over type-level colors
represent the color or colors of this chain.
The names~\oc|'lc| and \oc|'rc|
stand for ``left color'' and ``right color''.
In the case of an empty chain,
% they are set to \oc|green|.
% fp: I have removed this constraint
these parameters are unconstrained.
In the case a single chain,
they both represent
the~color of the root packet of the chain.
In the case of a pair chain,
they represent
the~colors of the two root packets.
% that is,
% the root packet of the left single chain
% and
% the root packet of the right single chain.

\subsubsection{Stored triples}

The definition of the type~\oc|'a stored|
appears in \fref{fig:cd_ocaml_stored}.
A~stored triple is either \emph{big} or \emph{small}:
this corresponds to
the cases~\ref{size:stored:big}
and~\ref{size:stored:small}
in~\sref{sec:cadeque:treerep:buffers:size}.

A~big stored triple
consists of a prefix buffer, a child chain, and a suffix buffer,
where each buffer contains at least three elements.
As indicated in~\sref{sec:cadeque:treerep:buffers:trees},
recursive slowdown takes place at this point:
in a big stored triple
whose elements have type~\oc|'a|,
the child chain contains
elements of type~\oc|'a stored|.
%
% We might wish to explain why the child chain is \oc|only|.
%      This reflects the fact that it is a unique child
%      (its father is a stored triple)
%      as opposed to a left child or a right child.
% The reason is the same as for toplevel cadeques:
% see \sref{sec:cadeque:ocaml:chains:main}.

A~small stored triple consists of a single buffer
that contains at least three elements.

\subsubsection{Bodies}

Next in \fref{fig:cd_ocaml_body}
comes the definition of the type~\oc|('a, 'b, 'hk, 'tk) body|.
%
% fp: a reminder:
From
\sref{sec:cadeque:pointerrep:nobuffers:packets},
recall that
a~body and a~tail node, together, form a~packet.
One can think that the tail~node is
``placed in the hole''
at the end of the body.

The parameter~\oc|'a| is the type of the elements of this body.
The parameter~\oc|'b| is the type of the elements
of the tail~node
that follows this body.
These parameters
play the same role as the parameters by the same names
in the type \oc|('a, 'b, 'c) packet|
of non-catenable deques
(\fref{fig:ncd_ocaml_types}).

The parameter~\oc|'hk|,
for ``head kind'',
is the kind of the head node of this body,
if there is such a node,
that is,
if this body is nonempty.
Otherwise, it is the kind of the tail node
that follows this body.
The parameter~\oc|'tk|,
for ``tail kind'',
is the kind of the tail~node that follows this body.

The four constructors
correspond to the four cases
in the description of bodies
that was given
in~\sref{sec:cadeque:pointerrep:nobuffers:packets}.
\begin{itemize}

\item
The constructor \oc|Hole|
represents an empty body.
In this case,
because the ``head node'' is the tail node,
the parameters~\oc|'a| and \oc|'b| are equated,
and
the parameters~\oc|'hk| and \oc|'tk| are equated.

\item
The constructor \oc|Single_child|
carries a root node of arity~1, % \oc|eq1|
which must be orange or yellow, % \oc|nogreen * _ * _ * nored|
and a child body,
which represents the continuation of the parent body.
Because the~child body is the only child
of its parent,
its head kind must be \oc|only|.
As one moves from parent to child,
recursive slowdown takes place:
whereas the parent body has elements of type~\oc|'a|,
the child body has elements of type~\oc|'a stored|.

\item
The constructor \oc|Pair_orange|
carries a root node of arity~2, % \oc|eq2|
which must be orange,
and two children.
The left child,
which is the non-preferred child,
is a green chain of arity~1; % \oc|single|, \oc|green|
the~right child,
which is the preferred child,
is a body,
which represents the continuation of the parent body.
Because the child chain is the left child
of its parent,
its kind must be \oc|left|.
Because the child body is the right child
of its parent,
its head kind must be \oc|right|.
As one moves from parent to child,
recursive slowdown takes place:
both children
have elements of type~\oc|'a stored|.

\item
The constructor \oc|Pair_yellow| is symmetric
but places
no color constraint
on the child~chain.

\end{itemize}

\subsubsection{Packets}

The definition of the type
\oc|('a, 'b, 'arity, 'kind, 'c) packet|
is shown in \fref{fig:cd_ocaml_pkt}.

The parameter \oc|'a|
is the type of the elements of this packet.
The parameter~\oc|'b|
is the type of the elements
of the chain that follows this packet.
%
% This will be evident in the constructor \oc|Single|
% of the type \oc|chain|.
%
The parameter~\oc|'arity| is
the arity of this packet.
By definition
(\sref{sec:cadeque:pointerrep:nobuffers:packets}),
it is the arity of the tail node of this packet.
It is also the arity
of the chain that follows this packet.
The parameter~\oc|'kind|
is the kind of the root node
of this packet.
The parameter~\oc|'c|
is the color of this packet.
By definition
(\sref{sec:cadeque:pointerrep:nobuffers:packets}),
it is the color of the tail node of this packet.

A packet is
a pair of a body
and a tail node,
which must be \greenorred.
The pattern \oc|_*noyellow*noorange*_| encodes ``\greenorred''.
Writing \oc|_*noyellow*noorange*_ as 'c| allows us to simultaneously
name this color \oc|'c| and require it to match this pattern.
%
% We could say more by describing in detail how the parameters are
% instantiated in these three lines, but we would end up paraphrasing.
% They are instantiated in the only correct way, given the meaning
% that we have assigned to each parameter.
%

\subsubsection{Chains}

% LATER
% the terminology empty chain, single chain, pair chain
% is not used much,
% so we might wish to remove it.
% It's also OK to keep it.

The definition of the type~\oc|chain|
is the last one in \fref{fig:cd_ocaml_chain}.
There are three constructors,
which correspond to chains of arity~0, 1, and~2,
as announced in
\sref{sec:cadeque:pointerrep:nobuffers:packets}.
\begin{itemize}

\item
The constructor~\oc|Empty|
constructs a chain of arity~0,
that is,
an \emph{empty chain}.

\item
The constructor \oc|Single|
constructs a chain of arity~1,
that is,
a \emph{single chain}.
It carries a~root packet
and a child chain.
The kind~\oc|'kind| and color~\oc|'c|
of the chain % I mean, the parent chain
are the kind and color of its root packet.
The arity~\oc|'arity1|
and
the element type~\oc|'b|
of the child chain
are dictated by the root packet.
A~regularity constraint
relates
the color of the root packet
and the colors of the root packets of the child chain.
%
%      We have not defined "the colors of a chain".
%
As indicated in
\sref{sec:cadeque:pointerrep:nobuffers:packets},
this regularity constraint
states that if the root packet is red
then the following packets must be green.
This corresponds to constraint \ref{cd_reg:red}
in \fref{fig:cd_reg}.
A detailed commentary
of the type \oc|regularity|
is deferred to
\sref{sec:cadeque:ocaml:chains:regularity}.

\item
The constructor \oc|Pair|
carries two chains of arity~1 % \oc|single|
% (that is, two single chains)
and constructs a chain of arity~2,
that is,
a \emph{pair chain}. % \oc|pair|
The left child chain has kind~\oc|left|;
the right child chain has kind~\oc|right|.
The colors~\oc|'lc| and \oc|'rc|
respectively
describe the left child chain
and the right child chain.

\end{itemize}

% ------------------------------------------------------------------------------

% Regularity.

\subsubsection{Regularity}
\label{sec:cadeque:ocaml:chains:regularity}

\begin{figure}[t]
\begin{lstlisting}[language=ocaml,backgroundcolor=\colorocaml]
(* A relation between the color 'c of a packet and the
   color(s) 'lc and 'rc of the following packet(s). *)
type ('c, 'lc, 'rc) regularity =
  | G : (green,     _,     _) regularity
  | R : (  red, green, green) regularity
\end{lstlisting}
\caption{\cdocaml: regularity constraints}
\label{fig:cd_ocaml_reg}
\end{figure}

The color of a packet
must be \greenorred
(\sref{sec:cadeque:treerep:nobuffers:regularity}).
Therefore,
in the definition of the data constructor \oc|Single|
(\fref{fig:cd_ocaml_chain}, \lineref{Single:regularity}),
the color~\oc|'c| must be \greenorred.
In view of this remark,
it is evident that merely
two constructors
suffice for the type \oc|regularity|,
namely \oc|G| and \oc|R|
(\fref{fig:cd_ocaml_reg}).
%
\begin{comment}
Utilising the type parameters
\oc|'c|, \oc|'lc1| and \oc|'rc1|
of \fref{fig:cd_ocaml_chain} (\lineref{Single:regularity}),
one can sum up
the definitions of these constructors as follows:
\begin{itemize}
\item if \oc|'c| is green then \oc|'lc1| and \oc|'rc1| are unconstrained;
\item if \oc|'c| is red then \oc|'lc1| and \oc|'rc1| must be green.
\end{itemize}
\end{comment}
%
Translated into informal English,
the types of these constructors can be read as follows:
\begin{itemize}
\item if the root packet is green, the following packets can have any color;
\item if the root packet is red, the following packets must be green.
\end{itemize}
%
\begin{comment}
As announced earlier,
the latter implication corresponds to constraint \ref{cd_reg:red}
in \fref{fig:cd_reg}.
\end{comment}

\subsubsection{\Cdeques}
\label{sec:cadeque:ocaml:chains:main}

\begin{figure}
\begin{lstlisting}[language=ocaml,backgroundcolor=\colorocaml]
type 'a cadeque =
  T :
     ('a, _, only, green, green) chain
  -> 'a cadeque
\end{lstlisting}
\caption{\cdocaml: the type of catenable deques}
\label{fig:cd_ocaml_cadeque}
\end{figure}

Finally,
the type of \cdeques is defined
in \fref{fig:cd_ocaml_cadeque}.
For a chain
to form a stand-alone \cdeque,
this chain must be green,
and its kind must be \oc|only|.
The first requirement
reflects
the regularity constraint
\ref{cd_reg:top}
(\fref{fig:cd_reg}).
The second requirement
admits
chains of arity~0, % empty chains
chains of arity~1 whose root node has kind \oc|only|, % only single chains
and chains of arity~2. % pair chains
It rules out
``left single chains'' and ``right single chains'',
that is,
chains of arity~1 whose root node has kind \oc|left| or \oc|right|.
Indeed, these are not meant to exist in isolation.

This completes our presentation of the data structure:
at~this point,
all of the type definitions that describe the
structure of \cdeques have been presented.
There remains to briefly describe
our implementation of the operations.

% The latest section is commented out because it is more useful for the
% implementation of the five operations rather than the implementation of
% cadeques.

% ------------------------------------------------------------------------------

% Conclusion.

\subsection{Operations}
\label{sec:cadeque:ocaml:operations}

After defining several auxiliary data types and functions,
we implement \opush, \opop, \oinject, \oeject, and \oconcat.
\gitrepofile{lib/cadeque_core.ml}
The definition of buffers (that is, size-indexed \sdeques)
occupies 190 lines of code.
The auxiliary data types occupy 90 lines.
The auxiliary functions and the main five operations occupy 570 lines.

\section{Obstacles and workarounds} % or: Limitations of \Coq and workarounds
\label{sec:obstacles}

We would now like to port our OCaml definition of catenable deques,
which comprises OCaml type definitions and OCaml code,
to \Coq.
Because we have used only the purely functional subset of OCaml,
one might expect a~literal translation from OCaml to \Coq to be possible.
Unfortunately,
this is not the case.
Because \Coq's type system
is designed to enforce termination, % strong normalization,
it is in several ways significantly more restrictive
than OCaml's type system.
In particular,
in an inductive type definition,
\Coq requires
% the type of
every constructor
to satisfy a~\emph{positivity condition}.
Furthermore,
in an inductive function definition,
\Coq requires
the actual argument of every~recursive call
to be \emph{structurally smaller}
than the formal parameter of the function.

Because the formal definition of the positivity condition
is quite difficult to read and understand,
we~provide a simplified paraphrase of~it
(\sref{sec:obstacles:uniform,sec:obstacles:positivity}),
followed with three simple examples
of inductive type definitions
that are rejected by \Coq
(\sref{sec:obstacles:examples}).
Then,
we point out
why the OCaml definition of \cdeques
that we have presented in the previous section
is similar to each of these three examples
(\sref{sec:obstacles:similar}).
%
% Thus,
% one might say that
% there are three reasons
% why a~literal translation
% of OCaml to~\Coq fails!
%
Fortunately,
by parameterizing our type definitions with extra ``level'' and ``size'' indices,
it is possible to work around all three problems:
we sketch this technique,
which we later exploit
in a~revisited implementation of non-catenable deques
(\sref{sec:deque:coq:revisited})
and in an~implementation of catenable deques
(\sref{sec:cadeque:coq}).

We end this section with a discussion
of a situation
where
the  definition of a model function
is rejected by \Coq
because
it is not structurally recursive---
that is,
this inductive function definition
involves recursive calls
whose arguments are not structurally smaller
than the function's formal parameter.
We are able to work around this problem
by defining a~model function
at a more general type
(\sref{sec:obstacles:model}).
This idea is also
later exploited
in
\cref{sec:deque:coq:revisited,sec:cadeque:coq}.

% ------------------------------------------------------------------------------

% Two small subsections about (non-)uniform parameters
% and about the positivity condition.

% This may seem overkill,
% but it is better structured this way.

\subsection{Uniform and non-uniform parameters}
\label{sec:obstacles:uniform}

\Coq splits the parameters of an inductive type in two groups,
namely
the~uniform parameters and
the~non-uniform parameters.
%
% I omit the URL because it is not helpful:
%
% \footnote{\url{https://rocq-prover.org/doc/master/refman/language/core/inductive.html#coq:flag.Uniform-Inductive-Parameters}}
%
% In the manual, the notion of uniform (or "recursively uniform") parameter
% plays a role in the strict positivity condition, but this notion is never
% explicitly defined. The manual points to the command
% `Set Uniform Inductive Parameters`, which means that
% all parameters are by default considered uniform.
% But that does not tell us what a uniform parameter *is*.
% One understands that it is a parameter that does not vary,
% and (therefore) could be declared up front in a Section.
%
\begin{comment}
% This is somewhat irrelevant:
By convention,
the uniform parameters must appear before the non-uniform parameters;
and,
if desired,
the two groups can be separated by a~vertical bar \coqe+|+.
\end{comment}
%
The distinguishing feature of a~uniform parameter
is that
it~must remain unchanged
in every recursive occurrence
of the type that is being defined.

For example,~in
the usual inductive definition of lists,
which fits
in one line
under the form
\coqe+Inductive list A := nil : list A | cons : A -> list A -> list A+,
the parameter~\coqe|A| is uniform. % or: can be uniform

On the contrary,
in the definition of \coqe|plist A|,
which appears in \fref{fig:inductive:examples},
the parameter~\coqe|A| must be non-uniform.
Indeed,
in the type of the constructor \coqe|PCons|,
the parameter~\coqe|A|
is instantiated with~\coqe|A * A|.
Thus,
a~parameter that is subject to recursive slowdown
must be non-uniform.

A~type with a~non-uniform parameter,
such as \coqe|plist|,
is known in the literature
as a~\emph{non-regular data type},
or a~\emph{nested data type}.
The type \coqe|plist| is known as \emph{Nest}
by \citet{bird-meertens-98}
and as \emph{PList}, for \emph{power list},
by \citet{abel-matthes-uustalu-05}.

The notion of uniform parameter
plays a~role in the statement of
the positivity condition,
which we now recall.

% ------------------------------------------------------------------------------

\subsection{The positivity condition}
\label{sec:obstacles:positivity}

A~formal definition of the positivity condition appears in \Coq's
\href{https://rocq-prover.org/doc/master/refman/language/core/inductive.html\#well-formed-inductive-definitions}
{reference manual}.
It is difficult to read.
We offer the following simplistic paraphrase:
\begin{enumerate}[label=(PC)]
\item
\label{pc}
In the definition of an inductive type~$T$, \\
% or: of a parameterized inductive type family
if a constructor carries an argument of type~$U$, \\
then the type~$T$ may occur within the type~$U$
at the following positions, \\
and \emph{only} at these positions:
\begin{itemize}
\item at the root,
\item in the right-hand side of a universal quantifier,
\item inside a \emph{uniform} argument of a \emph{pre-existing} inductive type.
      % in which case one must this inline away inductive type,
      % that is, expand the definitions of its constructors,
      % and continue checking
      % using the "nested positivity" criterion.
\end{itemize}
\end{enumerate}

% Fix Indentation
\noindent
As a consequence,
the positivity condition
implies
the following restriction:
\begin{enumerate}[label=(PR)]
\item
\label{pcneg}
In the definition of an inductive type~$T$, \\
if a constructor carries an argument of type~$U$, \\
the type~$T$ \emph{cannot} occur within the type~$U$
\begin{enumerate}[label=(PR\arabic*)]
\item
\label{pc:nonuniform}
inside a~non-uniform argument of a~pre-existing inductive type,
\item
\label{pc:self}
inside an argument to itself.
\end{enumerate}
\end{enumerate}

%Fix Indentation
\noindent
Although these statements are admittedly very informal,
they help explain
why the examples
that we are about to present
are rejected by \Coq.
A better explanation of the strict positivity condition
and of its interaction with nested inductive types
is provided by \citet{lamiaux-forster-sozeau-tabareau-26}.
% They do not cover truly nested inductive types, which \Coq rejects.

% \citet[\S3.1]{coquand-paulin-88} prove that strict positivity, as opposed
% to just positivity, is required. Their example involves the fixed point
% of the functor F(X) = (X -> Prop) -> Prop. This example is in fact also
% presented in the \Coq manual.

% ------------------------------------------------------------------------------

\subsection{Examples of invalid inductive type definitions}
\label{sec:obstacles:examples}

\begin{figure}
\begin{lstlisting}[linerange=3-9,language=coq,backgroundcolor=\colorcoq]
Inductive tree (A : Type) : Type :=             (* accepted *)
  | Node : A -> list (tree A) -> tree A.

Inductive plist (A : Type) : Type :=            (* nested; accepted *)
  | PNil : plist A
  | PCons : A -> plist (A * A) -> plist A.
\end{lstlisting}
\begin{lstlisting}[linerange=10-19,firstnumber=8,language=coq,backgroundcolor=\colorcoqbroken]
Fail Inductive ptree (A : Type) : Type :=       (* rejected *)
  | PNode : A -> plist (ptree A) -> ptree A.

Fail Inductive bush (A : Type) : Type :=        (* truly nested; rejected *)
  | Leaf : bush A
  | Cons : A -> bush (bush A) -> bush A.

Fail Inductive nztree (A : Type) : Type :=      (* rejected *)
  | NZLeaf : nztree A
  | NZNode : A -> forall ts: list (nztree A), length ts > 0 -> nztree A.
\end{lstlisting}
\caption{Three examples of inductive types that \Coq rejects}
\label{fig:inductive:examples}
\end{figure}

\fref{fig:inductive:examples} presents two examples
of inductive type definitions
that \Coq accepts,
followed with three examples
that it rejects.

The type \coqe|tree| is a~fairly common type of trees
where each node carries
an~element of type~\coqe|A|
and a~list of subtrees.
% therefore, an arbitrary (finite) number of subtrees
Its definition is accepted.
The fact that \coqe|tree A|
occurs as an~argument of the pre-existing inductive type \coqe|list|
is not a~problem
because \coqe|A| is a~uniform parameter of \coqe|list A|.

The type \coqe|plist|,
already discussed earlier
(\sref{sec:obstacles:uniform}),
is the type of power lists.
Its definition is accepted.

% Invalid example: ptree.

The definition of \coqe|ptree|
is a~copy of
the definition of
\coqe|tree|,
where \coqe|list|
has been replaced with \coqe|plist|.
This definition is rejected
because it violates \ref{pc:nonuniform}.
Indeed,
\coqe|A| is a~non-uniform parameter of \coqe|plist A|.

% Invalid example: bush.

The definition of the type \coqe|bush|
can be viewed as
a~copy of
the definition of
\coqe|plist|
where the product \coqe|_ * _|
has been replaced with \coqe|bush| itself.
This definition is rejected
because it violates \ref{pc:self}.
As far as we know,
the type \Bush
has first been defined by \citet{bird-meertens-98}.
It is known by
\citep{abel-matthes-uustalu-05}
as a~\emph{truly nested data type}
and by
\citep{blanchette-meyer-popescu-traytel-17}
as a~\emph{self-nested data type}.

\begin{comment}
Dmitriy Traytel writes: (personal communication with FP, 05/01/2026)

Let me point out that our Isabelle commands easily support ptree and nztree as follows:

nonuniform_datatype 'a plist = PNil | PCons (phd: 'a) (ptl: "('a × 'a) plist")
nonuniform_datatype 'a ptree = PNode 'a "'a ptree plist"

typedef 'a nzlist = "{xs :: 'a list. xs ≠ []}" by auto
lift_bnf (no_warn_wits) 'a nzlist by auto
datatype 'a nztree = NZLeaf | NZNode 'a "'a nztree nzlist”
\end{comment}

% Invalid example: nztree.

The definition of the type \coqe|nztree|
can be viewed as
a~variant of
the definition of
\coqe|tree|,
where
every node is required to~carry a nonempty list of subtrees.
(A~new constructor for leaves is also introduced,
 so as to ensure that the type~\coqe|nztree A|
 is nonempty.)
This nonemptiness requirement is expressed
% via a subset type:
% \coqe+{ ts: list (nztree A) | length ts > 0 }+
% is the type of the nonempty lists of trees.
% This type can also be written under the more verbose form
% \coqe+@sig (list (nztree A)) (fun ts => @length (nztree A) ts > 0)+.
%
% The type \coqe|sig| is the subset type;%
% footnote{\url{https://rocq-prover.org/doc/v8.9/stdlib/Coq.Init.Specif.html}}
by letting the constructor~\coqe|NZNode|
carry an argument
of type~\coqe|length ts > 0|.
This type is convertible
with~\coqe|1 <= length ts|.
This is sugar
for \coqe|1 <= @length (nztree A) ts|,
where
the modifier~\coqe|@|
makes all arguments explicit.
In this form,
it is apparent
that
\coqe|nztree A|
occurs
inside the second argument
of the relation \coqe|<=|.
Furthermore,
the relation \coqe|<=| is inductively defined,
and a look at its definition
shows that its second parameter
is a~non-uniform parameter.
Therefore,
the definition of~\coqe|nztree| violates \ref{pc:nonuniform}.
It is rejected.

\subsection{Why our definitions exhibit similar features, and how to fix them}
\label{sec:obstacles:similar}

Our OCaml definition of \cdeques
(\sref{sec:cadeque:ocaml}),
which involves several mutually recursive data types,
exhibits similarities
with each of the~three examples
of invalid inductive definitions
that we have just presented.
% (\sref{sec:obstacles:examples})
%
We point out these similarities
and explain how we work around
the problematic limitations of \Coq.

\begin{figure}
\begin{lstlisting}[linerange=1-7,language=coq,backgroundcolor=\colorcoqbroken]
Fail Inductive stored A : Type :=
  | Small :           list A -> stored A (*@ \label{line:stored:small} *)
  | Big   : chain (stored A) -> stored A

with chain A : Type :=
  | Single:         stored A -> chain A.
\end{lstlisting}
\begin{lstlisting}[linerange=8-16,firstnumber=8,language=coq,backgroundcolor=\colorcoq]
(* [istored A l] is meant to be isomorphic to [stored^l A]. *)
Inductive istored A : nat -> Type :=
  | Ground     :                  A -> istored A 0
  | Small  {l} : list (istored A l) -> istored A (S l) (*@ \label{line:istored:small} *)
  | Big    {l} :     ichain A (S l) -> istored A (S l)

(*  [ichain A l] is meant to be isomorphic to [chain (stored^l A)]. *)
with ichain A : nat -> Type :=
  | Single {l} :    istored A (S l) -> ichain A l.
\end{lstlisting}
\caption{Encoding a truly nested data type via indexing by levels}
\label{fig:cdc_simple}
\end{figure}

% ------------------------------------------------------------------------------

\subsubsection{A \texttt{bush}-like example}
\label{sec:obstacles:similar:bush}

Let us begin by considering
the data types
shown in~\fref{fig:cdc_simple}.
The upper part of the figure presents an attempt to define
two mutually inductive types,
\coqe|stored| and \coqe|chain|.
This definition is rejected
by~\Coq
because,
like~\coqe|bush|
in~\fref{fig:inductive:examples},
these data types are truly nested:
in the type of the constructor \coqe|Big|,
the parameter~\coqe|A| is instantiated with \coqe|stored A|.

The data types~\coqe|stored| and \coqe|chain|
in~\fref{fig:cdc_simple}
can be viewed as
simplified versions
of the types that we wish to~define.
Indeed,
they are similar to
the OCaml data types
that we have presented in \fref{fig:cd_ocaml_central}.
The constructors \oc|Big| and \oc|Small|
are roughly the same in both figures.
Among the arguments of the constructor~\oc|Big|,
we have kept just the child chain.
In~the~argument of the constructor \oc|Small|,
we have represented the buffer as a list,
for simplicity.
%
% This makes the example intentionally simpler.
% We later discuss what happens when this list
% is replaced with a \sdeque.
%
In the definition of \coqe|chain|,
we have performed a massive simplification,
retaining just the fact that
\coqe|chain A|
depends on
\coqe|stored A|.
%
% LATER this point is nonobvious, I think:
%      it is difficult to see in \fref{fig:cd_ocaml_central}
%      why chain depends on stored.
%      I suppose this involves unfolding (inlining away) chain/packet/body,
%      but the details are unclear to me.
%

The lower part
of \fref{fig:cdc_simple}
shows how we work around the problem.
We exploit the fact that
the desired type definitions
have a simple pattern of recursive slowdown:
when one goes down by one level,
the type parameter \coqe|A|
is instantiated with \coqe|stored A|.
The key idea, then,
is to index every data type
with a natural number~\coqe|l|,
which we refer to as a~\emph{level}.
%
% LATER Comment on the connection with the known idea of
% defining all powers of the type \Bush at once.
% I think I have seen it in the slides by Ralph Matthes.
%
Intuitively,
this natural number indicates how many times
\coqe|stored|,
viewed as
a~function of types to types,
must be iterated.
Thus,
the types~\coqe|istored A l|
and~\coqe|ichain A l|
are intended to be isomorphic
to
% the types
\coqe|stored^l A|
and~\coqe|chain (stored^l A)|,
where
\coqe|stored| and \coqe|chain|
are the types that we wished to define in the first place,
and where
we informally write \coqe|_^_| for the usual power function.
A level can also be visually understood as a depth:
in this data structure,
at~level~0, one finds elements of type~\coqe|A|;
at~level~1, elements of type~\coqe|stored A|;
at~level~2, elements of type~\coqe|stored (stored A)|;
and so on.

The lower part of \fref{fig:cdc_simple}
is obtained from the upper part of this figure
via a systematic transformation.
In this transformation, the type \coqe|stored| plays
a~distinguished role because it is
the pattern of recursive slowdown;
in other words, it is the type function that is iterated.
Here is a precise informal description of this transformation:
\begin{itemize}
\item
In its header line,
every data type
is renamed
and becomes parameterized with a level.
\\
For example,
\coqe|chain A|
is renamed to
\coqe|ichain A|,
\\
and its result type
is changed
from~\coqe|Type|
to~\coqe|nat -> Type|.
\item
In the type of every data constructor,
the following transformations are successively applied:
\begin{itemize}
\item
A universal quantification \coqe|{l}| is introduced;
\item
the variable \coqe|A| is instantiated with \coqe|stored^l A|;
\item
every occurrence of
\coqe|stored (stored^l A)|
is replaced with
\coqe|stored^(S l) A|;
\item
% one must understand that \coqe|l| is locally quantified in this rule;
% so the rule can be applied with \coqe|l| or \coqe|S l|.
every occurrence of
\coqe|stored^l A|
is replaced with
\coqe|istored A l|;
\item
every occurrence of
\coqe|chain (stored^l A)|
is replaced with
\coqe|ichain A l|.
\end{itemize}
\item
In the distinguished data type \coqe|istored|, \\
the constructor~\coqe|Ground : A -> istored A 0| is introduced.
\end{itemize}

% At this point,
% we might wish to explain why the transformation is "correct",
% but this explanation would have to be informal,
% since the original types are rejected by Coq.
%
% Furthermore, I think it is obvious enough that it is "correct".
% No need to say more.

% Fix Indentation
\noindent
This transformation eliminates nesting:
\coqe|istored| and \coqe|ichain|
are ordinary (non-nested) inductive types.
The type parameter~\coqe|A|
is a uniform parameter of
\coqe|istored| and \coqe|ichain|.
Indeed, it~does not vary:
instead, only the level varies.
The definition of
\coqe|istored| and \coqe|ichain|
is accepted by \Coq.

Our \Coq implementation of
\kt's non-catenable deques
(\sref{sec:deque:coq:revisited})
relies on this transformation,
which we apply manually.
This is visible in \fref{fig:cdc_types},
where all types are indexed with levels,
and where the definition of the type \coqe|stored|
includes a constructor \coqe|Ground|.

% ------------------------------------------------------------------------------

\subsubsection{A \texttt{ptree}-like example}
\label{sec:obstacles:similar:ptree}

In \fref{fig:cdc_simple},
\coqe|list|
appears in the type of the constructor \coqe|Small|,
at \linerefs{stored:small}{istored:small}.
Let us now imagine that
\coqe|list|
is replaced with \coqe|deque|,
the type of non-catenable deques (\sref{sec:deque}).
Then,
% the definition in
the lower part of the figure
breaks.
Because \coqe|A| is a~non-uniform parameter of \coqe|deque A|,
\Coq does not allow \coqe|istored A l|
to appear as an argument of \coqe|deque|
on \lineref{istored:small}.

In this form,
this example is analogous to \coqe|ptree|
in \fref{fig:inductive:examples}.
Furthermore,
the types that we wish to define
exhibit a similar feature.
Indeed,
the constructor \coqe|Small|
in \fref{fig:cd_ocaml_central}
does carry
a \sdeque
as an~argument.%
\footnote{The abstract type \oc|('a, _ plus3) Buffer.t|,
once unfolded, is synonymous with \oc|'a deque|.}

Fortunately,
the technique of indexing with levels,
which has been described above
(\sref{sec:obstacles:similar:bush}),
allows working around this problem.
The type \coqe|deque A|,
where the parameter~\coqe|A| is non-uniform,
is transformed
into \coqe|ideque A l|,
where the parameter~\coqe|A| is uniform.
In the type of the constructor \coqe|Small|,
instead of \coqe|deque (istored A l)|,
one uses \coqe|ideque (istored A l) 0|.
%
% This is not clearly visible in \fref{fig:cdc_types}
% because the type \coqe|Deque.deque| or \coqe|suffix'|
% there is not indexed with a level.

% ------------------------------------------------------------------------------

\subsubsection{An \texttt{nztree}-like example}
\label{sec:obstacles:similar:nztree}

In the type of the constructor \coqe|Small|,
we need not only to indicate that this constructor carries a buffer,
but also to impose a size constraint:
this buffer must have size at least~3.
This is visible
in the type of
the constructor \coqe|Small|
in \fref{fig:cd_ocaml_central},
where the size of the buffer is \oc|_ plus3|.

Therefore,
in \Coq,
one is tempted to decorate
the constructor \coqe|Small|
(\fref{fig:cdc_simple}, \lineref{istored:small})
with a constraint of the form~\coqe|size d >= 3|,
where the function~\coqe|size| computes the size of a~buffer.
(Here, whether the buffer is a list or a \sdeque is irrelevant.)
However,
this addition
makes the lower part of \fref{fig:cdc_simple}
similar to \coqe|nztree| in \fref{fig:inductive:examples},
and causes it to be rejected by \Coq.

Fortunately,
there is
a~simple
% LATER is this a well-known workaround?
%       if so, what is the source?
% maybe Dybjer https://www.jstor.org/stable/2586554
workaround:
the type of buffers should be indexed with a~size.
Then, one can impose a size constraint
without referring to a \coqe|size| function.
This eliminates the problem.%
\footnote{In fact, in our OCaml implementation, the type \oc|('a, 'n) Buffer.t|
\emph{was} indexed with a lower bound \oc|'n| on the size of the buffer.
There, this was the only viable approach, because OCaml does not have
dependent types. In \Coq, indexing with sizes and relying on a \coqe|size|
function are two approaches that seem viable at first.
However, indexing turns out to be the only viable approach.}
Looking ahead,
this solution is visible
in the type of the constructor \coqe|Small|
in \fref{fig:cdc_types}.
There, the size index is instantiated with \coqe|3 + q|,
forcing the size of the buffer to be at least~3.

% ------------------------------------------------------------------------------

\subsection{Structural recursion and model functions}
\label{sec:obstacles:model}

\begin{figure}
\begin{lstlisting}[linerange=14-24,language=coq,backgroundcolor=\colorcoq]
(* A data type of buffers of exactly one element. *)
Inductive buffer1 (A : Type) : Type :=
  | B1 : A -> buffer1 A.

(* This turns a buffer into a mathematical sequence. *)
Equations buffer1_seq {A} : buffer1 A -> list A :=
buffer1_seq (B1 x) := [x].

(* This is analogous to tree in Fig. (*@\ref{fig:inductive:examples}*). *)
Inductive tree1 (A : Type) : Type :=
  | Node1 : A -> buffer1 (tree1 A) -> tree1 A.  (*@ \label{line:tree1} *)
\end{lstlisting}
\begin{lstlisting}[linerange=26-28,firstnumber=13,language=coq,backgroundcolor=\colorcoqbroken]
(* This model function is rejected. *)
Fail Equations tree1_seq : forall {A}, tree1 A -> list A := (*@ \label{line:fail27} *)
tree1_seq (Node1 x ts) := [x] ++ concat (map tree1_seq (buffer1_seq ts)). (*@ \label{line:fail28} *)
\end{lstlisting}
\begin{lstlisting}[linerange=30-58,firstnumber=17,language=coq,backgroundcolor=\colorcoq]
(* So far, we have been trying to define seq functions, whose type is: *)
Definition seq_fun (D : Type -> Type) : Type :=
  forall A,  D A -> list A.

(* We must in fact define concat_map_seq (cmseq) functions, whose type is: *)
Definition cmseq_fun (D : Type -> Type) : Type := (*@ \label{line:cmseq_fun1} *)
  forall T,
  (forall A,     T A  -> list A) ->
  (forall A,  D (T A) -> list A). (*@ \label{line:cmseq_fun4} *)

(* A cmseq function can be specialized to obtain a seq function. *)
Definition cmseq2seq {D} (cmseq : cmseq_fun D) : seq_fun D := (*@ \label{line:cmseq2seqa} *)
  cmseq (fun A => A) (fun _ x => [x]). (*@ \label{line:cmseq2seqb} *)

(* A cmseq function for buffer1. *)
Equations buffer1_cmseq : cmseq_fun buffer1 :=
buffer1_cmseq T f A (B1 x) := f _ x.

(* A cmseq function for tree1. *)
Equations tree1_cmseq : cmseq_fun tree1 :=
tree1_cmseq T f A (Node1 x ts) :=
  f _ x ++ buffer1_cmseq _ (tree1_cmseq _ f) _ ts. (*@ \label{line:miracle} *)

(* A seq function for tree1, obtained as a special case. *)
Definition tree1_seq : seq_fun tree1 := cmseq2seq tree1_cmseq.(*@ \label{line:tree1_seq} *)

(* tree1_cmseq is correct. *)
Lemma correct_tree1_cmseq T (f : forall A, T A -> list A) A (t : tree1 (T A)) :(*@ \label{line:correct_tree1_cmseq} *)
  tree1_cmseq T f A t = concat (map (f A) (tree1_seq (T A) t)).(*@ \label{line:correct_tree1_cmseq_2} *)
\end{lstlisting}
\caption{A model function that \Coq rejects, and a work-around}
\label{fig:model:obstacle}
\end{figure}

Each of the data structures considered in this paper
is intended to represent a sequence of elements.
To make this intent explicit,
for each such data structure,
we wish to define a \emph{model function}\index{model functions},
% The following has been said in the introduction already:
which maps
a~data structure
to the sequence
of elements
that this data structure represents.
%
% In other words,
% a model function computes the \emph{fringe} of a~data structure.
%
Since the shape of a data structure is usually described in \Coq
by an inductive type definition,
its model function
is usually defined
via an inductive function definition.
For this definition to be accepted by \Coq,
it must be
\href{https://rocq-prover.org/doc/master/refman/language/core/inductive.html\#fixpoint-definitions}
{structurally recursive}:
that is,
the actual argument of every~recursive call
must be a subterm
of the formal parameter of the function.

\newcommand{\footnoteotherapproach}{\footnote{%
There is an alternative approach to
defining the function \coqe|tree1_seq|.
One first equips the type \coqe|buffer1|
with a~function \coqe|buffer1_map|.
Then, one defines \coqe|tree1_seq (Node1 x ts)|
as \coqe|concat (buffer1_seq (buffer1_map tree1_seq ts))|.
% Equations tree1_seq : forall {A}, tree1 A -> list A :=
% tree1_seq (Node1 x ts) :=
%   concat (buffer1_seq (buffer1_map tree1_seq ts)).
%
% This definition is accepted by \Coq. % explain why?
Though it might be possible,
we have not been able
to scale this approach
to the complex data structures
considered in this paper.}}

Unfortunately,
when a data structure is described by a nested inductive type,
its model function typically involves nested recursive calls,
which \Coq does not accept.
\fref{fig:model:obstacle} offers a~simple example of this situation.
There, the inductive type \coqe|buffer1|
describes a~``buffer'' of exactly one element.
The model function \coqe|buffer1_seq|
turns a buffer into a sequence of elements:
the buffer \coqe|B1 x|
is turned into the~singleton sequence \coqe|[x]|.
Then, the inductive type~\coqe|tree1|
is defined.
It is identical to the type \coqe|tree|
in \fref{fig:inductive:examples},
except that its definition uses \coqe|buffer1|
where the definition of \coqe|tree| uses \coqe|list|.
At \linerefs{fail27}{fail28}
in \fref{fig:model:obstacle},
an~attempt is made
to define a~model function for the type \coqe|tree1|.%
\footnoteotherapproach{}
There, \coqe|ts| has type \coqe|buffer1 (tree1 A)|:
it is a buffer of trees.
Thus, \coqe|buffer1_seq ts|
is a list of trees.
The (list) \coqe|map| function
lets us
apply \coqe|tree1_seq|
to each tree in this list.
This yields a list of lists, % of elements
which
\coqe|concat| flattens
to a~list. % of elements.
Unfortunately,
this definition is rejected by \Coq.
The manner in which
\coqe|tree1_seq| is used
within its own definition
is not acceptable.
Indeed, via \coqe|map|,
\coqe|tree1_seq| is applied to every tree
in the list \coqe|buffer1_seq ts|.
Even though a human reader can see
that these trees are in fact subterms of \coqe|ts|,
they are not ``structurally smaller'' than \coqe|ts|
in the strict sense that \Coq requires.

The~reader might wonder whether,
instead of structural recursion,
our model functions
could be defined
by well-founded recursion.
One approach would be to rely on the usual ordering
of the natural numbers, which is well-founded.
We would then need to equip each of our data types
with a~\emph{measure function},
in such a~way that the measure of a~component
is strictly smaller than the measure of the whole.
Unfortunately, we would then run into another instance
of the same problem:
defining measure functions
requires structural recursion,
and \Coq is likely to reject
our measure functions
for the same reason
that it rejects our model functions.
Another approach might be
% to avoid going through a measure
to define a~well-founded ordering directly on our data structures,
and to define our model functions
by recursion over an accessibility witness~\cite{leroy-recursion-24}.
However, because our data types are non-uniform,
we would be unable to use a homogeneous ordering
of type \coqe|D A -> D A -> Prop|;
instead we might need to use a heterogeneous ordering
of type \coqe|D A -> D B -> Prop|.
We have not explored this avenue.

To work around this obstacle,
we define model functions of a more general type,
which removes the need for explicit uses
of list \coqe|concat| and list \coqe|map|.
So far,
for a data structure \coqe|D|,
such as \coqe|buffer1| or \coqe|tree1|,
we have defined model functions
of type \coqe|forall A,  D A -> list A|.
That is,
out of a data structure that contains elements,
a (plain) model function extracts a~sequence of elements.\index{model function!plain}
The idea is to
generalize this as follows
(lines~\ref{line:cmseq_fun1}--\ref{line:cmseq_fun4}).
For an arbitrary notion of ``troop'',
if
out of a troop of elements
one can extract a sequence of elements,
then
out of a data structure that contains troops of elements,
a \emph{generalized model function}\index{model function!generalized}
extracts a~sequence of elements.

A generalized model function
can be specialized
to obtain a plain model function:
to do so,
one lets a ``troop'' be a single element,
so the ``troop'' type constructor is \coqe|fun A => A|
and the sequence of elements of the troop \coqe|x|
is the singleton sequence \coqe|[x]|
(\lineref{cmseq2seqb}).

Defining generalized model functions
for the types \coqe|buffer1| and \coqe|tree1|
is straightforward.
% In particular,
In the definition of \coqe|tree1_cmseq|
(\lineref{miracle}),
the manner in which the recursive calls are nested
reflects
the manner in which the types \coqe|buffer1| and \coqe|tree1| are nested
in the definition of \coqe|tree1|
(\lineref{tree1}).
There is no longer a need to use \coqe|concat| and \coqe|map|,
as the effect of these functions has been built into the notion
of a generalized model function.
This definition is recognized by \Coq
as structurally recursive:
it is accepted.%
\footnote{%
% We believe that
In order to see that this definition is acceptable,
one must destruct the buffer \coqe|ts|
and inline away \coqe|buffer1_cmseq|.
Then, it becomes clear that \coqe|tree1_cmseq|
is applied to a subterm of \coqe|Node1 x ts|.}

Out of the generalized model function \coqe|tree1_cmseq|,
one recovers the plain model function \coqe|tree1_seq|
that we wished to define in the first place
(\lineref{tree1_seq}).
Furthermore,
one can prove that \coqe|tree1_cmseq| is \emph{correct}
in the sense that it is
equal to
a~composition of \coqe|concat|, \coqe|map|,
and \coqe|tree1_seq|
(\linerefs{correct_tree1_cmseq}{correct_tree1_cmseq_2}).
The proof requires an~induction principle
for the type \coqe|tree1|.

In this paper,
we define generalized model functions
for non-catenable deques
and for some of the data types
involved in the definition of catenable deques
(\sref{sec:cadeque:coq:model}).

\section{Non-catenable deques: revisited \Coq implementation}
\label{sec:deque:coq:revisited}

% How should this section be subdivided?
% We may wish to use \paragraph,
% but it currently causes a compilation error.
% We may wish to use \subsection*,
% which looks good,
% but it is strange to use \subsection* here
% and \subsection everywhere else.
\newcommand{\mysubsection}[1]{\subsection{#1}}

As explained in
\sref{sec:obstacles:similar},
the \Coq implementation of non-catenable deques
that we have presented in
\sref{sec:deque:coq}
does not form
a suitable building block
for an implementation
of catenable deques.
Therefore, it must be revisited:
every inductive type must be indexed
by a level~$l$ and/or by a size~$n$.
For example,
the type \coqe|chain A C| of \sref{sec:deque:coq},
where \coqe|A|~is the type of the elements
and \coqe|C|~is a color,
becomes
\coqe|chain A l n C|
in this section.
A chain of this type stores
$n$ elements of type $A^{2^l}$,
that~is,
$n \cdot 2^l$ elements of type $A$.

\begin{figure}[t]
\begin{lstlisting}[language=coq,backgroundcolor=\colorcoq]
Definition level := nat.
Definition size := nat.

Inductive prodN (A : Type) : level -> Type :=
  | prodZ     : A -> prodN A 0
  | prodS {l} : prodN A l -> prodN A l -> prodN A (S l).

Inductive buffer (A : Type) (l : level) : size -> color -> Type :=
  | B0         : buffer A l 0 red
  | B1 {y r}   : prodN A l -> buffer A l 1 (Mix NoGreen y r)
  | B2 {g y r} : prodN A l -> prodN A l -> buffer A l 2 (Mix g y r)
  | B3 {g y r} : prodN A l -> prodN A l -> prodN A l ->
                 buffer A l 3 (Mix g y r)
  | B4 {y r}   : prodN A l -> prodN A l -> prodN A l ->
                 prodN A l -> buffer A l 4 (Mix NoGreen y r)
  | B5         : prodN A l -> prodN A l -> prodN A l ->
                 prodN A l -> prodN A l -> buffer A l 5 red.

Inductive packet (A : Type) (l : level) : level -> size -> size -> color -> Type :=
  | Hole {n}                      :
      packet A l l n n uncolored
  | Packet {hl pn pktn sn hn C y} :
      buffer A l pn C ->
      packet A (S l) hl pktn hn (Mix NoGreen y NoRed) ->
      buffer A l sn C ->
      packet A l hl (pn + 2 * pktn + sn) hn C.(*@ \label{line:double_pkt} *)

Inductive chain (A : Type) (l : level) : size -> color -> Type :=
  | Ending {n C}          :
      buffer A l n C ->
      chain A l n green
  | Chain {hl n hn C1 C2} :
      regularity C1 C2 ->
      packet A l hl n hn C1 ->
      chain A hl hn C2 ->
      chain A l n C1.

Inductive deque (A : Type) (n : size) : Type :=
  | T {g y} : chain A 0 n (Mix g y NoRed) -> deque A n.
\end{lstlisting}
\caption{\ncdcoq: level- and size-indexed types for \sdeques}
\label{fig:ncdr_coq_types}
\end{figure}

The new (indexed) type definitions
appear in \fref{fig:ncdr_coq_types}.
They can be compared with
the earlier (non-indexed) type definitions
in \fref{fig:ncd_coq_types}.

These indexed definitions are obtained by manually
applying the transformation
that was sketched in \sref{sec:obstacles:similar:bush}.
% We cannot claim that the transformation is systematic,
% I believe, as we treat |packet| in a special way.
%
Here,
the ``pattern of recursive slowdown''
is the product type \coqe|_ * _|.
Indeed,
when one goes down by one level in the data structure,
the parameter~\coqe|A| is instantiated with \coqe|A * A|.
Thus, the index~$l$
indicates how many times
the product type constructor
is iterated.

\mysubsection{Iterated products}

\Coq's product type, written \coqe|A * B| or \coqe|prod A B|,
is an inductive type with one constructor,
\coqe|pair|.
By specializing it to the homogeneous case,
where both components have type \coqe|A|,
and by
applying the transformation
to~it,
one obtains
the inductive type \coqe|prodN|
in \fref{fig:ncdr_coq_types}.
This is an iterated product type:
a value of type \coqe|prodN A l| consists
of $2^l$ elements of type~\coqe|A|.
The constructor \coqe|prodZ| is introduced by the transformation:
it~corresponds to the \coqe|Ground| constructor
in \sref{sec:obstacles:similar:bush}.
The constructor \coqe|prodS|
is an indexed version of
the pre-existing constructor~\coqe|pair|.

\mysubsection{Buffers}

The type \coqe|buffer|
is transformed
in a~straightforward way.
It becomes
indexed by a~level and a~size
(\fref{fig:ncdr_coq_types}).
In comparison with
the non-indexed version of this type
in \fref{fig:ncd_coq_types},
an atomic element of type~\coqe|A| is replaced
with a composite element of type~\coqe|prodN A l|.
Furthermore,
the size index~$n$
reflects the number of composite elements
in the buffer.
Thus,
a~buffer of type~\coqe|buffer A l n c|
stores $n$ elements of type $A^{2^l}$.

\mysubsection{Packets}
\label{sec:deque:coq:revisited:packet}

The type \coqe|packet| is indexed with levels
in a slightly ad hoc manner.
In the original type \coqe|packet A B C|
(\fref{fig:ncd_coq_types}),
$A$~was the~type of the~elements at the~top level of the~packet
and
$B$~was the~type of the~elements at the~bottom level of the~packet.%
\footnote{The~parameter~$C$ is a color.
The treatment of colors is unaffected by the introduction of indexing.}
By definition of the type \coqe|packet|,
the parameter~$B$~was implicitly constrained
to be $A^{2^d}$,
where $d$ is the depth of the packet,
that is,
the difference between the bottom and top levels.
Here,
we choose to abandon the parameter~$B$
and instead
keep track of two levels,
namely
the top level
and the bottom level
of the packet.
In the~type \coqe|packet A l hl _ _ C|,
the index~\coqe|l| is the top level,
while~\coqe|hl| is the bottom level,
or \emph{hole level}.
The level~\coqe|hl|
is also the top level of the next packet in the chain;
this is visible in the type of the constructor \coqe|Chain|
(\fref{fig:ncdr_coq_types}).

Furthermore, we index the type \coqe|packet|
with two sizes.
In~\coqe|packet A l hl n hn C|,
the index~\coqe|n|
represents
the total size of this packet
and of the subchain that follows~it,
under the assumption
that the size of the subchain
is~\coqe|hn|.
%
% fp: This suggests that one could use just one size index instead of two,
%     namely the difference n - hn, which is the size of this packet.
%     But this intuition is wrong, because n and hn are expressed in
%     distinct units; the unit depends on the level.
%     This is explained below when we discuss the multiplication by two.
%
(As noted earlier,
by convention,
the ``size'' of a data structure whose top level is~\coqe|l|
is the number of composite elements of type~\coqe|prodN A l|
that this data structure contains.)
The formula~\coqe|pn + 2 * pktn + sn|
at~\lineref{double_pkt}
in~\fref{fig:ncdr_coq_types}
adds up
the size~\coqe|pn| of the prefix buffer,
the size~\coqe|pktn| of the subpacket and of the subchain that follows it,
and
the size~\coqe|sn| of the suffix buffer.
The multiplication by two
accounts for the fact
that the size~\coqe|pktn| is expressed in a different unit of measure:
whereas this packet has top level~\coqe|l|,
its subpacket has top level~\coqe|S l|.
Because one element of type \coqe|prodN A (S l)|
corresponds to two elements of type \coqe|prodN A l|,
the size of the subpacket must be multiplied by two
when it is viewed as a contribution to the size of the packet.
%
% This is reminiscent of the multiplication by two in \fref{fig:bc_coq_models}.

\mysubsection{Regularity}

The type \coqe|regularity| does not need to be indexed.
It remains the same as
in \fref{fig:ncd_coq_types}.
Therefore, it is not shown again in \fref{fig:ncdr_coq_types}.

\mysubsection{Chains}

The type \coqe|chain| becomes indexed by a~level
and by a~size
(\fref{fig:ncdr_coq_types}).
As announced earlier,
a chain of type
\coqe|chain A l n C|
stores
$n$ elements of type $A^{2^l}$.
In the type of the constructor~\coqe|Chain|,
at \lineref{hole_type_match} in \fref{fig:ncdr_coq_types},
the hole level and hole size of the first packet
are required to coincide with
the top level and size of the subsequent chain.

\mysubsection{Deques}

The type \coqe|deque|
becomes indexed by a size
(\fref{fig:ncdr_coq_types}).
For our purposes,
there is no need to index it with a level.
So,
we define a~deque
as a~chain whose top level is~zero.
%
% a green or yellow chain (the color constraints are unchanged)
%
If this chain has size~$n$,
then we declare that this deque has size~$n$.
This makes sense
because the top level of the chain is zero:
thus,
it does not matter whether the size is measured
as a~number of atomic elements of type~$A$
or as a~number of composite elements of type $A^{2^0}$.

\mysubsection{Code and proofs}

The introduction of level indices
does not require any additional functions, lemmas, or proofs.
However, our type definitions become more intricate:
their size grows from 30 to 60 lines.
For the sake of illustration,
in our online repository,
we provide an implementation of \sdeques
that is indexed with just levels.
\gitrepofile{theory/Deque/Deque_lvl.v}

In contrast,
the introduction of size indices
presents a greater challenge:
new proof obligations appear,
which are equalities between natural numbers.
As a consequence,
three new functions,
two new lemmas,
and
a~custom tactic
are required.
Our verified \Coq implementation
of non-catenable deques,
indexed with levels and sizes,
spans about 670 lines,
including
80~lines of type definitions
and
30~lines of model function definitions.
% and 560~lines for the rest.
\gitrepofile{theory/Deque/Deque.v}

\section{Catenable deques: \Coq implementation}
\label{sec:cadeque:coq}

In the previous section (\sref{sec:deque:coq:revisited}),
we have presented a~\Coq implementation of non-catenable deques.
From here on, we refer to the type of non-catenable deques
via~its qualified name, \coqe|Deque.deque A n|.
This type has two qualities
that are important for our purposes in this section:
its parameter~\coqe|A| is a~uniform parameter;
% (\sref{sec:obstacles:uniform})
%
and it is indexed with a~size parameter~\coqe|n|.
These qualities allow us to use non-catenable deques
(with size constraints)
at the very heart of the inductive definition
of catenable deques,
while avoiding the obstacles discussed in \sref{sec:obstacles}.
Furthermore,
for this definition
to be accepted by \Coq,
its main data~types
must be indexed with levels,
as explained in \sref{sec:obstacles}.
Up to these differences,
our \Coq implementation of catenable deques
closely follows the OCaml implementation of
\sref{sec:cadeque:ocaml}.

% ------------------------------------------------------------------------------

\subsection{Types}
\label{sec:cadeque:coq:types}

In a~catenable deque,
a~\emph{buffer} is a~non-catenable deque.
%
% fp: The following repeats information that has already been given
%     in the section header (above). We could keep it or cut it,
%     as you prefer.
%
To describe buffers,
we rely on the type~\coqe|Deque.deque A n|
of \sref{sec:deque:coq:revisited}.
As can be seen in \fref{fig:ncdr_coq_types},
although the definition of this type relies on
auxiliary data~types that are indexed with levels and sizes,
the type~\coqe|Deque.deque A n|
is indexed with just a~size~\coqe|n|.

The 4-color scheme is expressed in \Coq
in a~straightforward way
(\fref{fig:cdc_color}):
\gitrepofile{theory/Color/GYOR.v}
each of the four hues forms an inductive type
with two~constructors;
a~color is a~tuple of four hue bits.

As in~\sref{sec:cadeque:ocaml:nodes:parameters},
an~\emph{arity} is 0, 1, or~2,
and
a~\emph{kind} is ``only'', ``left'', or ``right''.
Here,
kinds form an inductive type
with three constructors,
namely \coqe|only|, \coqe|left|, and \coqe|right|
(\fref{fig:cdc_kind_arity}).
We~define \coqe|arity| as a~synonym for \coqe|nat|
and define \coqe|empty|, \coqe|single|, and \coqe|pair|
as short-hands for 0, 1, and~2
(\fref{fig:cdc_kind_arity}).

\begin{figure}[t]
\begin{lstlisting}[language=coq,backgroundcolor=\colorcoq]
(* Each of the four hues is an inductive type with two constructors. *)
Inductive green_hue  := SomeGreen  | NoGreen.
Inductive yellow_hue := SomeYellow | NoYellow.
Inductive orange_hue := SomeOrange | NoOrange.
Inductive red_hue    := SomeRed    | NoRed.

(* A color is a quadruple of four hue bits. *)
Inductive color    :=
  Mix : green_hue -> yellow_hue -> orange_hue -> red_hue -> color.

Notation green     := (Mix SomeGreen   NoYellow   NoOrange   NoRed).
Notation yellow    := (Mix   NoGreen SomeYellow   NoOrange   NoRed).
Notation orange    := (Mix   NoGreen   NoYellow SomeOrange   NoRed).
Notation red       := (Mix   NoGreen   NoYellow   NoOrange SomeRed).
Notation uncolored := (Mix   NoGreen   NoYellow   NoOrange   NoRed).
\end{lstlisting}
\caption{\cdcoq: types for colors}
\label{fig:cdc_color}
\end{figure}

\begin{figure}[t]
\begin{lstlisting}[language=coq,backgroundcolor=\colorcoq]
(* An arity is 0, 1, or 2. *)
Notation arity  := nat.
Notation empty  := 0.
Notation single := 1.
Notation pair   := 2.

(* A kind is only, left, or right. *)
Inductive kind : Type := only | left | right.

(* A level is a natural integer. *)
Definition level := nat.

(* A size is a natural integer. *)
Definition size := nat.
\end{lstlisting}
\caption{\cdcoq: arities, kinds, levels, and sizes}
\label{fig:cdc_kind_arity}
\end{figure}

\begin{figure}[t]
\begin{lstlisting}[language=coq,backgroundcolor=\colorcoq]
(* Convenient synonyms. *)
Definition prefix' := Deque.deque.
Definition suffix' := Deque.deque.

(* Constraints on buffer sizes, arity, and color. *)
Inductive node_coloring : size -> size -> arity -> color -> Type :=
  | EN {qp qs}   : node_coloring (0 + qp) (0 + qs) 0     green
  | GN {qp qs a} : node_coloring (3 + qp) (3 + qs) (S a) green
  | YN {qp qs a} : node_coloring (2 + qp) (2 + qs) (S a) yellow
  | ON {qp qs a} : node_coloring (1 + qp) (1 + qs) (S a) orange
  | RN {qp qs a} : node_coloring (0 + qp) (0 + qs) (S a) red.

(* Nodes. *)
Inductive node' (A : Type) : arity -> kind -> color -> Type :=
  | Only {qp qs a C}  :
      node_coloring qp qs (S a) C ->
      prefix' A (5 + qp) ->
      suffix' A (5 + qs) ->
      node' A (S a) only C
  | Only_end {q}      :
      prefix' A (S q) ->
      node' A 0 only green
  | Left {qp qs a C}  :
      node_coloring qp qs a C ->
      prefix' A (5 + qp) ->
      A * A ->
      node' A a left C
  | Right {qp qs a C} :
      node_coloring qp qs a C ->
      A * A ->
      suffix' A (5 + qs) ->
      node' A a right C.
\end{lstlisting}
\caption{\cdcoq: types for nodes}
\label{fig:cdc_col_node}
\end{figure}

The definitions of the types
\coqe|node_coloring| and \coqe|node|
appear in \fref{fig:cdc_col_node}.
These definitions are analogous
to our earlier OCaml definitions
(\fref{fig:cd_ocaml_node}).
%
% We define
% \coqe|prefix'| and \coqe|suffix'|
% as synonyms for \coqe|Deque.deque|.
%
The \Coq types \coqe|prefix'|, \coqe|suffix'|, and \coqe|node'|
in \fref{fig:cdc_col_node}
correspond to
the OCaml types \oc|prefix|, \oc|suffix|, and \oc|node|
in \fref{fig:cd_ocaml_node}.
The reason why we use ``primed'' names
% \coqe|prefix'|, \coqe|suffix'| and \coqe|node'|
is that we later define
\coqe|prefix A~l|, \coqe|suffix A~l|, and \coqe|node A~l|
as abbreviations for
\coqe|prefix' (stored A~l)|,
\coqe|suffix' (stored A~l)|, and
\coqe|node' (stored A~l)|.
This is convenient in our code,
but is not visible in the paper.

\begin{figure}[t]
\begin{lstlisting}[language=coq,backgroundcolor=\colorcoq]
Inductive regularity : color -> color -> color -> Type :=
  | G {lC rC} : regularity green lC    rC
  | R         : regularity red   green green.
\end{lstlisting}
\caption{\cdcoq: regularity constraints}
\label{fig:cdc_reg}
\end{figure}

\begin{figure}[p]
\begin{lstlisting}[language=coq,backgroundcolor=\colorcoq]
Inductive stored (A : Type) : level -> Type :=
  | Ground                :
      A -> stored A 0
  | Big {l qp qs a lC rC} :
      prefix' (stored A l) (3 + qp) ->
      chain A (S l) a only lC rC ->
      suffix' (stored A l) (3 + qs) ->
      stored A (S l)
  | Small {l q}           :
      suffix' (stored A l) (3 + q) ->
      stored A (S l)

with body (A : Type) : level -> level -> kind -> kind -> Type :=
  | Hole {l k}                    :
      body A l l k k
  | Single_child {hl tl hk tk y o}:
      node' (stored A hl) 1 hk (Mix NoGreen y o NoRed) ->
      body A (S hl) tl only tk ->
      body A hl tl hk tk
  | Pair_yellow {hl tl hk tk C}   :
      node' (stored A hl) 2 hk yellow ->
      body A (S hl) tl left tk ->
      chain A (S hl) single right C C ->
      body A hl tl hk tk
  | Pair_orange {hl tl hk tk}     :
      node' (stored A hl) 2 hk orange ->
      chain A (S hl) single left green green ->
      body A (S hl) tl right tk ->
      body A hl tl hk tk

with packet (A : Type) : level -> level -> arity -> kind -> color -> Type :=
  | Packet {hl tl a hk tk g r} :
      body A hl tl hk tk ->
      node' (stored A tl) a tk (Mix g NoYellow NoOrange r) ->
      packet A hl (S tl) a hk (Mix g NoYellow NoOrange r)

with chain (A : Type) : level -> arity -> kind -> color -> color -> Type :=
  | Empty {l k lC rC}          :
      chain A l empty k lC rC
  | Single {hl tl a k C lC rC} :
      regularity C lC rC ->
      packet A hl tl a k C ->
      chain A tl a only lC rC ->
      chain A hl single k C C
  | Pair {l k lC rC}           :
      chain A l single left lC lC ->
      chain A l single right rC rC ->
      chain A l pair k lC rC.
\end{lstlisting}
\caption{\cdcoq: types for stored triples, bodies, packets, chains} % this caption must fit in one line
\label{fig:cdc_types}
\end{figure}

% fp: Because the above figure is too tall, I have moved the definition of the
% type cadeque back to a separate figure. An alterative would be to place
% several arguments on the same line in the types of Single and Pair.

\begin{figure}
\begin{lstlisting}[language=coq,backgroundcolor=\colorcoq]
Inductive cadeque : Type -> Type :=
  | T {A a} : chain A 0 a only green green -> cadeque A.
\end{lstlisting}
\caption{\cdcoq: the type of \cdeques}
\label{fig:cdc_types:main}
\end{figure}

The
% definition of the
type \coqe|regularity|
is easily translated
from OCaml (\fref{fig:cd_ocaml_reg})
to \Coq (\fref{fig:cdc_reg}).

To translate
the definition of the four inductive types
\coqe|stored|, \coqe|body|, \coqe|packet|, and \coqe|chain|
from OCaml (\fref{fig:cd_ocaml_central})
to \Coq (\fref{fig:cdc_types}),
indexing with levels
is used,
as described in \sref{sec:obstacles:similar:bush}.
Here, the pattern of recursive slowdown is \coqe|stored _|.
Thus, a~new constructor \coqe|Ground : A~-> stored A~0|
appears in the definition of the type \coqe|stored|.
The type \oc|packet|
is translated in the same ad hoc way
as in \sref{sec:deque:coq:revisited:packet},
so it becomes indexed with two levels.
The last definition is
the definition of the type \coqe|cadeque|
(\fref{fig:cdc_types:main});
it is analogous to the type \oc|cadeque|
in \fref{fig:cd_ocaml_cadeque}.
% with level 0

Our type definitions
include
24 types in total
and
span
190 lines of code.
\gitrepofile{theory/Cadeque/Types.v}

% ------------------------------------------------------------------------------

\subsection{Model functions}
\label{sec:cadeque:coq:model}

In order to be able to state the correctness
of the operations on catenable deques,
we must define model functions.\index{model functions!cadeques}

Following the approach outlined in \sref{sec:obstacles:model},
for the non-catenable deques of \sref{sec:deque:coq:revisited},
we define the generalized model function
\coqe|Deque.deque_cmseq|
(not shown).
\gitrepofile{theory/Deque/Deque.v}
For the type \coqe|node'| in \fref{fig:cdc_col_node},
we also define a~generalized model function,
\coqe|node'_cmseq|
(not shown).
\gitrepofile{theory/Cadeque/Models.v}
For the four mutually inductive types
\coqe|stored|, \coqe|body|, \coqe|packet|, and \coqe|chain|,
four mutually recursive model functions are needed.
For these types,
although we could define
generalized model functions,
it is possible and sufficient
to define plain model functions,
shown in \fref{fig:cdc_model}.
%
% By applying \coqe|cmseq2seq|
% (\fref{fig:model:obstacle}, \linerefs{cmseq2seqa}{cmseq2seqb}),
% from these generalized model functions,
% we obtain plain model functions.

\begin{figure}
\begin{lstlisting}[language=coq,backgroundcolor=\colorcoq]
Equations stored_seq A l (st : stored A l) : list A
by struct st :=
stored_seq A l (Ground a) :=
  [a];
stored_seq A l (Small s) :=
  Deque.deque_cmseq stored_seq s;
stored_seq A l (Big p child s) :=
  Deque.deque_cmseq stored_seq p ++
  chain_seq child ++
  Deque.deque_cmseq stored_seq s

with body_seq {A hl tl hk tk} (b : body A hl tl hk tk) : list A -> list A
by struct b :=
body_seq Hole accu :=
  accu;
body_seq (Single_child hd b) accu :=
  node'_cmseq stored_seq hd (body_seq b accu);
body_seq (Pair_yellow hd b cr) accu :=
  node'_cmseq stored_seq hd (body_seq b accu ++ chain_seq cr);
body_seq (Pair_orange hd cl b) accu :=
  node'_cmseq stored_seq hd (chain_seq cl ++ body_seq b accu)

with packet_seq {A hl tl ar k C} : packet A hl tl ar k C -> list A -> list A :=
packet_seq (Packet b tl) accu :=
  body_seq b (node'_cmseq stored_seq tl accu)

with chain_seq {A l ar k lC rC} (c : chain A l ar k lC rC) : list A
by struct c :=
chain_seq Empty := [];
chain_seq (Single _ pkt rest) := packet_seq pkt (chain_seq rest);
chain_seq (Pair lc rc) := chain_seq lc ++ chain_seq rc.

Equations cadeque_seq {A} : cadeque A -> list A :=
cadeque_seq (T c) := chain_seq c.
\end{lstlisting}
\caption{\cdcoq: model functions} % for stored triples, bodies, packets, chains, and \cdeques
\label{fig:cdc_model}
\end{figure}

\enlargethispage{\baselineskip}

The definitions of the model functions
for catenable deques
amount to 110~lines in~total.
\gitrepofile{theory/Cadeque/Models.v}

\begin{comment}
% fp: remove this comment; checking the whole project takes only about 3-4
%     minutes with Rocq 9.2. I don't know whether Rocq 9.3 will be faster.
Coq 8.20 type-checks them
in approximately 8 minutes
on a~modern laptop computer.
We~have reported this problem%
\footnote{\url{https://github.com/mattam82/Coq-Equations/issues/620}}
% https://github.com/mattam82/Coq-Equations/issues/620
% https://github.com/rocq-prover/rocq/issues/21526
% Merge PR #21666: Use automata rather than regular trees for guard checking.
% This representation prevents an exponential blowup when nesting inductive
% types. It makes guard checking basically instantaneous on some examples
% that used to take about 30s.
% fp: as far as I can tell, this PR has been merged after V9.2.0 was fixed
%     so will be released as part of 9.3.
and have been informed
that \Coq 9.3
(which at the time of writing is not yet released)
will be much faster on this example,
thanks to fixes and improvements in the guard checker.
%
Unfortunately, we~could not yet experiment with \Coq~9 because we rely on
\Hammer~\citep{czajka-20}, which at the time of writing is not available for
\Coq 9.
\end{comment}

% ------------------------------------------------------------------------------

\subsection{Code}
\label{sec:cadeque:coq:code}

The operations on catenable deques
are ported
from OCaml to \Coq
in a~rather straightforward way.
About 60 auxiliary functions are involved,%
\gitrepofile{theory/Cadeque/Core.v}
on top of which
the main five operations,
\opush, \oinject, \opop, \oeject, and \oconcat,
% plus the constant empty
are easily defined.
\gitrepofile{theory/Cadeque/Cadeque.v}

The correctness of each auxiliary function or operation
is expressed by decorating its definition with a~postcondition
whose statement uses suitable model functions.
For example,
the correctness of \opush
is expressed
by decorating the definition of \coqe|push x d|
with the postcondition
\coqe|cadeque_seq d' = [x] ++ cadeque_seq d|.

The auxiliary functions occupy approximately 1000 lines of code,
while the main five operations span about 60 lines of code.
All of these definitions are type-checked
and verified by \Coq in a~few minutes.
% (obviously NOT counting the 8 minutes mentioned above)

As indicated at the beginning of the paper
(\sref{sec:intro:code}),
\gitrepofile{theory/Signatures.v}
the file \texttt{Signatures.v}
sums~up the types and operations
that must be offered by an implementation of catenable deques,
as well as the correctness properties that these operations must satisfy;
and the file \texttt{Cadeque/Summary.v}
\gitrepofile{theory/Cadeque/Summary.v}
proves that we have implemented these operations
and established their correctness properties.

% README:
% to update these graphs:
% - go to the repository that contains the code (see \gitrepo in macros.tex).
% - run `cd bench && make clean && make run && make view`.
% - in Jupyter, click "Run All", then "Save".
% - commit.
% - then come back to the paper's repository and run `make import`.
%   (if you are not me, you may need to set REPO while running `make import`)

\begin{figure}[p]
\begin{subfigure}[b]{0.496\linewidth}
\includegraphics[width=\linewidth]{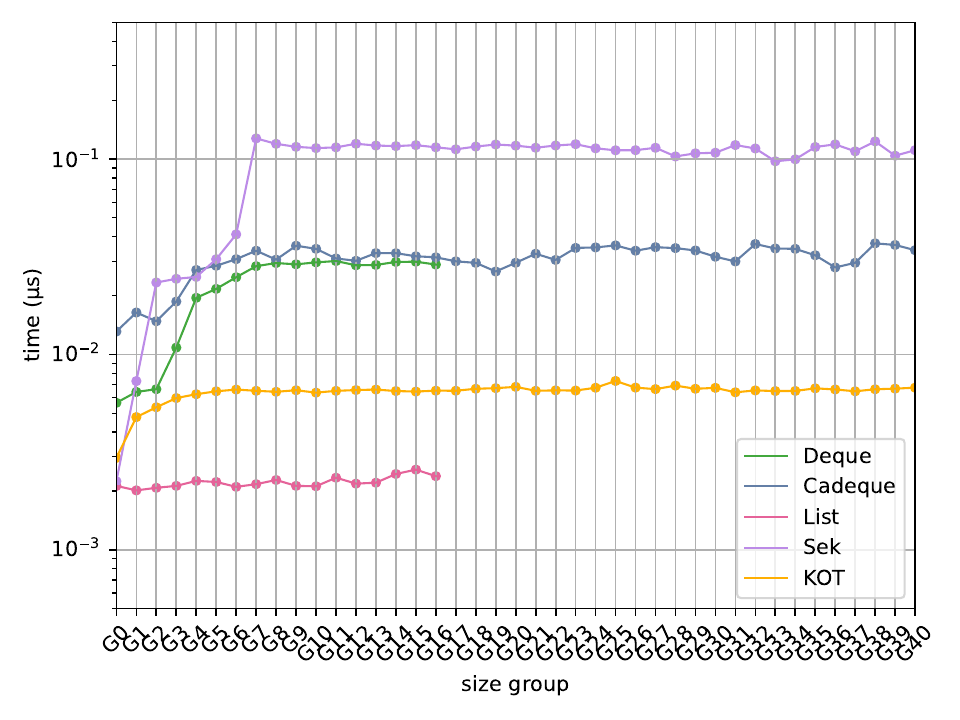}
\caption{\opush}
\label{fig:benchmark:push}
\end{subfigure}
\begin{subfigure}[b]{0.496\linewidth}
\includegraphics[width=\linewidth]{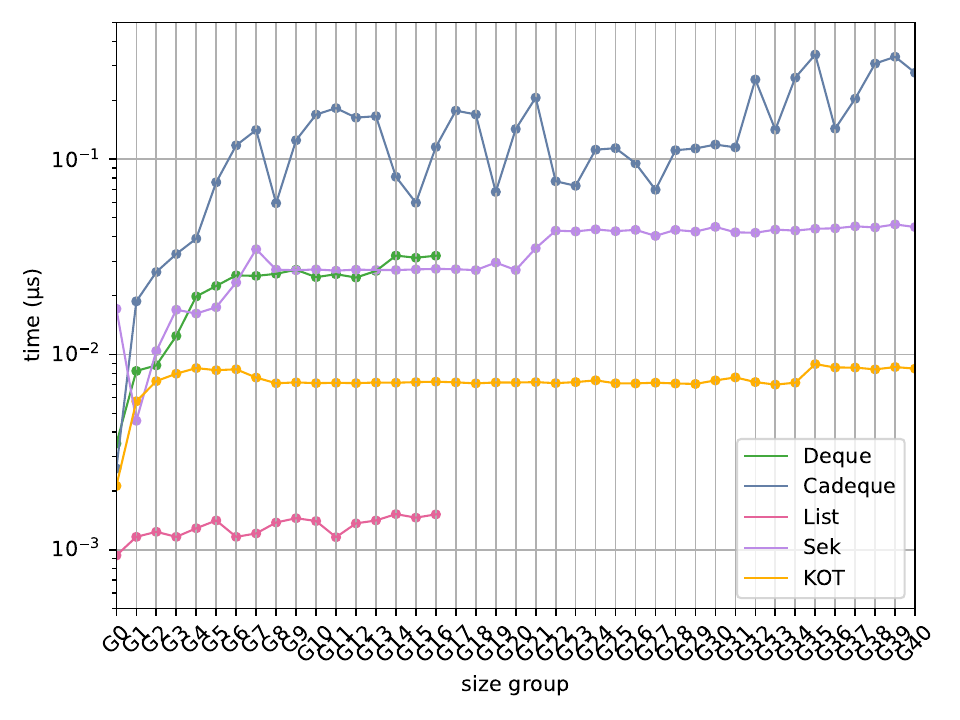}
\caption{\opop}
\label{fig:benchmark:pop}
\end{subfigure}

\vspace{8mm}

\begin{subfigure}[b]{0.496\linewidth}
\includegraphics[width=\linewidth]{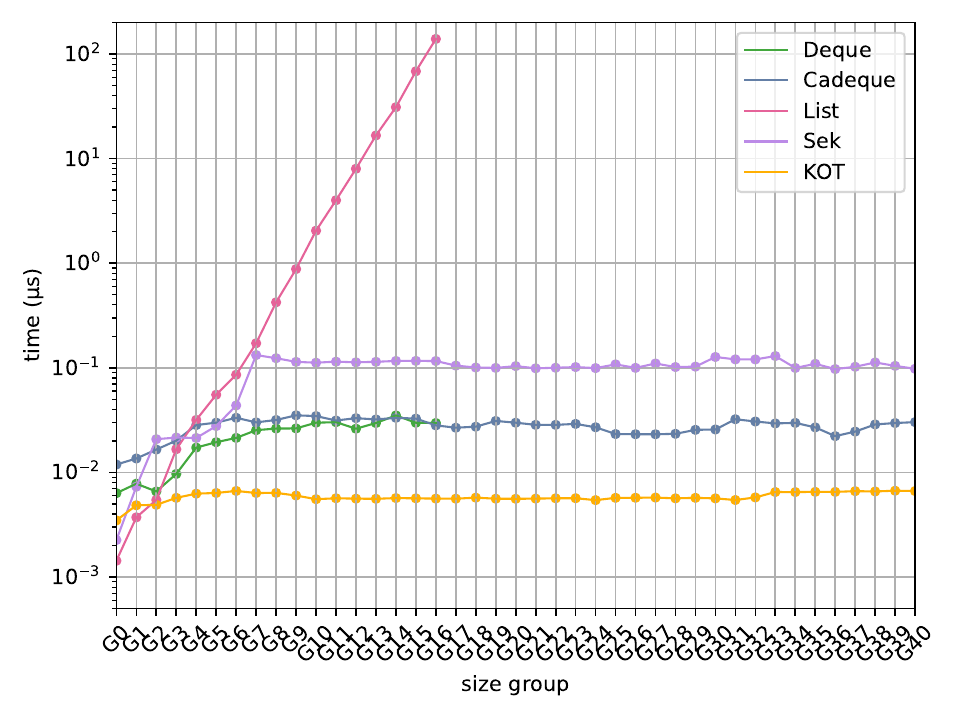}
\caption{\oinject}
\label{fig:benchmark:inject}
\end{subfigure}
\begin{subfigure}[b]{0.496\linewidth}
\includegraphics[width=\linewidth]{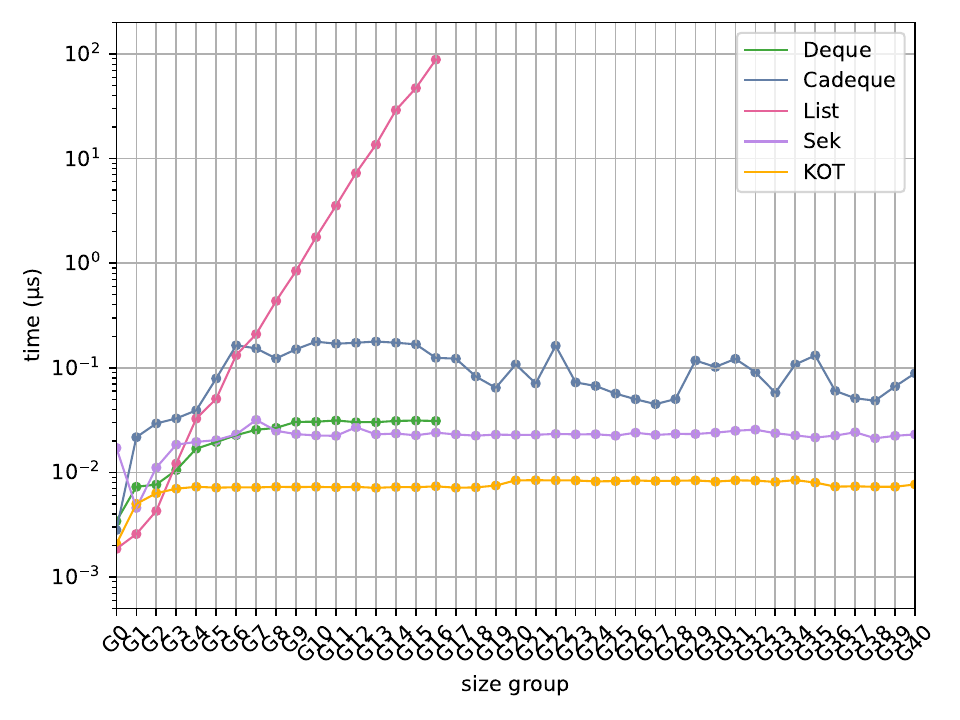}
\caption{\oeject}
\label{fig:benchmark:eject}
\end{subfigure}

\vspace{8mm}

\begin{subfigure}[b]{0.496\linewidth}
\includegraphics[width=\linewidth]{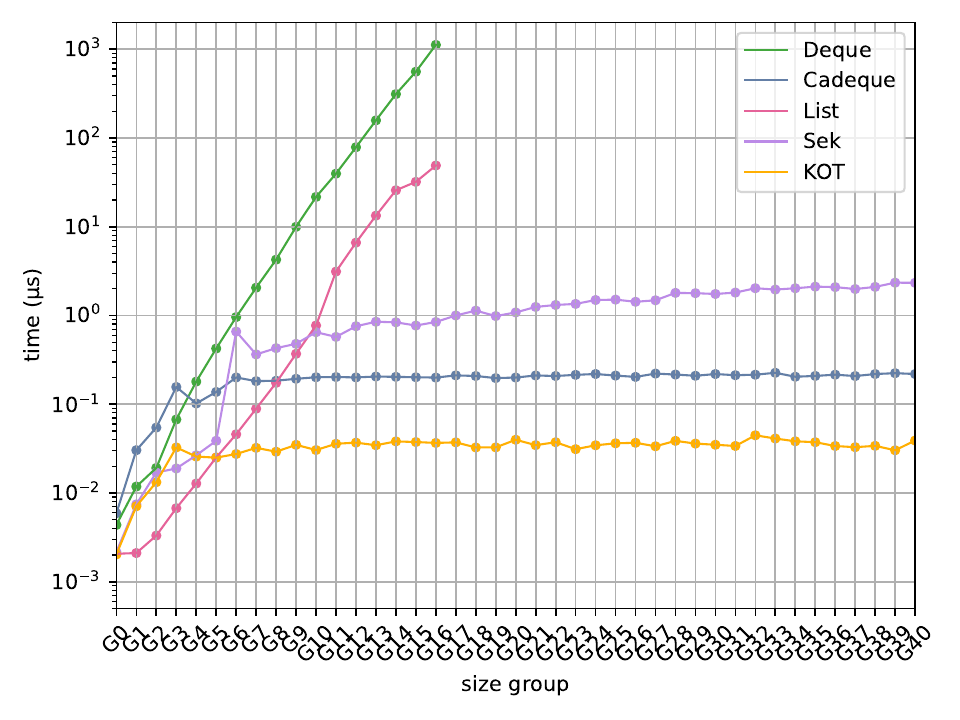}
\caption{\oconcat, with arguments of roughly similar sizes}
\label{fig:benchmark:concat}
\end{subfigure}
\caption{Experimental evaluation}
\label{fig:benchmark}
\end{figure}

\section{Executing our code} % was: Experimental evaluation
\label{sec:eval}

In this section, we discuss
some practical aspects of our OCaml and \Coq
implementations of catenable deques.
%
% fp: we benchmark catenable deques only (right?)
%     we could benchmark non-catenable deques, too,
%     but that would add confusion.
%
We explain how we have tested our OCaml implementation
and provide an evaluation of its performance
(\sref{sec:eval:ocaml}).
Our \Coq implementation can in principle be executed
either inside \Coq
or outside \Coq, via a translation to OCaml
(\sref{sec:eval:coq}).
Both approaches suffer from serious limitations,
which destroy the worst-case constant-time complexity bound of our code
and have a dramatic negative impact on performance.

% ------------------------------------------------------------------------------
% ------------------------------------------------------------------------------

\subsection{OCaml implementation}
\label{sec:eval:ocaml}

Although our OCaml code is not verified,
we have experimentally validated its functional correctness
via heavy random testing.
Using Monolith~\cite{pottier-monolith-21},
we have built millions of well-typed scenarios,
that~is, sequences of operations
of bounded length.
%
% We have used the default bound of 128.
%
We have run each scenario
using two distinct implementations of catenable deques,
namely
our candidate implementation
and
a~simplistic reference implementation,
% (based on lists)
%
and we~have checked that all observable results are the same.
No discrepancies have been found.

Furthermore, we have experimentally evaluated the performance of our OCaml
code. For comparison, we have evaluated four persistent data structures that
represent sequences:
\begin{enumerate}
\item
  Kaplan and Tarjan's purely functional non-catenable deques
  (\sref{sec:deque:ocaml})
  (``Deque'');
\item
  Kaplan and Tarjan's  purely functional catenable deques
  (\sref{sec:cadeque:ocaml})
  (``Cadeque'');
\item
  ordinary linked lists
  % from OCaml's standard library
  (``List'');
\item
  Charguéraud and Pottier's
  persistent chunked sequences~\cite{chargueraud-pottier-26}
  (``Sek'');
\item
  Kaplan, Okasaki, and Tarjan's simple persistent catenable deques
  \cite{kaplan-okasaki-tarjan-00},
  which have been implemented and verified
  by Ponsonnet and Pottier
  \cite{ponsonnet-pottier-26}
  (``KOT'').
\end{enumerate}
% Should we also show Steque (not discussed in this paper)?

% Fix Indentation
\noindent
Our methodology is as follows. Because the data structures that we wish to
evaluate are persistent, we do not restrict our attention to
``single-threaded'' scenarios, where each data structure is used at most once
as an operand. We consider complex scenarios, where the operand or operands of
each operation can be chosen in an arbitrary manner among the results of the
previous operations. Furthermore, because we wish to exhibit the relationship
between the cost of an~operation and the logical length of its result, we need
to build sequences whose logical lengths span a large interval and are roughly
evenly distributed across this interval.
We use the interval $\semiopen{0}{\pow{40}}$.
Therefore, we construct sequences
whose logical lengths range up to roughly one trillion,
that is, $10^{12}$.
We divide this interval into 41 disjoint subintervals, or \emph{bins},
namely
$\semiopen 01$,
$\semiopen 12$,
$\semiopen 24$,
and so on,
up to
$\semiopen{\pow{39}}{\pow{40}}$.
Through constrained random search,
we construct a scenario
of 2050 operations ($2050 = 41 \times 50$)
in such a~way
that each~bin is populated with 50 inhabitants,
  % not necessarily distinct!
  % note, though, that two sequences of length 1 are not necessarily equal;
  % they can have different internal structure.
that~is,
50 data structures
whose logical length falls within this bin.
Each operation in this scenario
is one of the five operations
$\opush$, $\opop$, $\oinject$, $\oeject$, and $\oconcat$.
(We use a~smaller proportion of $\oconcat$ operations
because they tend to produce sequences whose logical length
is huge.)
We measure how much time
each operation in this scenario
requires,
and display the results,
summarizing each bin as one data point.
Because these data structures are persistent,
an~operation can be played and replayed
in isolation as many times as one wishes.
We take advantage of this feature
to measure the cost of each operation independently
and reliably.

If a~persistent data structure has internal mutable state
then this method is somewhat problematic:
as an operation is replayed,
its result remains the same,
but its cost can vary.
In our benchmark, this issue potentially affects Sek and KOT,
both of which have internal mutable state.
Sek is optimized to perform best under single-threaded scenarios,
so we believe that, by replaying an operation several times,
we are likely to observe the worst-case complexity.
% e.g., one might push into a sequence
%       whose chunk has already been used for pushing;
%       this prevents a monotonic update and requires copying the chunk
%       (Sek does not compare elements to see that the element is in fact the same!)
%
KOT is a self-adjusting data structure,
which may update its internal state
before answering a~query;
therefore we believe that
the best-case complexity will likely be observed.
We accept these shortcomings of our methodology
because the main subjects of our benchmarks are
List, Deque and Cadeque,
which are purely functional.
Our evaluation of Sek and KOT is provided for comparison
but should not be considered fully reliable.

The benchmarks have been executed on a 2023 Apple MacBook Pro,
equipped with an~Apple M2 Max processor
and 64 gigabytes of RAM.
We used OCaml 5.4
and set the minor heap size to 512 million words,
that is, 4 gigabytes.
(This setting speeds up the benchmarks that allocate a lot of memory,
such as the List benchmark, by a factor of up to~12.)

Because the more naive data structures, namely List and Deque,
do not scale well to sequences of length $\pow{40}$,
we stop measuring their performance at length $\pow{16}$.

Our results appear in \fref{fig:benchmark}.
They correspond to commit
\texttt{444358e
}in our \href{\gitrepo}{repository}.
In each graph,
the horizontal axis shows the logical length of the result
of an~operation; a base-2 logarithmic scale is used.
The vertical axis shows the time consumed by this operation,
in microseconds; a base-10 logarithmic scale is used.
Here is our reading of these graphs:
\begin{itemize}
\item
As expected,
each of the five data structures under test
appears to perform a~$\opush$ operation in constant time.
Lists are cheapest,
at~about 2 nanoseconds per operation.
% (A list $\opush$ requires allocating one memory block of size~3.)
KOT comes second, at approximately 6 nanoseconds per operation.
Deque and Cadeque come third, at~about 36 nanoseconds per operation.
Sek is most expensive, at about 120 nanoseconds per operation.
One can see that Sek uses a~different representation of short
sequences up to size 64.
At size 64 and above, Sek represents a~sequence as a~tree
of \emph{chunks}, which are arrays of size 128. When a~$\opush$
operation is repeatedly applied to the same sequence, each
operation requires allocating a new chunk and copying data into~it;
this explains why Sek appears slowest in this benchmark.
\item
As expected,
each of the five data structures
appears to perform $\opop$ in constant time.
In~this benchmark,
catenable deques appear more expensive
than non-catenable deques,
and their behavior seems somewhat more erratic.
On catenable deques,
a~$\opop$ operation
can cost up to 330 nanoseconds;
$\opop$ appears to be significantly more expensive than $\opush$.
% List pop is super-cheap, as (we think) it involves two reads and a test.
% Sek pop is less expensive than Sek push,
% probably because it does not require copying a chunk;
% it creates a more restricted view on an existing chunk.
\item
As expected,
lists
perform
$\oinject$ and $\oeject$
in linear time.
The other three data structures
perform these operations
in constant time,
with costs that seem roughly comparable
to the costs of $\opush$ and $\opop$.
\item
We have measured the cost of $\oconcat$ for two arguments
of arbitrary (independent) lengths.
However, this information is difficult to visualize.
Therefore, we show the diagonal only:
the graph in \fref{fig:benchmark:concat}
represents the cost of concatenating two sequences
of roughly similar lengths.
As expected,
lists and non-catenable deques
perform $\oconcat$ in linear time.
Sek visibly does not have constant time complexity:
this seems consistent with its documentation,
which indicates that the complexity of concatenation
is $O(\log n)$ in ``common cases''
and $O(\log^2 n)$ in the worst case.
The remaining two data structures
perform concatenation in constant time.
%
% KOT is fastest, at about 40 nanoseconds per operation.
% But, again, it is possible that this benchmark measures
% its best case behavior.
%
Catenable deques behave consistently,
at roughly 220 nanoseconds per operation.
\end{itemize}

% Fix Indentation
\noindent
In summary,
although catenable deques can be up to 100 times slower than lists
on $\opush$ and $\opop$ operations,
they appear to perform all five operations
in constant time, as expected,
at a~speed of
roughly one to ten~million operations
per second on this machine.

% ------------------------------------------------------------------------------
% ------------------------------------------------------------------------------

\subsection{\Coq implementation}
\label{sec:eval:coq}

There are two main ways of executing a piece of code that is written in the
programming language \Coq. One approach is to run it inside \Coq: several
commands exist for this purpose. The other approach is to use \Coq's
\emph{extraction} mechanism,
%       there are by now two versions of this extraction mechanism
which translates \Coq code into OCaml code.
This OCaml code can then be linked
with other (verified or unverified) OCaml code
and compiled
% by the OCaml compiler
into executable code.
The first approach must be followed
if one wishes to use our catenable deques
as part of a verified algorithm
(say, a decision procedure)
and execute this algorithm as part of a proof that \Coq can check.
The second approach can be used
if one wishes to use our catenable deques
as part of a~(verified or unverified) program
that does not need to run inside \Coq.
We discuss both approaches in turn
(\sref{sec:eval:ocaml}, \sref{sec:eval:coq}).

% We might wish to explain that when computing inside \Coq,
% \Coq is able to ignore (not reduce) the terms whose sort is \coqe|SProp|;
% and when extracting to OCaml,
% \Coq is able to erase the terms whose sort is \coqe|SProp| or \coqe|Prop|.
% Is this correct? Is it relevant?

We must point out that
at this time,
regardless of which approach is chosen,
our \Coq implementation is \emph{not} efficient.
Indeed, our data types involve levels and sizes,
which \Coq is unable to ignore or erase,
because they are natural numbers,
whose sort is \coqe|Type|.
Although we like to think of levels and sizes as ``indices''
that have no influence on the computation,
% they are used only in the computation of other indices
we currently have no way of expressing this intent
and of taking advantage of it.
Because of this, our operations on catenable deques
perform
(intuitively useless, yet costly!)
computations on levels and sizes,
therefore do \emph{not} have
worst-case constant time complexity.
%
% To make matters worse, levels and sizes are natural numbers in Peano's
% representation, that is, in unary representation.
%
% We might wish to indicate that there is ongoing research work on ghost
% data in \Coq; this is discussed in the conclusion.

% Also: colors are not erased. That said, 1- I am not sure that colors are
% actually irrelevant; it might be the case that the code performs case
% analyses on colors, which cannot be erased; and 2- computing colors adds
% only O(1) overhead, which is acceptable.

% ------------------------------------------------------------------------------

\subsubsection{Executing the verified code inside \Coq}
\label{sec:eval:coq:execute}

We have experimentally verified that our \Coq implementation is able to
``compute inside \Coq''.
\gitrepofile{test_coq/reduction/}
In other words,
%
% fp: Armaël wrote:
%     it is possible to build a~deque by using the operations of the library
%     and to compute its result (its normal form)
% fp: but actually, if the result is a deque, then it is intuitively useless
%     to ask for its normal form, as it is supposed to be an abstract type
%     anyway. What we should care about is the observable result of a piece
%     of code that uses deques and eventually produces a concrete output.
%
% LATER because of this remark, in test_coq/reduction,
%      I would suggest extending the test to produce
%      an ordinary list of the elements of the deque test12
%      and check that it is the list that we expect.
%
given a~closed \Coq code fragment that uses our catenable deques
and has a~concrete type such~as (say) \coqe|list nat|,
\Coq can compute a~normal form,%
\footnote{This computation can be requested via the command \coqe|Eval vm_compute|.}
and this normal form is a~canonical form for the type \coqe|list nat|:
it is a~list of natural numbers.

This result may seem trivial,
but in fact it is not.
We found that,
in our initial attempt,
reduction could be blocked by opaque terms.%
\footnote{A \Coq~definition is
\href{https://rocq-prover.org/doc/master/refman/proofs/writing-proofs/equality.html\#controlling-reduction-strategies-and-the-conversion-algorithm}
{either \emph{transparent} or \emph{opaque}}.
An opaque definition cannot be unfolded by \Coq's reduction engine.
By default, all proofs are opaque.
This is a~good default
because, in most uses of \Coq,
there is no need to reduce proofs.
Making proofs opaque saves time
by forbidding reductions inside proofs,
which would be pointless and could be costly.
Here, though,
the code that we~wish to execute
contain casts,
which contain proofs of equality.
These opaque proofs block execution.}
% fp: there is some redundancy between this footnote and the text,
%     but never mind
%
Our analysis revealed that
these opaque terms could appear
as part of proofs of equality
between indices,
such as levels and sizes.
%
% Such proofs of equality are required (and frequently appear)
% because our data types are indexed with levels and sizes.
%
For example,
viewing a~deque of size~$2n+2$
as a~deque of size~$2(n+1)$
requires a~cast,
which itself requires
a~proof of the equality $2n+2 = 2(n+1)$.
When \Coq attempts to compute the~normal form of a~cast,
it~attempts to first reduce this proof of equality to a~normal form,%
\footnote{This seems silly, as \Coq's meta-theory guarantees that
every \emph{closed} proof of equality can be reduced to
\coqe|eq_refl|. However, as far as we understand, letting \Coq's type-checker
assume that \emph{every} proof of equality is convertible with
\coqe|eq_refl| would break strong normalization and logical consistency.}
and can then reduce the cast itself.
If for some reason this proof of equality contains an opaque term,
then the whole process becomes blocked.

In our initial attempt, we constructed equality proofs by using off-the-shelf
\Coq tactics, such as \coqe|lia|, a decision procedure for integer linear
arithmetic.
However, the proof terms produced by \coqe|lia| contain opaque terms. In fact,
even the basic arithmetic lemmas found in \Coq's standard library are declared
opaque, because they are not intended to be used in computations.
One could in principle define transparent copies of all of these lemmas,
effectively duplicating part of the standard library, and either avoid the use
of \coqe|lia| or define a copy of \coqe|lia| that uses our transparent copies
of the basic lemmas, but that would be unbearably tedious.
Clearly, we must use the standard library and \coqe|lia| without modification.

\begin{figure}[t]
\begin{lstlisting}[language=coq,backgroundcolor=\colorcoq]
Equations comp_eq {A} {eq_dec: EqDec A} {x y : A} (eq : x = y) : x = y :=
comp_eq eq with eq_dec x y => {
  | left e => e;
  | right ne => _ }.
\end{lstlisting}
\caption{Turning a proof of equality into a transparent proof of equality}
\label{fig:coq_comp_eq}
\end{figure}

To solve this issue,
we use a clever trick
that was communicated to us by Guillaume Melquiond.
The key idea is to define a function,
named \coqe|comp_eq| (\fref{fig:coq_comp_eq}),\gitrepofile{theory/Utils/comp_eq.v}
which transforms an~arbitrary proof of the equality \coqe|x = y|,
which may contain opaque terms,
into another proof of \emph{the same equality},
which does not contain opaque terms,
so that \Coq is able to reduce this new proof to a~normal form.
%
% fp: this is silly anyway since (intuitively) x and y are indices
%     and should not even exist at runtime...
%
The implementation of \coqe|comp_eq| is simple but clever.
If \coqe|x| and \coqe|y| have type \coqe|A|,
then one requires the type \coqe|A|
to be equipped with an equality test, \coqe|eq_dec|,
whose type is \coqe|forall x y : A, {x = y} + {x <> y}|. % ≠ does not work
One applies this equality test to \coqe|x| and \coqe|y|.
Crucially, the term \coqe|eq_dec x y|
\emph{can} (usually) be reduced to a~normal form
without becoming blocked somewhere along the way,
because an equality test is intended for this very purpose.
If this normal form is \coqe|left e|,
where \coqe|e| is a~proof of the equality \coqe|x = y|,
% or: a witness
then we return~\coqe|e|.
% whose normal form (if we are working with closed terms) will be \coqe|eq_refl|, right?
% instead of returning \coqe|e|, could we return \coqe|eq_refl|? (maybe not)
%
Furthermore,
the case where this normal form is \coqe|right ne|,
where \coqe|ne| is a proof of the disequality \coqe|x <> y|,
cannot arise!
Indeed, in this case,
by confronting the equality proof~\coqe|eq|
and the disequality proof~\coqe|ne|,
we are able to obtain a contradiction.
Thus, the second branch in the definition of \coqe|comp_eq| is dead.
Because the equality proof \coqe|eq| is used only in this branch,
% (where it is used implicitly)
there is never a~need to reduce it to a~normal form.
Therefore,
even though it~might contain opaque subterms,
it cannot block reduction.

Thanks to \coqe|comp_eq|, we can continue to use \coqe|lia| and other
off-the-shelf tactics to prove equalities between levels and sizes. We simply
need to wrap the resulting proof with \coqe|comp_eq|. This amounts to ignoring
the computational content of the equality proof produced by \coqe|lia| and
replacing it with a direct equality test, which we know will always succeed.

Of course, this state of things is not entirely satisfactory. First, it
requires the type of indices (in our case, \oc|nat|) to be equipped with
an~equality test. Second, it forces us to actually execute this equality test
(and pay its runtime cost), even though we have proved that it cannot fail.
Finally, it requires the indices \coqe|x| and \coqe|y| to exist at runtime,
even though we would like to think of them as computationally irrelevant.

Finally, we have no rigorous way of making sure that we have found all of the
places where execution can be blocked by an opaque term. As far as we know,
this can be determined only via testing.

% ------------------------------------------------------------------------------

\subsubsection{Extracting the verified code from \Coq to OCaml}
\label{sec:eval:coq:extract}

We have checked that our verified implementation can be translated to OCaml
using \Coq's extraction mechanism.
\gitrepofile{extraction/}
This attests that our \Coq formalization is indeed executable.
We have submitted this OCaml code to random testing:
as expected and hoped, no bug was detected.

The resulting OCaml code has same the computational content as the \Coq code.
During extraction,
all proofs
% (of sort \coqe|Prop|)
are erased.
In particular,
every call to \coqe|comp_eq| (\sref{sec:eval:coq:execute})
is erased,
because its result is a proof of equality.
% This result is destruct in a~\coqe|match| construct
% where there is just one branch, for \coqe|eq_refl|.
% But this remark is irrelevant, I believe.
Unfortunately,
all indices (colors, levels, and sizes) remain,
requiring computations whose results are ultimately useless.
Thus, the OCaml code that is obtained via extraction is not efficient and
does not have worst-case time complexity~$O(1)$.
In practical applications, our hand-written OCaml code should be used.

% Encouraged by one of the anonymous reviewers,
We have compared the performance
of the~hand-written code
and
of the extracted code
in a micro-benchmark that involves $N$~\opush operations
followed with $N$~\opop operations.
As~expected, the observed performance of the hand-written code is a linear
function of~$N$: on~average, a \opush or \opop operation requires
about 60 nanoseconds on our benchmark machine.
The observed performance of the extracted code is much worse,
and is super-linear in~$N$.
At $N=500$,
the extracted code is about 300 times slower than the hand-written code;
at $N=5000$
it is about 4500 times slower.
This confirms that the extracted code does not perform a \opush or
\opop operation in constant time.

% See the subdirectory minibenchmark.
% Each mini-benchmark involves N push operations followed with N pop operations.
% Benchmark 1 (hand-written OCaml code) is repeated 100 times.
%   At n =    500 it consumes 7ms.
%   At n =   5000 it consumes 58ms.
%   At n = 100000 it consumes 1180ms.
%     (thus, about one million push/pop operations in 58ms, that is, one operation in about 60 nanoseconds)
% Benchmark 2 (extracted OCaml code) is repeated only 1 time.
%   At n =  500 it consumes 22ms. (slowdown = 316x)
%   At n = 5000 it consumes 2.7s. (slowdown = 4600x)

\section{Related work}
\label{sec:related}

% https://cstheory.stackexchange.com/questions/1539/whats-new-in-purely-functional-data-structures-since-okasaki/1550#1550

% ------------------------------------------------------------------------------

\subsection{Families of persistent data~structures}

% Situate and Kaplan~and Tarjan's paper in the landscape.

\kt's purely functional, real-time catenable deques
are part of a~broader landscape
of persistent data~structures.
For an~overview of the previous work in this area,
we~refer the reader
to their paper~\cite{kaplan-tarjan-99}.
Let us merely indicate that
the persistent data~structures
in the literature
can~be classified in several families,
which rely on different implementation techniques.

%
% Family 1.
%
One family of data~structures,
which includes
Driscoll~et~al.'s
persistent lists with catenation
\cite{driscoll-sleator-tarjan-94}
and
Buchsbaum and Tarjan's
persistent deques~\cite{buchsbaum-tarjan-95},
achieve persistence
by relying on
imperative updates
and explicit management of version numbers.
Due to the use of imperative updates,
it seems safe to say that
the data~structures in this family
are comparatively difficult to verify.
%
% In fact, as far as we know, none have been verified.
% -- maybe such a~claim would be a~bit strong
% -- perhaps the work of Sebastian Burckhardt et al.
%    on Concurrent Revisions could be relevant
% -- e.g. "Prettier Concurrency", which I have not read
%
% Some form of Separation Logic
% would likely be required.
%

% Family 2.
%
A~second family,
which includes
Okasaki's
simple queues and deques~\cite{okasaki-95},
his
catenable deques~\cite{okasaki-97},
as well as a~large variety of data~structures
described in his book~\cite{okasaki-book-99},
involve suspensions---%
% also known as thunks
simple mutable data~structures
that offer a~basic form of memoization.
They can~be easily and elegantly implemented
in lazy functional programming languages,
such as Haskell,
where suspensions are implicit.
These data~structures
are known as ``purely functional''
because
memoization
influences their time and space complexity
but not their correctness.
Therefore, in a~proof of correctness,
the presence of memoization,
and the fact that memoization involves imperative updates,
can~be ignored.

A~somewhat atypical member of this family
is Kaplan, Okasaki, and Tarjan's
simple catenable deques~\cite{kaplan-okasaki-tarjan-00}.
Instead of relying on suspensions,
whose mutable field can~be written at most once,
this data~structure
involves mutable fields
that can~be written several times.
However,
the authors ensure that
every update is logically unobservable,
because
the overwritten value
and
the new value
are
two data~structures
whose logical models are equal.
Again,
these restricted updates
influence the data~structure's time and space complexity
but not its correctness.
The correctness and time complexity of this data structure
have been verified by Ponsonnet and Pottier~\cite{ponsonnet-pottier-26}.

By ignoring the presence of imperative updates,
the data~structures in this second family
can in principle be
% relatively easily
verified within a~proof assistant
based on type theory, such as \Coq.
However,
if the code is executed
(either inside \Coq or via~extraction to OCaml)
without imperative updates
then its time complexity will be incorrect.

%
% Family 3.
%
A~third family of data~structures
are ``strictly functional'':
that is,
they rely strictly on immutable heap-allocated memory blocks
and pointers between these blocks.
No imperative updates,
whether explicit or implicit,
are involved.
Examples of this family
include
% also:
% many data~structures in Okasaki's book
% many data~structures in the FP literature (AVL trees, heaps, Patricia~trees, tries, finger trees, etc.)
% myers-83
Kaplan~and Tarjan's
persistent catenable stacks~\cite{kaplan-tarjan-95}
as well as the data~structure
that is
considered in this paper,
namely,
Kaplan~and Tarjan's
persistent real-time catenable deques~\cite{kaplan-tarjan-99}.
Many more examples can be found
in the books by Appel~\cite{appel-vfa},
by Nipkow et al.~\cite{nipkow-fav},
and
by Nipkow~\cite{nipkow-fdsa}.
The data~structures in this third family
can~be
% relatively easily
implemented and
verified within a~proof assistant
based on type theory, such as \Coq.
Furthermore,
there is hope that, someday,
these data structures can~be efficiently executed
inside \Coq or via~extraction to OCaml.
This requires irrelevant proofs and indices to be erased,
which, today, does not yet seem possible (\sref{sec:eval}, \sref{sec:conclusion}).

\subsection{Kaplan~and Tarjan's deques}

In this paper,
we have described OCaml and \Coq implementations
of \kt's real-time catenable deques \cite{kaplan-tarjan-99},
a~strictly functional data~structure.
\kt provide only an~English description of the data~structure:
therefore,
the type definitions, the~code,
and the proofs of correctness
are contributions of this paper.

A~set of unpublished course notes by \citet{mihaescu-tarjan-03} sketches
a~variation of \kt's non-catenable and catenable deques.
In comparison with \kt,
Mihaescu and Tarjan
appear to impose stricter size constraints on buffers.
Their description is informal and brief.
% English
% 4 pages
We~have not studied it in depth.
It would be interesting to implement it
in OCaml, \Coq,
or some other typed functional programming language.
The techniques that we have used
can~in principle be applied to~it.

An~online repository
by~Thomas Refis,\footnote{\url{https://github.com/trefis/deques/}}
dated 2013,
contains a~\Coq implementation
of \kt's non-catenable deques.
% This was a~master's internship (TRE), supervised by Yann Régis-Gianas
% Thomas Refis sent us his report (in French) (private communication)
%
% Refis uses Program, and explains (in the report) that he had to work
% around some bugs/limitations of Program, which would generate ill-typed
% proof obligations.
% The workaround uses an~auxiliary inductive type, regularisation_cases.
%
% The repository contains 1000 lines of specs and 650 lines of proofs
% according to coqwc, and does not include a~proof of correctness.
% Our version is 400 lines of specs and under 100 lines of proofs. For the
% size- and level-indexed version, these numbers go up to 600 and 300.
The data~structure is represented as chain of packets,
%
% a~list of "stacks" of "levels",
% where a~level is a~pair of a~prefix buffer and a~suffix buffer.
% so stack = packet
%
where the type of packets carries two parameters~\coqe|A| and \coqe|B|,
as in our code (\fref{fig:ncd_coq_types}).
The data~structure's invariant,
which involves sizes and colors,
is expressed a~posteriori. % whereas we express it a~priori, using extra~indices
There is a~proof that the four operations
preserve this invariant.
There is no proof of functional correctness.
% In his report, Refis writes that this proof is not necessary because
% "le type de nos opérations est suffisament précis pour encoder leur spécifications".
% I do not agree.
% In fact, in a~personal communication,
% Refis agrees that the proof of correctness is missing.
% Oh, in fact, the report does indicate that the proof of correctness
% is missing: "En faisant passer les éléments d’un niveau à l'autre,
% nous avons par exemple peut-être inversé, par inattention, l'ordre
% de deux éléments".

% should we say something about this repository?
% Pierre Pradic
% https://gitlab.aliens-lyon.fr/ppradic/functional-steak
% (Armael has contributed to it.)
% Last commit in 2018.
% It seems to contain OCaml implementations of non-catenable and catenable deques.
% It is not clear whether the code is complete.
% (A~comment in the README suggests that it is not.)
% It is not clear whether the code has been tested.

\citet{schaub-23} describes a~Haskell implementation of \kt's
non-catenable deques.
This implementation is not verified.
Schaub describes the data~structure with
a~plain algebraic data~type:
it is not a~generalized or nested algebraic data~type.
As~a~result,
the~content of a~buffer must be described in a~coarse way:
it is a~``binary tree''
where every node
is either a~leaf (which carries an~element)
or a~binary internal node.
To distinguish between these~cases,
every node carries a~tag,
and runtime tests are needed.
If the code contained a~conceptual mistake,
these tests could fail.
We~do~not need such tags or runtime tests:
by exploiting nested and generalized algebraic data~types,
we are able to represent the content of a~buffer
as nested pairs of type~$A^{2^l}$.

% We are aware of a~few attempts to implement \kt's catenable deques.
%
Two online repositories
by~Alex Lang\footnote{\url{https://github.com/alang9/deque/}}
and
by~Edward Kmett,\footnote{\url{https://github.com/ekmett/deque/}}
which are related to one another
and are
both dated 2015,
appear to
contain implementations of \kt's non-catenable and catenable deques.
This code is not commented and appears to be untested.%
\footnote{Alex Lang, private communication, August 2024.}
It has not been released or verified.
%
\begin{comment}
Kmett's repository is a~fork of Lang's repository.
The two repositories seem to have incomparable commits.
There is a~package description file,
but this package does not seem to exist on Hackage.
We have emailed Alex Lang and Edward Kmett to ask them about the status of this work.
As far as I can~tell, Edward Kmett did not reply.
Alex Lang sent a~brief reply:
\begin{quote}
Sorry I've been sick for the last couple of days. Apologies as well for the
lack of comments in these repos. I \emph{think} the code in the repos was
based on this algorithm from Kaplan~and Tarjan:
\url{http://www.math.tau.ac.il/~haimk/adv-ds-2000/jacm-final.pdf}.
It's been a~while since I've thought about these deques and it looks like
there's no tests in the repo so I can't really confirm that the code is
correct.
\end{quote}
\end{comment}

% ------------------------------------------------------------------------------

% Data-Structural Bootstrapping is the subject
% of Chapter~10 in Okasaki's book \cite{okasaki-book-99}.
%
% The name (and idea) is attributed to Buchsbaum (1993).
% Two techniques are identified:
%
% + structural decomposition
%   obtaining a~complete data~structure out of an~incomplete one
%   e.g. queues of unbounded size, out of queues of bounded size
%   this seems to often involve non-uniform recursion
%
% + structural abstraction
%   collections that contain collections as elements
%   (so concat becomes insert!)

% ------------------------------------------------------------------------------

% Connections with numerical systems

\subsection{Connections with numeral systems}

The ties between numeral systems and data~structures have been pointed
out and exploited by many researchers,
including
\citet{clancy-knuth-77},
\citet{vuillemin-78},
\citet{myers-83},
\citet{hinze-tr-98},
% Pierre-Évariste that says Hinze's TR served as inspiration for
% "Calculating Datastructures"
% https://link.springer.com/chapter/10.1007/978-3-031-16912-0_3
\citet[Chapter~9]{okasaki-book-99}.
%
% also: Implicit Recursive Slowdown is the subject
% of Chapter~11 in Okasaki's book \cite{okasaki-book-99}.
%
\citet{elmasry-katajainen-22} % fp: I have not been able to obtain this article
survey
the use of regular numeral systems
in the design of data~structures.
In
\kt's
data~structures~\cite{kaplan-tarjan-95,kaplan-tarjan-99},
the redundant binary representation
serves as a~source of inspiration,
but in the end,
the connection between the numeral system
and the data~structure appears somewhat loose.
In~particular, a~data~structure
that represents a~sequence of $n$~elements
does not have the same shape as
a~representation of the number~$n$.

% Swierstra. Heterogeneous binary random-access lists.

\begin{comment}
Pierre says:
``Étude du lien systématique entre représentation numérique et
  système numérique, motivé précisemment pour éviter le passage par
  les nested types''
\citet{dagand-letouzey-taghayor-22}
Dagand et al.\ appear to criticize the ``nested data~type'' approach
to representations of numbers (and data~structures based on numerical systems).
However, their criticism is difficult to understand:
``[This approach] forcefully smashes intricate data~containers (leaf binary
trees in this case) down to a~hodgepodge of binary products, making proper
algorithmic work on the underlying data~container meaningless''.
\end{comment}

% ------------------------------------------------------------------------------

\subsection{Nested data~types}
\label{sec:related:nested}

Nested data~types~\cite{bird-meertens-98} play a~central role
in the description of the data~structures
considered in this paper
and of many other data~structures in the literature.
The need for nested data~types naturally arises
out of an~algorithmic design principle:
one typically wishes to organize
a~data~structure in several levels
and to~view a~group of elements at level~$l$
collectively
as a~single element at level~$l+1$.
Thus,
as one moves down from one level to the next,
the type parameter that describes ``elements''
must be instantiated
with ``group of elements''.
As a~suitable notion of~``group'',
one might wish to use a~pre-existing data~type,
such as ``pair'',
or
one of the data~types involved in
the data~structure that is being defined,
in which case a~\emph{truly nested} data~type~\cite{abel-matthes-uustalu-05} arises.
% also known as: self-nested \cite{blanchette-meyer-popescu-traytel-17}
%
The~data~type \Bush~\cite{bird-meertens-98}
is an example of a~truly nested data~type;
so are the data~types
of non-catenable deques
and catenable deques
considered in this paper.

As discussed in \sref{sec:obstacles},
defining and working with nested data~types
can be difficult.
Difficulties appear
not only in \Coq
but also in other proof assistants,
such as Agda.
In some cases,
the definition of a~nested data~type
is rejected by the proof assistant;
in other cases,
it~is accepted,
but the definition of an inductive function
over this data~type
is rejected.
To~work around these obstacles,
we have indexed several types
with levels and sizes
(\sref{sec:obstacles:similar})
and we have used
generalized model functions
instead of plain model functions
(\sref{sec:obstacles:model}).

We have sketched
a~systematic way of
transforming a~data~type
into a~level-indexed variant of this data~type
(\sref{sec:obstacles:similar:bush}).
Although
we could not find a~paper where
this transformation
is documented,
%
% montin-ledein-dubois-22 document it in the specific case of \Bush
%
the idea~of enriching nested data~types with extra~indices
does appear in several forms
in the literature.
%
% fp: I believe that Abel's indexing with "sizes"
%     is not the same thing as indexing with levels;
%     it is about counting the iterations of the functor that defines Bush,
%     not about counting the number iterations of Bush itself;
%     unless that is the same thing? I am confused.
%
For example, \citet{abel-06} presents a~system of ``sized types''
whose termination checker is type-based.
This contrasts with \Coq's termination checker,
which is based on an ad~hoc syntactic criterion.
Abel presents a~``size''-indexed variant of the type \Bush
and shows that he is able to define inductive functions
over this type.
Besides,
we~remark that
indexing \Bush with a~level
seems tantamount to
defining all powers of the type \Bush simultaneously,
an idea~that was reportedly suggested to \citet{matthes-11}
by Conor McBride.

In a~related vein,
\citet{fu-selinger-23}
show that,
although the type \Bush can be defined in Agda
(when the positivity check is turned off and when ``large indices'' are enabled),
% Fu and Selinger do not indicate that it is necessary
% to disable the positivity check. Yet Johann and Polonsky
% explicitly use {-# NO_POSITIVITY_CHECK #-} when defining
% the type Bush. I have checked that Agda 2.8.0 does indeed
% require NO_POSITIVITY_CHECK. It also requires --large-indices.
% Johann and Polonsky indicate that it is OK to turn off the
% positivity check by referring to their earlier paper
% "Higher-kinded data types: Syntax and semantics".
defining $\mathit{map}$ and $\mathit{fold}$ functions
for this type is difficult.
To resolve this difficulty,
they define $\mathit{map}$ and $\mathit{fold}$
simultaneously for all powers of the type \Bush.
%
%
% They use `NTimes n Bush`.
%
Thus, even though they do not index the type \Bush with a~level,
they do in some places need to explicitly work with the type $\Bush^n$,
which can be viewed as a~level-indexed variant of \Bush.

\citet{johann-polonsky-20} criticize Fu and Selinger for working with an
indexed variant of the type \Bush, instead of directly with the type \Bush
itself. Still using Agda, they propose a different approach, \emph{deep
induction}, which is applicable to all algebraic data types,
% but not term-dependent types; only type-indexed types
including nested types and truly nested types.
%
% I have not understood in detail *how* the deep induction principle
% is defined.
%
They claim that this lets them present
``the first-ever useful induction rule for bushes''.
% Johann and Polonsky criticize an earlier version of Fu and Selinger's work
% (2018). In short, they point out that Fu and Selinger's treat only the type
% Bush (as opposed to arbitrary truly nested data types). Furthermore they do
% not obtain an induction principle for Bush itself; instead they obtain one
% for NBush.
% Johann and Polonsky's code can be found at
% https://cs.appstate.edu/johannp/FoSSaCS20Code.html
% Some definitions of their code are decorated with TERMINATING,
% which bypasses the termination checker, I suppose. The paper
% does not mention this.

% Other work by Patricia Johann et al.:
% Johann and Ghiorzi (CPP 2022) extend deep induction to GADTs.
% Johann and Morehouse (WoLLIC 2025) extend deep induction to inductive families.
% Johann and Ghiorzi (LMCS 2021) study parametricity for nested ADTs and GADTs.

\newcommand{\LNDT}{\mathit{LNDT}}
With a~similar motivation,
\citet{montin-ledein-dubois-22}
present a~very general type, $\LNDT$,
of which many (but not all) nested data~types,
including \Bush,
are instances. % special cases
They then proceed to define
the functions
$\mathit{map}$ and $\mathit{fold}$,
and to establish properties of these functions,
in a~generic way,
for all instances of $\LNDT$ at once.
Unfortunately,
the cyclic definition $\Bush = \LNDT\;\Bush$
is rejected by \Coq,
and is also rejected by Agda,
unless its termination checker is turned off.
As a possible work-around,
they show that the type~\Bush
can be indexed with a level
(a~special case of the systematic transformation that we have sketched),
but they do not investigate this approach further.

Nested data types
can also exist in systems
where inductive data types
are not a~primitive feature,
but must explicitly encoded
in terms of lower-level concepts.
For example,
\citet{blanchette-meyer-popescu-traytel-17} encode nested data~types
in the proof assistant Isabelle/HOL,
which is based on
higher-order logic.
Following an idea by
\citet[\S10.1.1]{okasaki-book-99},
which we have already discussed (\sref{sec:intro:alternative}),
they perform a~two-step construction.
In~the first step, an~over-approximation of the desired nested data~type is
defined.
% (as an~ordinary data~type)
% One can~think of this as "dynamic typing": one gets a~data~structure
% with too many inhabitants and too many runtime tags.
%
In the second step, a~natural-number-indexed well-formedness predicate
is defined so as to select just the desired inhabitants.
Then, the desired type is obtained as a~subset type.
Furthermore, Blanchette et al.\ are able to automatically construct
induction and recursion principles for the type thus defined.
%
% This is quite involved.
%
% Their work is implemented in Isabelle/HOL.
% The relevant commands are `nonuniform_datatype`,
% `nonuniform_primrecursive`.
%
% They mention that "Agda~supports self-nested data~types" such as Bush.
%
They seem to indicate that their system does not support
truly nested data~types, such as \Bush. % self-nested
It is not clear to us
why this is~so
and
whether this is an~accidental or fundamental limitation.

\section{Conclusion}
\label{sec:conclusion}

% ------------------------------------------------------------------------------

% Summary of what we have achieved:

\subsection{Summary}

In this paper, we have presented OCaml and \Coq implementations of \kt's
non-catenable and catenable deques. Furthermore, we have verified the
functional correctness of our \Coq implementations. Our proofs of functional
correctness are mostly automated, therefore remarkably short: they occupy only
150~lines.

In this endeavor,
the main challenge
has been
to translate \kt's textual descriptions of the data~structures
% and of their invariants
into type definitions.
Beyond the initial hardship
of understanding these complex data~structures,
we~have faced the difficulty
of dealing with
the limited expressiveness of OCaml and \Coq.
In OCaml, the absence of dependent types
makes~it difficult to reason about integer sizes
at the type level;
as a~workaround,
in one place,
% (\sref{sec:cadeque:ocaml:buffers})
we~have used a~phantom type parameter.
In \Coq, the positivity condition
causes seemingly natural inductive type definitions
to be rejected;
as a~workaround,
we~have indexed several types with integer levels or sizes.

In the end,
thanks to
OCaml's
generalized algebraic data~types
and
\Coq's
indexed inductive types,
we~have been able
to formulate type definitions
that are precise,
in the sense that they fully express
the desired invariants.
As we~wrote the code
of the main operations
and
of the many auxiliary functions,
these type definitions offered an essential help:
they enabled the~early detection of
several kinds of mistakes,
including
violations of an~invariant
and
missing branches in case analyses.

% ------------------------------------------------------------------------------

% Lessons of our work, especially about Coq.

\subsection{Lessons about \Coq}
\label{sec:lessons}

From this work,
we~draw a~few lessons
about the \Coq proof assistant.
On the positive side,
the \Equations plugin
and the tactics~\Hammer and \AACTactics
are powerful and useful.
We~believe that,
thanks to them,
verifying purely functional data~structures
in \Coq
can be a~relatively pleasant experience.

On the negative side,
\Coq as a~programming language
suffers from a~certain lack of abstraction and modularity,
which stems from
the syntactic criteria
that \Coq applies
when determining
whether an inductive
type or function
definition
is acceptable.
For example,
in the course of this work,
after defining
what we~thought was a~nice abstract type \coqe|deque A|
of non-catenable deques,
we~discovered that this type could not be used as planned
inside the inductive definition of catenable deques.
There were two problems with~it.
First,
because
\coqe|A| was a~non-uniform parameter of \coqe|deque|,
the~inductive definition of catenable deques
was not allowed to exhibit a~cycle
% that goes
through this parameter.
Second,
even though we~could define a~function \coqe|size : deque A -> nat|,
we~were not allowed to use this function,
within
the~inductive definition of catenable deques,
to~impose a~constraint on the size of a~deque.
To work around these problems,
we~had to go back
and index the type of non-catenable deques
with integer levels and sizes.
%
% This is a~lack of modularity:
% depending on its intended use,
% the definition of a~type
% may or may not be acceptable.
%
In fact, the type of catenable deques
that we obtain in the end,
\coqe|cadeque A|,
suffers from a~similar problem:
it is not size-indexed.
If someone wished to use a~catenable deque,
subject to a~size constraint,
in the definition of a~new inductive type,
then they may have
  % (if it is a~deque of T's where T is the new type)
to go back
and define a~size-indexed variant of the type \coqe|cadeque A|.
For the sake of simplicity, we~have not done~this.

% A~similar problem exists for model functions:
% depending on how a~function is defined,
% Coq will recognize that its uses are acceptable or not.

% Furthermore, \Equations can require a~long type-checking time.
% And \Equations has quality-of-implementation problems
% (e.g. name collision bugs).

Another area~where improvement seems needed
is efficient execution of \Coq code,
either within \Coq,
or via~extraction to OCaml.
In either case,
the integer indices that appear in our type definitions
exist at runtime.
They are computed at runtime
and
stored in the data~structure.
Furthermore,
when executing the code inside Rocq,
casts between indices
are not erased
and have a~significant runtime cost,
as they require normalizing equality proofs.
Notions of ghost data~exists
in several proof assistants and program verification environments,
such as F*, Agda, Idris, and Why3.
There is hope that \Coq,
too, can be extended with such a~notion
\cite{winterhalter-24}.
We are very much looking forward
to such a~feature,
which seems necessary
for realistic (efficient) programming with indexed types.

% ------------------------------------------------------------------------------

% Directions for future work:

\subsection{Future work}

In future work,
it may be interesting to implement and verify
the variation on catenable deques
that is sketched in Mihaescu and Tarjan's course notes
\cite{mihaescu-tarjan-03}.
%
\begin{comment}
% fp: this was in the submission version
%     but has now been done!
It would also be interesting to verify
Kaplan, Okasaki, and Tarjan's catenable deques
\cite{kaplan-okasaki-tarjan-00}.
According to its authors,
this data~structure is simpler than \kt's catenable deques.
It relies on imperative updates
to support all five operations
with amortized time complexity $O(1)$.
%
Quite remarkably, these updates affect the data~structure's time complexity,
but not its correctness.
%
Thus,
one could set three separate goals:
verify the correctness of the data~structure, without imperative updates;
verify its correctness, with imperative updates;
verify its time complexity, with imperative updates.
%
% For the latter two goals,
% one cannot work with a pure program that is shallowly embedded in Coq.
% One must move to either a monadic program (?)
% or a deeply embedded program
% and use a program logic, such as Iris.
% But there is no point in saying too much here.
%
\end{comment}

In a~different direction,
it could be a worthwhile exercise
to use a program logic for OCaml,
such as Osiris~\cite{seassau-yoon-madiot-pottier-25}
or Zoo~\cite{allain-scherer-zoo-26},
to prove that our OCaml code
performs the same computations as our Rocq code.
In essence,
this would be a proof that the indices that exist in the Rocq code
truly are ``ghost'' indices
and are correctly erased in the hand-written OCaml code.
% which could be carried out without waiting for an implementation
% ghost index erasure in \Coq \cite{winterhalter-24}.

Another way of achieving a~similar result would be to implement
Kaplan and Tarjan's catenable deques using simple types,
as suggested in \sref{sec:intro:alternative},
as opposed to indexed types.
We have explored this route up to non-catenable deques
\gitrepofile{theory/Deque/DequeSimplyTyped.v}
but have not done catenable deques in this style.
The simply-typed approach can be expressed not only
in proof assistants based on dependent type theory,
such as \Coq, Lean, or Agda,
but also in proof assistants based on higher-order logic,
such as Isabelle/HOL.
This approach should allow obtaining executable code
that does not contain redundant computations of indices
and where each operation
therefore has worst-case time complexity $O(1)$.

% ------------------------------------------------------------------------------
% Bibliography.
% lmcs.cls sets this only when biblatex is loaded before the class.
% Loaded afterward, biblatex keeps the bibliography at body size.
\renewcommand*{\bibfont}{\normalfont\bibliofont}
\printbibliography

@String{acta   = "Acta Informatica"}

@String{cacm   = "Communications of the {ACM}"}

@String{computer = "Computer"}

@String{coqpl  = "Workshop on Coq for Programming Languages"}

@String{cpp    = "Certified Programs and Proofs (CPP)"}

@String{cup    = "Cambridge University Press"}

@String{fossacs = "Foundations of Software Science and Computation
                 Structures ({FOSSACS})"}

@String{icfp   = "International Conference on Functional Programming
                 (ICFP)"}

@String{ijcar  = "International Joint Conference on Automated
                 Reasoning"}

@String{ipl    = "Information Processing Letters"}

@String{jacm   = "Journal of the {ACM}"}

@String{jfla   = "Journées Françaises des Langages Applicatifs
                 (JFLA)"}

@String{jfp    = "Journal of Functional Programming"}

@String{lics   = "Logic in Computer Science (LICS)"}

@String{lncs   = "Lecture Notes in Computer Science"}

@String{mpc    = "Mathematics of Program Construction (MPC)"}

@String{msfp   = "Workshop on Mathematically Structured Functional
                 Programming (MSFP)"}

@String{pacmpl = "Proceedings of the ACM on Programming Languages"}

@String{pldi   = "{Programming Language Design and Implementation
                 (PLDI)}"}

@String{popl   = "Principles of Programming Languages ({POPL})"}

@String{scp    = "Science of Computer Programming"}

@String{siamjc = "SIAM Journal on Computing"}

@String{springer = "Springer"}

@String{stoc   = "Symposium on the Theory of Computing"}

@String{tapsoft = "Theory and Practice of Software Development
                 ({TAPSOFT})"}

@String{tcs    = "Theoretical Computer Science"}

@String{types  = "Types for Proofs and Programs"}

@String{wollic = "Workshop on Logic, Language, Information and
                 Computation (WoLLIC)"}

@InProceedings{abel-06,
  author       = "Andreas Abel",
  title        = "Towards Generic Programming with Sized Types",
  booktitle    = mpc,
  series       = lncs,
  volume       = "4014",
  pages        = "10--28",
  publisher    = springer,
  month        = jul,
  year         = "2006",
  URL          = "https://link.springer.com/chapter/10.1007/11783596_4",
}

@Article{abel-matthes-uustalu-05,
  author       = "Andreas Abel and Ralph Matthes and Tarmo Uustalu",
  title        = "Iteration and coiteration schemes for higher-order and
                 nested datatypes",
  journal      = tcs,
  volume       = "333",
  number       = "1--2",
  pages        = "3--66",
  year         = "2005",
  URL          = "https://www.cse.chalmers.se/~abela/tcs04.pdf",
}

@Article{allain-scherer-zoo-26,
  author       = "Clément Allain and Gabriel Scherer",
  title        = "{Zoo}: A framework for the verification of concurrent
                 {OCaml~5} programs using separation logic",
  year         = "2026",
  journal      = pacmpl,
  volume       = "10",
  number       = "{POPL}",
  month        = jan,
  URL          = "https://cambium.inria.fr/~scherer/research/multicore/allain-scherer-zoo-2026.pdf",
}

@InCollection{appel-vfa,
  author       = "Andrew W. Appel",
  title        = "Verified Functional Algorithms",
  booktitle    = "Software Foundations",
  volume       = "3",
  year         = "2025",
  shorthand    = "VFA",
  URL          = "https://softwarefoundations.cis.upenn.edu/vfa-current/index.html",
}

@InProceedings{bird-meertens-98,
  author       = "Richard Bird and Lambert Meertens",
  title        = "Nested Datatypes",
  booktitle    = mpc,
  pages        = "52--67",
  year         = "1998",
  volume       = "1422",
  series       = lncs,
  publisher    = springer,
  URL          = "http://www.cs.ox.ac.uk/richard.bird/online/BirdMeertens98Nested.pdf",
}

@InProceedings{blanchette-meyer-popescu-traytel-17,
  author       = "Jasmin Christian Blanchette and Fabian Meier and
                 Andrei Popescu and Dmitriy Traytel",
  title        = "Foundational nonuniform (Co)datatypes for higher-order
                 logic",
  booktitle    = lics,
  pages        = "1--12",
  month        = jun,
  year         = "2017",
  URL          = "https://matryoshka-project.github.io/pubs/nonuniform_paper.pdf",
}

@InProceedings{braibant-pous-11,
  author       = "Thomas Braibant and Damien Pous",
  title        = "Tactics for Reasoning Modulo {AC} in {Coq}",
  booktitle    = cpp,
  year         = "2011",
  pages        = "167--182",
  publisher    = springer,
  series       = lncs,
  volume       = "7086",
  URL          = "http://arxiv.org/abs/1106.4448",
}

@Article{chargueraud-pottier-26,
  author       = "Arthur Charguéraud and François Pottier",
  title        = "A Catenable, Splittable, Transient Sequence Data
                 Structure",
  journal      = pacmpl,
  year         = "2026",
  volume       = "10",
  number       = "{ICFP}",
  month        = aug,
  URL          = "https://cambium.inria.fr/~fpottier/publis/chargueraud-pottier-sek.pdf",
}

@InProceedings{czajka-20,
  author       = "Lukasz Czajka",
  title        = "Practical Proof Search for {Coq} by Type
                 Inhabitation",
  booktitle    = ijcar,
  series       = lncs,
  volume       = "12167",
  pages        = "28--57",
  publisher    = springer,
  month        = jul,
  year         = "2020",
  URL          = "https://www.mimuw.edu.pl/~lukaszcz/sauto.pdf",
}

@InProceedings{dagand-letouzey-taghayor-22,
  title        = "Rough Pearl: Manufacturing Cons-Cells",
  author       = "Pierre-Evariste Dagand and Pierre Letouzey and Ellenor
                 Fatemeh Taghayor",
  booktitle    = jfla,
  month        = jan,
  year         = "2024",
  URL          = "https://inria.hal.science/hal-04406422",
}

@Article{demaine-langerman-price-10,
  author       = "Erik D. Demaine and Stefan Langerman and Eric Price",
  title        = "Confluently Persistent Tries for Efficient Version
                 Control",
  journal      = "Algorithmica",
  volume       = "57",
  number       = "3",
  pages        = "462--483",
  year         = "2010",
  URL          = "https://erikdemaine.org/papers/ConfluentTries_Algorithmica/paper.pdf",
}

@Article{elmasry-katajainen-22,
  author       = "Amr Elmasry and Jyrki Katajainen",
  title        = "Regular numeral systems for data structures",
  journal      = acta,
  volume       = "59",
  number       = "2--3",
  pages        = "245--281",
  year         = "2022",
  doi          = {10.1007/s00236-021-00407-9},
}

@Misc{equations,
  author       = "{{The {Rocq} team}}",
  title        = "Documentation of {Equations}",
  URL          = "https://rocq-prover.org/docs/equations-docs",
  month        = jan,
  year         = "2026",
  shorthand    = "Equations",
  note         = "Online tutorials",
}

@InProceedings{fu-selinger-23,
  author       = "Peng Fu and Peter Selinger",
  title        = "Towards an Induction Principle for Nested Data Types",
  booktitle    = wollic,
  series       = lncs,
  volume       = "13923",
  pages        = "244--255",
  publisher    = springer,
  month        = jul,
  year         = "2023",
  URL          = "https://arxiv.org/pdf/2306.10124",
}

@Article{kaplan-okasaki-tarjan-00,
  author       = "Haim Kaplan and Chris Okasaki and Robert E. Tarjan",
  title        = "Simple Confluently Persistent Catenable Lists",
  journal      = siamjc,
  volume       = "30",
  number       = "3",
  pages        = "965--977",
  year         = "2000",
  URL          = "https://epubs.siam.org/doi/10.1137/S0097539798339430",
}

@InProceedings{kaplan-tarjan-95,
  author       = "Haim Kaplan and Robert Endre Tarjan",
  title        = "Persistent lists with catenation via recursive
                 slow-down",
  booktitle    = stoc,
  pages        = "93--102",
  month        = may,
  year         = "1995",
  doi          = {10.1145/225058.225090},
}

@Article{kaplan-tarjan-99,
  author       = "Haim Kaplan and Robert E. Tarjan",
  title        = "Purely functional, real-time deques with catenation",
  journal      = jacm,
  volume       = "46",
  number       = "5",
  year         = "1999",
  pages        = "577--603",
  URL          = "http://www.math.tau.ac.il/~haimk/adv-ds-2000/jacm-final.pdf",
}

@Article{lamiaux-forster-sozeau-tabareau-26,
  author       = "Thomas Lamiaux and Yannick Forster and Matthieu Sozeau
                 and Nicolas Tabareau",
  title        = "Nested Inductive Types: Justified and Usable Nested
                 Inductive Types in {Lean} and {Rocq}",
  year         = "2026",
  volume       = "10",
  number       = "PLDI",
  doi          = {10.1145/3808322},
  journal      = pacmpl,
  month        = jun,
}

@InProceedings{leijen-meijer-99,
  author       = "Daan Leijen and Erik Meijer",
  title        = "Domain-specific embedded compilers",
  booktitle    = "Conference on Domain-Specific Languages {(DSL)}",
  pages        = "109--122",
  month        = oct,
  year         = "1999",
  URL          = "https://www.usenix.org/legacy/publications/library/proceedings/dsl99/full_papers/leijen/leijen.pdf",
}

@InProceedings{leroy-recursion-24,
  author       = "Xavier Leroy",
  title        = "Well-founded recursion done right",
  booktitle    = "Workshop on Coq for Programming Languages {(CoqPL)}",
  month        = jan,
  year         = "2024",
  URL          = "https://xavierleroy.org/publi/wf-recursion.pdf",
}

@Article{matthes-11,
  author       = "Ralph Matthes",
  title        = "Map fusion for nested datatypes in intensional type
                 theory",
  journal      = scp,
  volume       = "76",
  number       = "3",
  pages        = "204--224",
  year         = "2011",
  doi          = {10.1016/j.scico.2010.05.008},
}

@Unpublished{mihaescu-tarjan-03,
  author       = "Radu Mihaescu and Robert E. Tarjan",
  title        = "Notes on Catenable Deques in Pure {Lisp}",
  note         = "Unpublished course notes, COS 528",
  year         = "2003",
  URL          = "https://www.cs.princeton.edu/courses/archive/fall03/cs528/handouts/Notes%20on%20Catenable%20Deques.doc",
}

@InProceedings{montin-ledein-dubois-22,
  author       = "Mathieu Montin and Amélie Ledein and Catherine
                 Dubois",
  title        = "{LibNDT}: Towards a Formal Library on Spreadable
                 Properties over Linked Nested Datatypes",
  booktitle    = msfp,
  series       = "EPTCS",
  volume       = "360",
  pages        = "27--44",
  month        = apr,
  year         = "2022",
  doi          = {10.4204/EPTCS.360.2},
}

@Article{myers-83,
  author       = "Eugene W. Myers",
  title        = "An Applicative Random-Access Stack",
  journal      = ipl,
  volume       = "17",
  number       = "5",
  pages        = "241--248",
  year         = "1983",
  doi          = {10.1016/0020-0190(83)90106-0},
}

@Misc{nipkow-fav,
  author       = "Tobias Nipkow and Jasmin Blanchette and Manuel Eberl
                 and Alejandro Gómez-Londoño and Peter Lammich and
                 Christian Sternagel and Simon Wimmer and Bohua Zhan",
  title        = "Functional Algorithms, Verified!",
  year         = "2021",
  month        = apr,
  URL          = "https://www21.in.tum.de/teaching/fds/SS21/assets/book-draft.pdf",
}

@Book{nipkow-fdsa,
  author       = "Tobias Nipkow",
  title        = "Functional Data Structures and Algorithms: A Proof
                 Assistant Approach",
  publisher    = "ACM Books",
  year         = "2025",
  month        = sep,
  URL          = "https://fdsa-book.net/functional_data_structures_algorithms.pdf",
}

@Misc{ocaml,
  author       = "Xavier Leroy and Damien Doligez and Alain Frisch and
                 Jacques Garrigue and Didier Rémy and Jérôme
                 Vouillon",
  title        = "The {OCaml} system, release 5.4",
  month        = oct,
  year         = "2025",
  shorthand    = "OCaml",
  URL          = "https://ocaml.org/manual/",
}

@Article{okasaki-95,
  author       = "Chris Okasaki",
  title        = "Simple and Efficient Purely Functional Queues and
                 Deques",
  journal      = jfp,
  volume       = "5",
  number       = "4",
  pages        = "583--592",
  year         = "1995",
  doi          = {10.1017/S0956796800001489},
}

@InProceedings{okasaki-97,
  author       = "Chris Okasaki",
  title        = "Catenable Double-Ended Queues",
  booktitle    = icfp,
  pages        = "66--74",
  month        = jun,
  year         = "1997",
  doi          = {10.1145/258948.258956},
}

@Book{okasaki-book-99,
  author       = "Chris Okasaki",
  title        = "Purely Functional Data Structures",
  publisher    = cup,
  year         = "1999",
  doi          = {10.1017/CBO9780511530104},
}

@Manual{omega,
  title        = "{${\Omega}$}mega",
  author       = "Tim Sheard",
  month        = nov,
  year         = "2005",
  URL          = "http://www.cs.pdx.edu/~sheard/Omega/",
}

@InProceedings{ponsonnet-pottier-26,
  author       = "Juliette Ponsonnet and François Pottier",
  title        = "Verified Persistent Catenable Deques",
  booktitle    = jfla,
  month        = jan,
  year         = "2026",
  URL          = "https://cambium.inria.fr/~fpottier/publis/ponsonnet-pottier-kot.pdf",
}

@InProceedings{pottier-monolith-21,
  author       = "François Pottier",
  title        = "Strong Automated Testing of {OCaml} Libraries",
  booktitle    = jfla,
  month        = feb,
  year         = "2021",
  URL          = "http://cambium.inria.fr/~fpottier/publis/pottier-monolith-2021.pdf",
}

@Book{real-world-ocaml,
  author       = "Yaron Minsky and Anil Madhavapeddy and Jason Hickey",
  title        = "Real World {OCaml}: Functional programming for the
                 masses, second edition",
  publisher    = cup,
  month        = oct,
  year         = "2022",
  URL          = "https://realworldocaml.org/",
}

@InProceedings{reynolds-85,
  author       = "John C. Reynolds",
  title        = "Three Approaches to Type Structure",
  booktitle    = tapsoft,
  month        = mar,
  series       = lncs,
  volume       = "185",
  publisher    = springer,
  year         = "1985",
  pages        = "97--138",
  doi          = {10.1007/3-540-15198-2_7},
}

@Manual{rocq,
  author       = "{{The {Rocq} team}}",
  title        = "The {Rocq} Prover",
  year         = "2026",
  shorthand    = "Rocq",
  URL          = "http://rocq-prover.org/",
}

@TechReport{schaub-23,
  author       = "Fabian Schaub",
  title        = "Purely Functional Real-Time Deques",
  institution  = "Universität Innsbruck",
  year         = "2023",
  type         = "Seminar Report",
  URL          = "http://cl-informatik.uibk.ac.at/teaching/ws22/rs/slides/reportFS.pdf",
}

@Article{seassau-yoon-madiot-pottier-25,
  author       = "Remy Seassau and Irene Yoon and Jean-Marie Madiot and
                 François Pottier",
  title        = "Formal Semantics and Program Logics for a Fragment of
                 {OCaml}",
  journal      = pacmpl,
  volume       = "9",
  number       = "{ICFP}",
  articleno    = "240",
  month        = aug,
  year         = "2025",
  URL          = "http://cambium.inria.fr/~fpottier/publis/seassau-yoon-madiot-pottier-osiris-2025.pdf",
}

@InProceedings{sozeau-finger-07,
  author       = "Matthieu Sozeau",
  booktitle    = icfp,
  pages        = "13--24",
  URL          = "http://mattam.org/research/publications/Program-ing_Finger_Trees_in_Coq.pdf",
  title        = "Program-ing Finger Trees in {Coq}",
  month        = sep,
  year         = "2007",
}

@InProceedings{sozeau-mangin-19,
  title        = "Equations reloaded: high-level de\-pen\-dent\-ly-typed
                 functional programming and proving in {Coq}",
  author       = "Matthieu Sozeau and Cyprien Mangin",
  booktitle    = pacmpl,
  year         = "2019",
  pages        = "1--29",
  volume       = "3",
  URL          = "https://sozeau.gitlabpages.inria.fr/www/research/publications/Equations_Reloaded-ICFP19.pdf",
}

@Article{vuillemin-78,
  author       = "Jean Vuillemin",
  title        = "A Data Structure for Manipulating Priority Queues",
  journal      = cacm,
  volume       = "21",
  number       = "4",
  pages        = "309--315",
  year         = "1978",
  doi          = {10.1145/359460.359478},
}

@Article{winterhalter-24,
  author       = "Théo Winterhalter",
  title        = "Dependent Ghosts Have a Reflection for Free",
  journal      = pacmpl,
  volume       = "8",
  number       = "{ICFP}",
  pages        = "630--658",
  year         = "2024",
  doi          = {10.1145/3674647},
}

@InProceedings{xi-chen-chen-03,
  author       = "Hongwei Xi and Chiyan Chen and Gang Chen",
  title        = "Guarded Recursive Datatype Constructors",
  booktitle    = popl,
  year         = "2003",
  month        = jan,
  pages        = "224--235",
  URL          = "https://hwxi.github.io/PUBLICATION/MYDATA/GRDT-popl03.pdf",
}

@TechReport{clancy-knuth-77,
  title       = "A programming and problem-solving seminar",
  author      = "Michael J. Clancy and Donald E. Knuth",
  institution = "Stanford University, Department of Computer Science",
  type        = "Seminar report",
  year        = "1977",
  URL         = "http://i.stanford.edu/pub/cstr/reports/cs/tr/77/606/CS-TR-77-606.pdf"
}

@TechReport{hinze-tr-98,
  author =       {Ralf Hinze},
  title =        {Numerical Representations as Higher-Order Nested Datatypes},
  institution =  {Institut für Informatik III, Universität Bonn},
  year =         1998,
  type =      {Technical Report},
  number =    {IAI-TR-98-12},
  month =     dec,
  URL = {https://www.cs.ox.ac.uk/people/ralf.hinze/publications/IAI-TR-98-12.ps.gz},
}

@article{driscoll-sleator-tarjan-94,
  author       = {James R. Driscoll and
                  Daniel Dominic Sleator and
                  Robert Endre Tarjan},
  title        = {Fully Persistent Lists with Catenation},
  journal      = jacm,
  volume       = {41},
  number       = {5},
  pages        = {943--959},
  year         = {1994},
  doi          = {10.1145/185675.185791},
}

@article{buchsbaum-tarjan-95,
  author       = {Adam L. Buchsbaum and
                  Robert Endre Tarjan},
  title        = {Confluently Persistent Deques via Data-Structural Bootstrapping},
  journal      = {Journal of Algorithms},
  volume       = {18},
  number       = {3},
  pages        = {513--547},
  year         = {1995},
  doi          = {10.1006/jagm.1995.1020},
}

@inproceedings{johann-polonsky-20,
  author       = {Patricia Johann and
                  Andrew Polonsky},
  title        = {Deep Induction: Induction Rules for (Truly) Nested Types},
  booktitle    = fossacs,
  series       = lncs,
  volume       = {12077},
  pages        = {339--358},
  publisher    = springer,
  month        = apr,
  year         = {2020},
  URL          = {https://libres.uncg.edu/ir/asu/f/Johann_Patricia_2020_Deep%20Induction_fossacs.pdf},
}

\end{document}